\documentclass[12pt,a4paper,oneside]{article}
\usepackage{graphicx,sectsty,amssymb,amsmath}
\usepackage{cite} 
\usepackage{setspace} 
\usepackage[linkcolor={blue},citecolor={blue},colorlinks=true,linktocpage,urlcolor={blue}]
{hyperref}
\usepackage[margin=2cm]{geometry}
\usepackage{multirow}

\usepackage[page,toc]{appendix}
\usepackage{amsmath}
\usepackage{tensor}
\usepackage{etoolbox}
\AtBeginEnvironment{align}{\setcounter{subeqn}{0}}
\newcounter{subeqn} %

\usepackage{pdfpages}
\numberwithin{equation}{section}
\usepackage{cancel}
\usepackage{subcaption}
\usepackage{xparse} 
\usepackage{empheq} 

\usepackage[normalem]{ulem} 
\usepackage{titling} 
\usepackage[affil-it]{authblk} 

\usepackage{enumitem} 
\usepackage{tocloft} 
\usepackage[utf8]{inputenc} 

\newcommand{\cO}{\mathcal{O}}
\newcommand{\pder}[2]{\frac{\partial#1}{\partial#2}} 

\renewcommand\Re{\operatorname{Re}}
\renewcommand\Im{\operatorname{Im}}
\DeclareDocumentCommand \at { o m }
{
           \IfNoValueTF {#1}
             {\big |_{#2}}
             {\left.#1\right|_{#2}}
}

\newcount\colveccount
\newcommand*\colvec[1]{
        \global\colveccount#1
        \begin{pmatrix}
        \colvecnext
}
\def\colvecnext#1{
        #1
        \global\advance\colveccount-1
        \ifnum\colveccount>0
                \\
                \expandafter\colvecnext
        \else
                \end{pmatrix}
        \fi
}

\newcommand{\vecx}{ \textbf{x} }
\newcommand{\vecy}{ \textbf{y} }
\newcommand{\vecz}{ \textbf{z} }

\newcommand{\vecp}{ \textbf{p} }

\newcommand{\mean}{\overline}
\newcommand{\dcft}{\bar d}
\newcommand{\qd}{v}

\newcommand{\cE}{\mathcal{E}}

\usepackage{feynmp} 
\usepackage{feynmp-auto}
\newcommand{\marrow}[5]{
    \fmfcmd{style_def marrow#1
    expr p = drawarrow subpath (0.4, 0.6) of p shifted 6 #2 withpen pencircle scaled 0.4; 
    label.#3(btex #4 etex, point 0.5 of p shifted 6 #2);
    enddef;}
    \fmf{marrow#1,tension=0}{#5}}
    
\title{The renormalization group flow in field theories with quenched disorder}
\author{Ofer Aharony
     \thanks{E-mail: \texttt{ofer.aharony@weizmann.ac.il}} }
\author{Vladimir Narovlansky
     \thanks{E-mail: \texttt{vladimir.narovlansky@weizmann.ac.il}} }
\affil{Department of Particle Physics and Astrophysics, \\ Weizmann Institute of Science, Rehovot 7610001, Israel}
\date{}

\makeatletter 
\let\@fnsymbol\@arabic
\makeatother

\begin{document}

\begin{titlingpage}
    \maketitle
    \begin{abstract}
    
    In this paper we analyze the renormalization group (RG) flow of field theories with quenched disorder, in which the couplings vary randomly in space. We analyze both classical (Euclidean) disorder and quantum disorder, emphasizing general properties rather than specific cases. The RG flow of the disorder-averaged theories takes place in the space of their coupling constants and also in the space of distributions for the disordered couplings, and the two mix together. We write down a generalization of the Callan-Symanzik equation for the flow of disorder-averaged correlation functions. We find that local operators can mix with the response of the theory to local changes in the disorder distribution, and that the generalized Callan-Symanzik equation mixes the disorder averages of several different correlation functions. For classical disorder we show that this can lead to new types of anomalous dimensions and to logarithmic behavior at fixed points. For quantum disorder we find that the RG flow always generates a rescaling of time relative to space, which at a fixed point generically leads to Lifshitz scaling. The dynamical scaling exponent $z$ behaves as an anomalous dimension (as in other non-relativistic RG flows), and we compute it at leading order in perturbation theory in the disorder for a general theory.
Our results agree with a previous perturbative computation by Boyanovsky and Cardy, and with a holographic disorder computation of Hartnoll and Santos. We also find in quantum disorder that local operators mix with non-local (in time) operators under the RG, and that there are critical exponents associated with the disorder distribution that have not previously been discussed.
In large $N$ theories the disorder averages may be computed exactly, and we verify that they are consistent with the generalized Callan-Symanzik equations.
    
    \end{abstract}
\end{titlingpage}

\singlespacing
\tableofcontents
\onehalfspacing

\section{Introduction}

Quenched disorder is of widespread interest, studied in many branches of physics including statistical physics, condensed matter and theoretical high-energy physics. 
A physical motivation for studying disorder comes from the fact that real systems are not pure.\footnote{A nice review of quenched disorder in classical statistical systems appears in chapter 8 of \cite{Cardy:1996xt}. } 
We expect that always, in addition to the basic homogeneous elementary matter and its interactions, there will also be impurities, or non-constant background fields, which will modify the microscopic interactions within the substance. In some cases these impurities may be treated as non-dynamical; this is called quenched disorder, and we will focus on this case here (as opposed to annealed disorder in which the impurities are dynamical). The impurities then correspond to changes in the local couplings of the system, and they are equivalent to non-homogeneous background fields. In general, in such a situation, all couplings which are allowed by the symmetries will vary in space. In many cases the impurities (or background fields) are random, so that they can be effectively described by randomly varying couplings in our original homogeneous system, with some probability distribution for finding specific space-dependent couplings. We will be interested in the behavior of these systems at long distances, and in particular at much larger distances than the scale of variation of the couplings (which is typically a microscopic scale like the lattice spacing). Thus we can approximately take the scale of this variation to zero, such that the couplings at different points vary randomly and independently.


There are various systems for which this idea applies. We can consider a statistical mechanics system near a second order phase transition, which is described by a Euclidean theory in the spatial directions; this is referred to as classical disorder. A specific example involves random variations in the temperature, where disorder couples to the `energy operator' $\mathcal{E}(x)$.
We will also consider quenched disorder in a quantum system (at zero temperature), which has also a time direction, and this is referred to as quantum disorder. In particular one can look for random quantum critical points. We will not discuss here disordered quantum systems at finite temperature, except for a few comments in section \ref{subsubsection:Lifshitz_non_connected_correlators}.

Note that even a small amount of disorder can lead to significant changes in the long-distance behavior. For instance, in a statistical mechanics system for which the disorder generates a relevant operator (in the renormalization group sense), it grows with the distance scale, and can lead to a flow to a different, random, fixed point at long distances, or to no fixed point at all, such that the second order phase transition disappears (see \cite{Domb:1976bk} and references therein).

In order to describe the different versions of disorder under the same framework, the basic quantity we will use is the action. For classical disorder, this action will be Euclidean and will stand for the Hamiltonian (more precisely the reduced Hamiltonian which includes the inverse temperature). For quantum disorder this action will be Lorentzian. However, to have a uniform description we will analytically continue it to Euclidean space. Our basic setup is therefore the following. We are given some `pure' system described by an action $S_0$. We can begin by considering the simplest case where the disorder affects a single coupling constant, namely it couples to a single scalar operator; in most cases the lowest dimension operator has the largest effect on the long distance physics, and the generalization of this (which will be generated under the renormalization group flow) is straightforward. The disorder field will be denoted by $h(x)$ (which can be, for example, the distribution of impurities, or a background magnetic field), and it will couple to some interaction term $\cO _0(x)$. Because of the correspondence between classical statistical mechanics and quantum mechanics (through the path integral formalism), we will refer to $\cO _0(x)$ as a local operator in all cases. The total action is thus schematically of the form
\begin{equation}
S= S_0 + \int h(x) \cdot  \cO _0(x) .
\end{equation}
For quantum disorder $h(x)$ will vary in space but not in time.
The partition function is as usual
\begin{equation}
Z[h(x)] = \int D\mu \, e^{-S} 
\end{equation}
where $D\mu $ stands for the appropriate path integral measure. 

We will treat the process of formation of the disorder as a random process, with probability distribution $P[h(x)]$ to have a specific disorder configuration $h(x)$. Namely, we will consider an ensemble of many different systems, whose disorder is drawn from the distribution with the appropriate probability. We can then compute the distributions for various physical measurements, such as thermodynamic quantities or correlation functions of local operators. In some systems, called self-averaging, the distributions of global observables become narrower as the system becomes large, such that all the systems in the ensemble have the same long-distance properties (depending on the probability distribution). Other systems are not self-averaging, and in particular this is the case for some observables when the system flows to a random fixed point of the renormalization group, which is scale-invariant \cite{wiseman1995lack,aharony1996absence,wiseman1998finite}.
Such a fixed point is characterized by some probability distribution for the disorder, and also for any observable. The properties of the fixed point cannot be measured using a single realization of the disorder, whose observables will be taken from that distribution, but only by measuring many different systems with different realizations of the disorder. 
If we look at a specific disorder realization, it will have a phase transition at a critical temperature, but its properties at the critical temperature at long distances will not be translation-invariant or scale-invariant. When we sample different subregions or different scales, we will obtain different results, that are all drawn from the same probability distribution characterizing the random fixed point.
As practically we never sit exactly at a critical point, self-averaging depends on the size of the system; that is, we will have self-averaging whenever the linear size of the system is much larger than the correlation length.

The explicit dependence of the action on the disorder $h(x)$ implies that in both classical and quantum disorder, translation invariance is broken. However, assuming that the disorder distribution is invariant under translations, translation invariance will be restored in the averages (and higher statistical moments) of any physical observables.
In particular we can consider the free energy averaged over the disorder distribution $P[h]$, which restores the translation invariance. The disorder averages of all thermodynamic quantities which are derivatives of the free energy can be computed from derivatives of the disorder-averaged free energy. 

In this paper we are interested in the renormalization group flow of various quantities, averaged over the disorder distribution. More generally, we are interested in the probability distribution of these quantities. Disordered systems, and in some cases their renormalization group (RG) flow, were extensively analyzed in the statistical mechanics and condensed matter literature before, but usually for specific models;\footnote{A general analysis of the expansion around a pure fixed point is given in \cite{shapir1981rigorous}.} here we will analyze completely general field theories, giving a few specific examples to illustrate our results.
We focus on a general discussion of the RG flow of correlation functions of local operators, and of their behavior at random fixed points (one particular case was studied in \cite{pelcovits1985structure}). For pure systems this is governed by the Callan-Symanzik (CS) equation, and we are interested in the generalization of this equation to averaged quantities of random systems (this was previously done in a specific example in \cite{Dotsenko:1997wf}).

One of the main tools we use in our analysis is the replica trick \cite{deGennes:1972zz,Emery:1975zz,Grinstein:1976zz,Ma:unpublished}. For classical disorder this describes averaged disordered systems as limits of standard `pure' systems, whose renormalization group behavior is well-understood; formally any disordered computation can be viewed in this way as a limit of standard computations. For quantum disorder the disorder-averaged quantities are related to a limit of non-local field theories, but we argue that one can still use renormalization group methods. In some cases taking limits of correlation functions may be subtle for dynamical reasons (such as replica symmetry breaking). However, since we will be interested in general features of the renormalization group flow of couplings and of local operators, that are independent of the specific dynamics, the replica analysis will be good enough for our purposes; in particular we show that it gives results that are consistent with a standard Wilsonian renormalization of the disordered theories. Since we use such general methods, we will not be able to say what the flow leads to in any specific system. However, whenever a random fixed point exists, our analysis shows what it will behave like.

\subsection{Summary of main new results, outline and open questions}

Since this paper is rather long, and contains some sections reviewing known results, let us summarize the main new results of this paper.

In \autoref{section:RG_classical_disorder} we analyze classical disorder:
\begin{itemize}
\item We write down the generalized Callan-Symanzik (GCS) equations governing the RG evolution of disorder-averaged correlation functions of local operators for classical disordered systems. For connected correlation functions we show that there are new contributions mixing these correlators with products of connected correlators, and that these lead to new types of anomalous dimensions at disordered fixed points. Correlators of local operators do not have simple scaling behavior  even at random fixed points.
This can be traced to the fact that during the RG flow in random systems, the distribution of the disordered couplings changes and mixes with the standard coupling constants, which implies that the local operators do not diagonalize the matrix of anomalous dimensions.

\item In addition, for general (not necessarily connected) correlation functions $\mean{\langle {\cal O}_1(x_1) \cdots {\cal O}_k(x_k) \rangle}$ we find an extra ``mixing'' contribution to the CS equation. For two-point functions at disordered fixed points this new contribution gives a logarithmic behavior previously found in \cite{Gurarie:1999yx,Gurarie:2004ce,Cardy:2013rqg}, and we analyze its implications. We show that, surprisingly, some of the long-distance observables related to such two-point functions become less and less self-averaging as the volume grows (namely, their normalized variance grows logarithmically with the volume), such that they are not (approximately) self-averaging even when the disorder at the fixed point is small.
\end{itemize}
At the end of this section, we relate our results to those in the statistical mechanics literature.

In \autoref{section:classical_disorder_large_N} we discuss the specific example of large $N$ systems. In this case one can perform explicit computations that can be used to illustrate our general methods, but it turns out that this case is actually more subtle than the generic case, because of degeneracies in the large $N$ spectrum of operators.

Then, we move on to quantum disorder in \autoref{section:RG_quantum_disorder}, with the following main results:
\begin{itemize}
\item We write down the same GCS equations also for systems of quantum disorder, noting the new contributions there. A new element that arises in this case is a mixing of local operators with non-local operators under the RG flow; for instance an operator ${\cal O}_1(\vecx,t)$ can mix with ${\cal O}_2(\vecx,t) \int dt'\, {\cal O}_3(\vecx,t')$.

\item It is sometimes stated that at disordered quantum critical points there is no independent critical exponent associated with the disorder (and with the running of the disorder distribution), unlike in the case of classical disorder (where this exponent is usually called $\phi$). We show that there is in fact such a critical exponent also at quantum critical points, and illustrate this explicitly in an example.

\item We show that renormalization group flows for quantum disorder always generate a coupling $h_{00} \int d\vecx dt\, T_{00}(\vecx,t)$, where $T_{\mu \nu}$ is the energy-momentum tensor.
That is, the coefficient of the Hamiltonian $h_{00} \int dt\, H$ in the action is running.
This can be interpreted as a stretching of the time dimension compared to the spatial dimensions. We show that the beta function of the coupling $h_{00}$ is precisely the deviation of the dynamical scaling exponent $z$ from $1$, such that these theories generally flow to Lifshitz-type fixed points, with a different scaling for the time and the space dimensions.\footnote{These statements are valid also in non-relativistic RG flows in pure theories, and some of them were discussed before in that context (see, for example, \cite{Korovin:2013bua}). However, the arguments used there do not directly apply for disordered theories.}

\item In the case of a weakly disordered fixed point, obtained by RG flow from a Lorentz-invariant conformal field theory, we show that this dynamical exponent $z$ has a universal value at leading order in the disorder, given by the formula \eqref{eq:Lifshitz_z_exp2}. We compare this result with two previous computations : a weakly coupled scalar theory analyzed by Boyanovsky and Cardy \cite{Boyanovsky:1982zz} and a holographic model analyzed by Hartnoll and Santos \cite{Hartnoll:2014cua}.
\end{itemize}
The results of this section are summarized in the companion letter \cite{letterversion}.

We end with various appendices containing technical details.

There are many interesting questions that we leave open. In our replica analysis we assume that the replica symmetry is not broken; it would be interesting to understand how such breaking modifies the RG flow of correlation functions. We discuss only disorder coupled to scalar operators, and it would be interesting to generalize our analysis to anisotropic disorder, and to consider in such cases correlation functions of non-scalar operators, which may be related to additional critical exponents. Similarly one can have anisotropies in internal symmetries. We focus on a general analysis and do not discuss any specific examples; it would be interesting to use our analysis for specific systems, and to see if our results have any measurable implications. On the more theoretical level, we analyze some properties of disordered fixed points when these exist, but it is not clear if these fixed points have conformal symmetry or only scaling symmetry (in fact, as mentioned above, even scaling symmetry is sometimes broken by logarithms). Related to this it would be nice to understand if disordered renormalization group flows obey any generalization of a $c$-theorem that would constrain the flow (see e.g \cite{Gurarie:1999bp}).

\section{The renormalization group for classical disorder} \label{section:RG_classical_disorder}

In this section we consider the quenched disorder version corresponding to classical statistical mechanics. The setup and the notations we will use are as follows. We have a statistical system in $d$ spatial Euclidean dimensions (with no time direction) described by the action $S_0$ which stands for the reduced Hamiltonian. We will restrict ourselves to the most interesting case in which the $d$-dimensional field theory described by $S_0$ is a conformal field theory (CFT), corresponding to a critical statistical system describing physics at a second order phase transition.  Applying quenched disorder to this system,
the disorder field is denoted by $h(x)$ and it is a function of the $d$-dimensional Euclidean space parametrized by $x$. The disorder is coupled to an operator $ \cO_0$ such that the action is
\begin{equation} \label{eq:RGnq_coupling_disorder_action}
S= S_0 + \int d^dx\,  h(x) \cO _0(x) .
\end{equation}
We assume for simplicity that disorder couples to a single scalar field, though, as we will see, this can change under the renormalization group flow. Usually one operator coupled to disorder will dominate the long distance behavior.

We will think of the disorder field as taken from a probability distribution $P[h]$ (which is a functional of $h(x)$), and in this paper we will be interested in disorder-averaged quantities. In some cases (called self-averaging) these will be typical values for large-volume systems (the thermodynamic limit), while in other cases, such as disordered fixed points with strong randomness, the variances may be large, and measuring the averages requires an ensemble of many systems with different values of the disorder. We denote disorder averages by
\begin{equation}
\mean{X}\equiv \int Dh\, P[h] X .
\end{equation}
A distribution which is frequently used is a Gaussian probability distribution in which
\begin{equation} \label{gaussian}
P[h] \propto \exp \left(- \frac{1}{2\qd} \int d^dx\, h^2(x)\right),
\end{equation}
normalized so that the sum of probabilities is $1$.
For this distribution
\begin{equation}
\mean{h(x)h(y)}= \qd \delta (x-y),
\end{equation}
and $\mean{h(x_1)\cdots h(x_k)}$ are given by Wick's theorem as the sum over contractions with $\mean{h(x) h(y)}$.
%
We will be interested in the disorder-averaged correlation functions defined by
\begin{equation}
\mean{\langle \cO _1(x_1) \cO _2(x_2) \cdots \rangle}= \int Dh\, P[h] \, \frac{\int D\mu \, \cO _1(x_1) \cO _2(x_2) \cdots e^{-S_0-\int d^dx\,  h(x) \cO _0(x)} }{\int D\mu \, e^{-S_0 - \int d^dx\,  h(x) \cO _0(x)} }  .
\end{equation}

Applying quenched disorder to the critical system described by $S_0$ can be thought of as a perturbation of the system, similar to adding some interaction. A well-known argument in the context of the renormalization group (RG) in disordered systems is the Harris criterion\cite{Harris:1974zz}. If the dimension of $\cO_0$ is $\Delta _0$, then the width of the Gaussian disorder has dimensions $[\qd]=d-2\Delta _0$. Then, by dimensional arguments we expect that for $\Delta _0 < d/2$ disorder will be relevant, for $\Delta _0>d/2$ it will be irrelevant, and for $\Delta_0 =d/2$ disorder is marginal. When we are close to the marginal case we can often use perturbation theory in the strength of the disorder and in the difference between its dimension and the marginal value (as in the $\epsilon$ expansion, see e.g. \cite{Komargodski:2016auf}), but we will not assume this and our final results for the RG flow will be general.

Our goal is to study the renormalization group for classical disorder. In pure systems, the RG provides us with a description of the theory at different scales.
The coupling constants $\lambda _i$ are thought of as running with the RG scale $M$ (which for Wilsonian RG is the defining scale, while for renormalized RG it is the renormalization scale). The basic RG coefficients are the beta functions $\beta _{\lambda _i} (\lambda_j)$ 
that represent the running of $\lambda _i$, and the gamma functions $\gamma $ that give anomalous dimensions. It is desirable to describe the RG flow of correlation functions of local operators as these appear in various physical observables. This is achieved for pure systems using the Callan-Symanzik (CS) equation\cite{Callan:1970yg,Symanzik:1970rt}. 
In the simplest form of the CS equation, a correlation function $G^{(k)} \equiv \langle \cO(x_1) \cdots \cO(x_k) \rangle$ of $k$ operators $\cO(x)$ (for instance, in a scalar theory we can choose $\cO (x)= \varphi(x)$) satisfies
\begin{equation} \label{eq:RGnq_CS_eq}
\left( M \pder{}{M} + \sum _i \beta _{\lambda _i} \pder{}{\lambda _i} + k \gamma \right) G^{(k)} = 0,
\end{equation}
including a sum over all the couplings $\lambda _i$ in the theory, where $\gamma $ is the anomalous dimension of the operator.

A natural question is what is the generalization of this to disordered systems. This is non-trivial since in the presence of disorder the couplings become space-dependent.
In addition, it was noticed in \cite{Aharony:2015aea} that some disorder-averaged correlation functions do not appear to satisfy standard CS equations \eqref{eq:RGnq_CS_eq}. We could then wonder whether there exists a simple way to express the RG for disordered systems. We will find that in fact there are generalized CS equations that are obeyed by the different types of disorder-averaged correlation functions. One way in which disorder manifests itself in these equations is that the disorder strength (or, more generally, the parameters of the disorder distribution) enters the equations as if it was a new coupling constant, to which we may associate a beta function. However, this is not the full story, and the form of the equations is modified compared to those of pure systems. In this section we will derive the generalized Callan-Symanzik (GCS) equations in a general setting, and in the next section we apply them to a simple example.

Disordered theories may flow at low energies to standard fixed points, or to disordered fixed points characterized by some non-trivial disorder distribution.
These fixed points are known to obey unusual properties. Pure fixed points are described by conformal field theories, where 2-point functions at large distances have a power law behavior. However, in disordered fixed points logarithms can appear in disorder-averaged 2-point functions \cite{Cardy:2013rqg,Gurarie:1999yx,Gurarie:2004ce}, as in logarithmic CFTs (see e.g. \cite{Gurarie:1993xq,Hogervorst:2016itc}). There are additional asymptotic behaviours of correlation functions that were computed in \cite{Aharony:2015aea}, which appear to be incompatible with usual scale invariance. We will use the RG of disordered systems to approach the fixed points and to study them. We will see where and how logarithms can appear in this language, and we will see that other 
non-standard behaviors of disordered systems at large distances are also compatible with the GCS equations.

There are two main methods that we will use. We begin by discussing the local Wilsonian renormalization approach. This approach is very physical, but it is complicated to perform computations with it, so we will use it just to gain physical intuition about the expected form of the renormalization of the couplings and of the local operators. We will then discuss the replica approach, which allows for a more precise general analysis of the possibilities for renormalization group flow. Our GCS equations are written directly in terms of the disordered theory, so they do not depend on the replica approach.

\subsection{The local renormalization group and the flow of the disorder distribution} \label{subsection:local_RG_approach}

In this subsection we use the local renormalization group to directly analyze the space-dependent couplings and their flow.
The averaged correlation functions are given by
\begin{equation}
\mean{\langle \cO (x_1) \cdots \cO (x_k) \rangle} = \int Dh\, P[h] \langle \cO (x_1) \cdots \cO (x_k)\rangle_{h(x),g _i} \,,
\end{equation}
where the subscripts on the right-hand side signify that we calculate the correlation function in the presence of the disorder configuration $h(x)$ and some other couplings $g _i$.
For a given disorder configuration, using Wilsonian RG we can change the cutoff $\Lambda $ in the theory, and the physics will be the same if the couplings appropriately run with the scale $h(x,\Lambda )$, $g_i(\Lambda )$. Under the RG, some of the couplings that were constant will become inhomogeneous, so we need to take $g_i(x,\Lambda)$, and we have an RG flow with inhomogeneous couplings which is referred to as local RG\cite{Osborn:1991gm}, and is demonstrated perturbatively in \autoref{section:local_RG_demo}. The different disorder configurations were originally endowed with some probability distribution, and as the disorder field changes under the RG, the distribution changes accordingly (namely, its moments will be given by those of the disorder configurations at the new scale after the flow), such that the disorder-averaged physics is the same. Because of the mixing between couplings under the RG, statistical correlations among the different inhomogeneous couplings will be induced. Therefore what we get is a set of inhomogeneous couplings,
including all couplings that are consistent with the symmetries of the disordered theory,
and a joint probability distribution on them that flows under the RG:
\begin{equation}
\mean{\langle \cO (x_1) \cdots \cO (x_k) \rangle}= \int Dh(\Lambda ) Dg_i(\Lambda ) P[h(x,\Lambda ),g_i(x,\Lambda ),\Lambda ] \langle \cO (x_1) \cdots \cO (x_k) \rangle ^{(\Lambda )} _{h(x,\Lambda ),g_i(x,\Lambda )}  .
\end{equation}
The description of the disordered theory at different scales is then given by the joint probability distribution $P[h,g_i,\Lambda ]$ at different scales. Note that the average values of different couplings also flow, such that the constant modes of the couplings mix with the disorder distribution.
This picture of the flowing probability distribution was used long ago for local disorder using block spin transformations (see e.g. \cite{harris1974renormalization,lubensky1975critical,andelman1984scale}, and see \cite{andelman1985critical} for the quantum disorder case). Computations using this local RG method are technically complicated, but should lead to the same results that we compute below using the replica trick (as we verify in simple cases).


As usual in RG, the form of the running distribution is restricted by symmetries. For instance, if we describe the Ising model in terms of a $\varphi ^4$ field theory, then in the random-bond Ising model in which disorder is coupled to $\varphi ^2$, no coupling to $\varphi$ (constant or varying) will be induced under the RG (this is clear e.g.\ from \eqref{eq:local_RG_demo_couplings_flow}). 
In the random-field Ising model, where disorder is coupled to $\varphi $, if we start from a distribution invariant under $h(x) \to -h(x)$ then this will be preserved under the RG flow.

\subsubsection{Operator mixings in the local RG}

In a standard renormalization group flow, the coupling constants run and the local operators mix with each other. If, under an RG step, an operator $\cO_i(x)$ mixes with $\cO_j(x)$, then the CS equation for the flow of the correlation function $\langle \cO_i(x) \cdots \rangle$ includes a term proportional to $\langle \cO_j(x) \cdots \rangle$, with a coefficient depending on the coupling constants at our RG scale $M$. 

In a disordered theory, such operator mixings are not uniform but depend on the local couplings around the point $x$, so the coefficient of the term with $\cO_j$ above will depend on these local couplings.
Therefore for a specific realization of the disorder, we have usual CS equations, with the only modification that the RG coefficients depend locally on the inhomogeneous couplings.
When we average over the disorder, we will have some contributions in which we separately average over the coefficients and over the correlation functions, and these contributions to the running of a disorder-averaged correlation function will be the same as in the standard CS equation. However, there are also new contributions that will appear when the disorder averaging mixes the correlation functions with the mixing coefficients.

Let us analyze this here in the simplest case, in which we have a single disordered coupling $h(x)$ chosen from a Gaussian distribution \eqref{gaussian}, and we expand at leading order in the disorder. Later we will see the same effects in the exact theory using the replica approach. In our approximation, in addition to a constant mixing of $\cO_i(x)$ with $\cO_j(x)$, we can have a mixing with $h(x) \cO_j(x)$. Before the disorder averaging, this simply mixes $\langle \cO_i(x) \cdots \rangle$ with $h(x) \langle \cO_j(x) \cdots \rangle$. However, when we average over the disorder, 
this means that $\mean{\langle \cO_i(x) \cdots \rangle}$ mixes with
\begin{equation} \label{local_mixing}
\begin{split}
\int Dh\, P[h] h(x) & \langle \cO_j(x) \cdots \rangle_{h(x)} = \int Dh\, (-\qd {\delta  \over {\delta  h(x)}} P[h])  \langle \cO_j(x) \cdots \rangle_{h(x)} = \cr
& \qd \int Dh\, P[h] {\delta  \over {\delta  h(x)}} \langle \cO_j(x) \cdots \rangle_{h(x)} = - \qd \left( \mean{ \langle \cO_0(x) \cO_j(x) \cdots \rangle} - \mean{ \langle \cO_0(x) \rangle \langle \cO_j(x) \cdots \rangle} \right).
\end{split}
\end{equation}
In the first term we can replace the product of two operators at the same point, using the OPE, by a set of other local operators, so this looks like a standard operator mixing. However, the second term gives a new type of mixing which is not present in standard field theories.

If instead of starting with the general correlation function we begin with the connected correlation function $\mean{\langle \cO_i(x) \cdots \rangle_{conn}}$, then the same arguments imply that it mixes with the connected correlation function
$\mean{\langle \cO_0(x) \cO_j(x) \cdots \rangle_{conn}}$. 
However, now we cannot simply use the OPE for the products of two operators at the same point, because the expansion of the connected correlation function in terms of general correlation functions involves both terms of the form $\langle \cO_0(x) \cO_j(x) \cdots \rangle \langle \cdots \rangle \cdots$, where we can use the OPE, and terms of the form $\langle \cO_0(x) \cdots \rangle \langle \cO_j(x) \cdots \rangle \langle \cdots \rangle \cdots$ where we cannot use it. Taking this into account, we find that the correlator $\mean{\langle \cO_0(x) \cO_j(x) \cdots \rangle_{conn}}$ appearing in the mixing should be understood as
\begin{equation} \label{eq:local_RG_diff_conn}
\begin{split}
& \mean{\langle \cO_0(x) \cO_j(x) \prod _{l=1} ^k \cO _{j_l}(x_l) \rangle_{conn}} = \mean{\langle \big[ \cO _0(x) \cO _j(x) \big] \prod _{l=1} ^k \cO _{j_l} (x_l) \rangle_{conn}}- \\
& \qquad - \sum _{\substack{\text{Partitions of } \{1,\dots ,k\} \\ \text{into } S_1,S_2 }} \mean{\langle \cO _{0} (x) \prod _{l \in S_1} \cO _{j_l} (x_l)  \rangle_{conn}  \langle\cO _{j} (x) \prod _{l \in S_2} \cO _{j_l} (x_l)\rangle _{conn}} ,
\end{split}
\end{equation}
where $[ \cO _0(x) \cO _j(x)]$ is a short-hand notation for the combination of local operators appearing in the OPE.
The second term on the right-hand side of \eqref{eq:local_RG_diff_conn} gives us a new type of mixing with other kinds of correlation functions (rather than the usual mixing with the same correlation function of different operators).


There is a special case of the discussion above where $\cO_j(x)$ is simply the identity operator; in particular there is always such a mixing with the identity for the operator $\cO_i(x)=\cO_0(x)$ itself (the RG flow mixes this with $h(x)$ times the identity operator). In this case, using \eqref{local_mixing}, the non-connected correlation function $\mean{\langle \cO_i(x) \cdots \rangle}$ mixes with $\mean{\langle \cO _0(x) \cdots \rangle} - \mean{\langle \cO _0(x)\rangle \langle \cdots \rangle}$. While the first term is of the usual kind of mixing, the second term gives us a non-trivial contribution to the flow of the correlation function, involving again a product of correlators. This mixing with the identity operator does not give additional non-trivial contributions of this sort to the flow of connected correlation functions in \eqref{eq:local_RG_diff_conn}  ($\cO _j$ being the identity), because connected correlation functions of the identity operator vanish.

The general analysis of the flow of correlation functions will be performed below using the replica trick, and we will see that it will agree with our expectations from the analysis of this section.
This should help in removing doubts about the validity of the replica trick due to the $n \to 0$ analytic continuation, at least as long as the replica symmetry is unbroken.

\subsection{The replicated theory} \label{subsection:RGnq_replicated_theory}

The main tool that we will use for explicit computations is the replica trick; we review it and its relation to disordered correlation functions here. 

To obtain a generating functional for correlation functions we couple sources $J^i(x)$ to various operators $\cO _i(x)$. The partition function in the disordered theory is then
\begin{equation}
Z[h,J^i]=e^{W[h,J^i]} =\int D\mu \, e^{-S_0- \int d^dx\, h(x) \cO _0(x)+ \sum _i \int d^dx\, J^i(x) \cO _i(x) } .
\end{equation}
It is useful to define $W_D[J^i]= \mean{W[h,J^i]}$, which is the averaged free energy. By construction this generates the disorder-averaged connected correlation functions
\begin{equation} \label{eq:RGnq_WD_relation_to_connected}
\at[ \frac{\delta  W_D[J^i]}{\delta J^{i_1} (x_1) \delta J^{i_2} (x_2) \cdots } ]{J^i=0} = \mean{\langle \cO _{i_1} (x_1) \cO _{i_2} (x_2) \cdots \rangle_{conn} } .
\end{equation}

The basic motivation for the replica trick is that while it is natural to average the partition function, this is not the case for averaging the free energy. To simplify this task, the replica trick expresses the free energy in terms of a power of the partition function, which itself behaves as a partition function of a slightly different theory, the averaging of which can then be performed.

In the replica trick we use the fact that $W = \log (Z) = \lim _{n \to 0} \left( \pder{Z^n}{n} \right)$. $Z^n$ can be thought of as $n$ copies of the same theory. Using this, introduce\cite{deGennes:1972zz,Emery:1975zz,Grinstein:1976zz}
\begin{equation} \label{eq:RGnq_W_n_def}
\begin{split}
W_n[J^i] &= \int Dh\, P[h] Z^n[h,J^i] = \\
&= \int Dh\, P[h] \int \prod _{A=1} ^n D\mu _A\, e^{-\sum _A S_{0,A} - \sum _A \int d^dx\, h(x)\cO _{0,A} (x) + \sum _{i,A} \int d^dx\, J^i(x) \cO _{i,A} (x)} \equiv \\
& \equiv  \int \prod _{A=1} ^n D\mu _A\, e^{-S_{replica} + \sum _{i,A} \int d^dx\, J^i(x) \cO _{i,A} (x) }
\end{split}
\end{equation}
where in the last line we have performed the integration over $h$, and $A=1,\dots ,n$. In this equation $n$ is treated as a non-negative integer, and in the replica trick we will treat analytic expressions in $n$ as if $n$ is real; in particular $W_D[J^i]= \lim _{n \to 0} \pder{W_n[J^i]}{n} $.  For a given disordered theory, \eqref{eq:RGnq_W_n_def} defines the associated replicated theory (which is local at least perturbatively in the cumulants of the probability distribution).

In examples we will focus on the case of Gaussian disorder. It is then straightforward to perform the $h$ integration in \eqref{eq:RGnq_W_n_def} to obtain
\begin{equation} \label{eq:RGnq_replicated_Gaussian_disorder}
S_{replica} = \sum _A S_{0,A} - \frac{\qd}{2} \sum _{A,B} \int d^dx\, \cO _{0,A} (x) \cO _{0,B} (x) .
\end{equation}
Several remarks are in order. First, note that for $A=B$ the product of the same operator at coincident points diverges in the absence of an ultra-violet cutoff. Using the operator product expansion (OPE), each such term can be replaced by the operators appearing in the OPE. The relevant operators must anyway be introduced once we deform the pure CFT we began with, since in general they will be generated under the RG,
so the $A=B$ terms do not need to be considered separately from other operators in $S_{0,A}$, and the sum can be restricted to $A \neq B$. In specific pure CFTs (free theories and large $N$ theories), there is a special operator appearing in a non-singular way in the OPE of $\cO _0$ and $\cO _0$, which is denoted by $ \cO_0 ^2$, and it will be marginal when the disorder is marginal, so that it is natural to include 
the term with $A=B$ in \eqref{eq:RGnq_replicated_Gaussian_disorder}.

Second, the sign of the term proportional to $\qd$ seems to give an unbounded potential from below (since $\qd \ge 0$). However, in general there will be terms in $S_0$ that will stabilize the potential. In addition, there are $n(n-1)$ terms in this sum over $A \neq B$ and so as we take $n \to 0$ the sign of this potential changes \cite{Cardy:1996xt}.

For any $P[h]$, there are several relations between correlation functions in the disordered theory and in the replicated theory. Using the relation of $W_D$ to $W_n$, \eqref{eq:RGnq_WD_relation_to_connected} and \eqref{eq:RGnq_W_n_def}, the disorder-averaged connected correlation functions can be obtained from the replicated theory as:\footnote{
Note that $\lim _{n \to 0} W_n=1$. There is an additional term on the RHS of $\lim _{n\to 0}  \langle \sum _A \cO _{i_1,A} (x_1) \cdots \rangle  \pder{}{n} W_n$, but as will be explained in a moment, the correlation function inside the limit is proportional to $n$ and thus this term vanishes as $n \to 0$. }
\begin{equation} \label{eq:RGnq_connected_coorelator_relation_replica}
\mean{\langle \cO _{i_1} (x_1) \cO _{i_2} (x_2) \cdots \rangle_{conn} }  = \lim_{n \to 0} \pder{}{n} \langle  \sum _A \cO _{i_1,A} (x_1) \sum _B \cO _{i_2,B} (x_2)  \cdots \rangle^{replicated},
\end{equation}
where the right-hand side is evaluated in the replicated theory, and the correlation function appearing there is a-priori not the connected one.  Actually, since as will be explained in a moment any correlation function of the form $ \langle \sum _A \cO _{i_1,A} (x_1) \cdots \rangle$ is proportional to $n$ as $n \to 0$, only the connected part contributes to \eqref{eq:RGnq_connected_coorelator_relation_replica}, so that we can take the connected correlation function in the replica side as well.

Equation \eqref{eq:RGnq_connected_coorelator_relation_replica} can be expressed in another form. A useful symmetry of the replicated theory is an $S_n$ permutation symmetry of the replicas (the $n$ copies of the theory), as can be seen by the definition of the replicated theory. We will assume that this `replica symmetry' is not broken spontaneously; when it is broken it is more difficult to relate the correlation functions of the replicated theory to those of the disordered one.\footnote{Note that even if this symmetry is spontaneously broken, this will not affect the renormalization group flow of the coupling constants and the local operators of the replica theory, and the flow in the replica theory as $n\to 0$ is the same as the one we saw in the local RG picture in \autoref{subsection:local_RG_approach} (including the flow of the probability distribution). However, such a breaking may affect the correlation functions and the equations that they obey.} From the $S_n$ symmetry it follows that in the correlation function $\langle \sum _{A_1} \cO _{i_1,A_1}(x_1) \sum _{A_2} \cO _{i_2,A_2}(x_2) \cdots \rangle$, the contribution of two different $A_1$'s is the same. Therefore $ \langle \sum _{A_1} \cO _{i_1,A_1}(x_1) \sum _{A_2} \cO _{i_2,A_2}(x_2) \cdots \rangle = n \langle \cO _{i_1,1}(x_1) \sum _{A_2} \cO _{i_2,A_2}(x_2) \cdots \rangle$ and
\begin{equation} \label{eq:RGnq_connected_coorelator_relation_replica2}
\mean{\langle \cO _{i_1} (x_1) \cO _{i_2} (x_2) \cdots \rangle_{conn} }  = \lim_{n \to 0} \langle \cO _{i_1,1} (x_1) \sum _{A_2} \cO _{i_2,A_2} (x_2)  \cdots \rangle^{replicated} .
\end{equation}

We can also extract non-connected correlation functions (by which we mean a general correlation function including both connected and disconnected contributions) from the replicated theory (see \cite{Cardy:2013rqg} and references therein). For any positive integer $n$ we have
\begin{equation}
\mean{\langle \cO _{i_1}(x_1) \cO _{i_2}(x_2) \cdots \rangle}= \int Dh\, P[h] \, \frac{\int D\mu \, \cO _{i_1,A_1=1}(x_1) \cO _{i_2,A_2=1}(x_2) \cdots e^{-\sum _A S_{0,A}-\sum _A \int d^dx\,  h(x) \cO _{0,A}(x)} }{Z[h]^n }  .
\end{equation}
Suppose that this equation can also be continued to $n \to 0$. Then (using $\lim _{n \to 0} W_n=1$)
\begin{equation} \label{eq:RGnq_disconnected_coorelator_relation_replica}
\mean{\langle \cO _{i_1} (x_1) \cO _{i_2} (x_2) \cdots \rangle }  = \lim_{n \to 0} \langle \cO _{i_1,1} (x_1)  \cO _{i_2,1} (x_2)  \cdots \rangle^{replicated} .
\end{equation}

This relation can also be generalized straightforwardly to include operators from different replicas, giving a similar relation which holds for all integer $n$ large enough. Assuming we can continue the obtained relation to $n \to 0$, we get\footnote{Note that we could add to the replica theory couplings which vanish as $n \to 0$, and still get the relations \eqref{eq:RGnq_connected_coorelator_relation_replica2}, \eqref{eq:RGnq_disconnected_coorelator_relation_replica_generalized} to the disordered theory.
This is useful since there are cases in which the original replicated theory is not well-defined (for instance due to the presence of tachyons), but such modifications can cure the problem. Such a situation was noticed and analyzed in \cite{Brezin:1998fb}.}
\begin{equation} \label{eq:RGnq_disconnected_coorelator_relation_replica_generalized}
\begin{split}
& \mean{\langle \cO _{i_1} (x_{i_1}) \cO _{i_2} (x_{i_2}) \cdots \rangle \langle \cO _{j_1} (x_{j_1} ) \cO _{j_2} (x_{j_2} ) \cdots \rangle \cdots } = \\
& \qquad \qquad \qquad = \lim _{n\to 0} \langle \cO _{i_1,1} (x_{i_1}) \cO _{i_2,1} (x_{i_2} ) \cdots \cO _{j_1,2} (x_{j_1} ) \cO _{j_2,2} (x_{j_2} ) \cdots \rangle^{replicated} .
\end{split}
\end{equation}

There is an important comment regarding the relation \eqref{eq:RGnq_disconnected_coorelator_relation_replica_generalized} (and the previous ones), which is strongly related to the renormalization group in these theories (the guiding idea will appear later in more detail, while the comment here is brief). This relation holds for bare operators, but it is not necessarily true for renormalized operators. The operators $\cO _A$ mix among themselves, so the renormalization is not multiplicative. We define the renormalized correlation functions in the disordered theory through this relation, with the operators on the replica side being the renormalized ones, ensuring that we get finite answers.
In particular, the correlators $\mean{\langle \cO \cdots \cO \rangle}$ mix with correlators of the form $\mean{\langle \cO \cdots \cO \rangle \langle \cO \cdots \cO \rangle \cdots }$ where the total number of operators is the same, implying that we cannot renormalize all of $\mean{\langle \cO \cdots \cO \rangle}$ just by a redefinition $\cO = \sqrt{Z} \cO ^R$.

There is another kind of replica correlation function whose relation to the disordered theory will be needed. It has the form $\langle \cO _{i_1,1}(x_1) \cdots \cO _{i_I,I}(x_I) \sum _1^n \cO _{j_1,A_1}(y_1) \cdots \sum _1^n \cO _{j_k,A_k}(y_k) \rangle$, where the subscripts $A_j=1,\dots ,n$ ($j=1,\dots,k$) are replica indices. As $n \to 0$ it  is given in the disordered theory by
\begin{equation} \label{eq:RGnq_product_connected_coorelator_relation_replica}
\begin{split}
& \lim _{n \to 0} \langle \cO _{i_1,1}(x_1) \cdots \cO _{i_I,I}(x_I) \sum _{A_1=1}^n \cO _{j_1,A_1}(y_1) \cdots \sum _{A_k=1}^n \cO _{j_k,A_k}(y_k) \rangle ^{replicated} = \\
& \qquad = \sum _{\substack{\text{Partitions of } \{1,\dots ,k\} \\ \text{into } S_1,\dots ,S_I }} \mean{\langle \cO _{i_1} (x_1) \prod _{l \in S_1} \cO _{j_l} (y_l)  \rangle_{conn}  \cdots \langle\cO _{i_I} (x_I) \prod _{l \in S_I} \cO _{j_l} (y_l)\rangle _{conn} } .
\end{split}
\end{equation}
In the sum on the right-hand side we are going over all the partitions of $\{1,\dots ,k\}$ into $I$ sets (where some of them can be empty). This is in fact a generalization of \eqref{eq:RGnq_connected_coorelator_relation_replica2}. Equation \eqref{eq:RGnq_product_connected_coorelator_relation_replica} is derived in \autoref{section:derivation_of_product_connected_correlator}.

For the validity of the relations above, we have assumed that there is a stable vacuum with no replica symmetry breaking. We also assume that the $n\to 0$ limit exists; this is true in any perturbative expansion in the disorder, where only polynomials in $n$ appear, so we believe that it should not affect the form of the GCS equations that we will derive.
Indeed, in simple cases we have confirmed that these equations can be derived from the original disordered theory (\autoref{subsection:local_RG_approach}), with no reference to the replica trick.

In the replica theory, the parameters of the disorder distribution become standard coupling constants as in \eqref{eq:RGnq_replicated_Gaussian_disorder}, so it is clear that they run like any other couplings, and mix with the other couplings, consistently with the description in \autoref{subsection:local_RG_approach}. In \eqref{eq:RGnq_W_n_def} we wrote a disorder distribution only for a single coupling $h(x)$, but an RG flow will generate such a distribution also for other couplings; in the replica theory this will happen by generating additional couplings between different replicas, beyond the ones present in \eqref{eq:RGnq_W_n_def}. Generically at the UV cutoff scale (e.g. the lattice scale) there will be some disorder distribution for all the couplings, and its parameters will then flow under the renormalization group, together with the flow of the standard coupling constants.

If we have a symmetry $G$ of our original theory that is unbroken by the coupling to the disorder, namely it is unbroken in \eqref{eq:RGnq_coupling_disorder_action}, then in the replica theory \eqref{eq:RGnq_W_n_def} (without the sources) we get $n$ copies of this symmetry ($G^n$), one for each replica. This does not happen when the symmetry is broken by the coupling to the disorder. However, we can sometimes restore the symmetry by making the disordered coupling $h(x)$ in \eqref{eq:RGnq_coupling_disorder_action} also transform, if the disorder distribution is invariant under this transformation. In such a case the symmetry $G$ will still be there in the disorder-averaged correlation functions, and the replica theory \eqref{eq:RGnq_W_n_def} that we obtain after the integral over $h$ still has one copy of this symmetry. This happens in particular for translations and rotations, and in some cases also for internal symmetries. 

As an example, the random-bond Ising model above with a random coupling ${\tilde h}(x) \varphi^2(x)$ preserves the $\mathbb{Z}_2$ symmetry taking $\varphi \to -\varphi$, so this becomes a $(\mathbb{Z}_2)^n$ symmetry in the replica theory.
A varying coupling for $\varphi $ corresponds in the replica to a $\sum _{A \neq B} \varphi _A \varphi _B$ coupling, that is allowed by the overall $\mathbb{Z} _2$ symmetry, but is forbidden by the $\left(\mathbb{Z} _2\right)^n$ symmetry. This agrees with our discussion in \autoref{subsection:local_RG_approach}.
On the other hand, the random-field Ising model with $h(x) \varphi(x)$ breaks the $\varphi  \to -\varphi $ symmetry. But if the distribution of the background field is Gaussian we have a new $(h(x), \varphi(x)) \to -(h(x), \varphi(x))$ ``symmetry'' of the disordered theory, which is a symmetry of the disorder-averaged correlation functions, and which becomes a single $\mathbb{Z}_2$ symmetry in the replica theory.

\subsection{The generalized Callan-Symanzik equations} \label{subsection:RGnq_RG_equations}

\subsubsection{Connected correlation functions}

The replicated theory is a standard QFT. 
In principle, we should include in it all the deformations which are relevant or marginal, and are consistent with the symmetries preserved by the CFT we began with and the term(s) induced by the disorder. It will be assumed that there is a finite number of such couplings. We can also discuss a Wilsonian renormalization group flow, in which we keep all deformations during the flow (relevant and irrelevant); we will take a prescription in which operators can mix with other operators of lower dimension, but not with operators of higher dimension.

Consider correlation functions of an operator $\cO (x)$, which at first is taken to be of lowest dimension (we will generalize this below).  Examples of such operators are scalar fields in weakly coupled quantum field theories constructed by a deformation of a free field theory in the ultra-violet. Even though in the pure CFT $\cO $ does not mix with other operators because of the dimension condition, in the replicated theory the corresponding operators $\cO _A$ mix among themselves.
This is reflected in the Callan-Symanzik equation for a correlation function of $k$ $\cO $'s by promoting the anomalous dimension $\gamma $ to a matrix $\gamma _{AB} $:
\begin{equation} \label{eq:RGnq_CS_replica}
\begin{split}
& M \pder{}{M} \langle \cO _A(x_1) \cO _B(x_2) \cdots  \rangle + \beta_{\bar \lambda _i} \pder{}{\bar \lambda _i} \langle \cO _A(x_1) \cO _B(x_2) \cdots \rangle  + \\
&\qquad\qquad+ \gamma _{AA'}  \langle\cO _{A'} (x_1) \cO _B(x_2) \cdots \rangle + \cdots = 0
\end{split}
\end{equation}
(we sum over repeated indices such as $i$ and $A'$ here). Here $\bar \lambda_i$ are all the couplings of the replica theory, and the dots stand for terms similar to the last one, for each of the other operators $\cO_B(x_2),\cdots $.
Next, sum over $A,B, \dots $ and take the derivative with respect to $n$:
\begin{equation}
\begin{split}
& M \pder{}{M} \pder{}{n}  \langle \sum _A\cO _A(x_1) \sum _B\cO _B(x_2) \cdots \rangle + \pder{\beta _{\bar \lambda _i}}{n} \pder{}{\bar \lambda _i} \langle \sum _A\cO _A(x_1) \sum _B\cO _B(x_2) \cdots \rangle + \\
& + \beta _{\bar \lambda _i} \pder{}{\bar \lambda _i} \pder{}{n} \langle \sum _A\cO _A(x_1) \sum _B\cO _B(x_2) \cdots \rangle + \pder{}{n} \sum _{A,A'} \gamma _{AA'}  \langle \cO _{A'}(x_1) \sum _B\cO _B(x_2) \cdots \rangle + \cdots = 0 .
\end{split}
\end{equation}
The $S_n$ replica symmetry implies that
\begin{equation}\label{eq:RGnq_mixing_O_A}
\gamma _{AB} = \gamma ' \delta _{AB} +\gamma  '' .
\end{equation}
Taking the limit of $n \to 0$ (assuming that the $\beta $ and $\gamma $ functions are smooth as $n \to 0$, which is guaranteed in perturbation theory), using \eqref{eq:RGnq_connected_coorelator_relation_replica}, and noting again that $\langle \sum _A \cO _A (x_1) \sum _B \cO _B(x_2) \cdots \rangle$ vanishes as $n \to 0$, we get
\begin{equation}
\left( M \pder{}{M} + \at[\beta _{\bar \lambda _i}]{n=0} \pder{}{\bar \lambda _i} +k \at[\gamma ']{n=0} \right) \mean{\langle \cO (x_1) \cdots \cO (x_k) \rangle_{conn} } =0.
\end{equation}
We have shown that the disordered connected correlation functions of such operators $\cO (x)$ satisfy a standard Callan-Symanzik equation with $\at[\beta (\bar \lambda _i)]{n=0}$ and $\at[\gamma ']{n=0}$.

In the case of Gaussian disorder, the disorder strength $\qd$ should be treated as one of the couplings ${\bar \lambda_i}$ in \eqref{eq:RGnq_CS_replica} (see \eqref{eq:RGnq_replicated_Gaussian_disorder}), in addition to any other couplings $\lambda _i$ of the original theory (these correspond to the constant part of what were called $g_i$ above). Additional moments of the disorder distribution should also be included in principle, but if the disorder is close to being marginal, the higher moments will correspond to irrelevant operators, see \autoref{section:Ph_non_Gaussian}, and these do not have to be included.
The CS equation in the disordered theory is thus
\begin{equation} \label{eq:RGnq_RG_eq_conn}
\left( M \pder{}{M} + \beta _{\qd} \pder{}{\qd} + \beta _{\lambda _i} \pder{}{\lambda _i} +k \gamma ' \right) \mean{\langle \cO (x_1) \cdots \cO (x_k) \rangle_{conn} } =0
\end{equation}
where the beta and gamma functions in this equation are the replica beta and gamma functions at $n=0$. Equation \eqref{eq:RGnq_RG_eq_conn} suggests that these should be called the disordered beta and gamma functions; they are not the same as those of the pure system.

It is useful to take advantage of the $S_n$ replica symmetry as was done above, since it restricts the possible mixing. Operators transforming as the same irreducible representation can mix with each other under the RG, but not with operators transforming as a different irreducible representation. Therefore it is favorable to work in a basis of operators forming irreducible representations of $S_n$. Operators with a single replica index $\cO _A$ decompose into two irreducible representations
\begin{equation}
\begin{split}
\tilde \cO &= \sum _A \cO _A ,\\
\tilde \cO _A &= \cO _A - \frac{1}{n} \sum _A \cO _A ,
\end{split}
\end{equation}
the first being a singlet (invariant under permutations). Disordered connected correlation functions are related to replica correlation functions of $\tilde \cO $ by \eqref{eq:RGnq_connected_coorelator_relation_replica}. In the simplest case considered above, $\tilde \cO $ was the only singlet, which then did not mix with other operators, and this is why a simple CS equation was obtained.

For a general operator $\cO_*(x)$ of the original theory, the corresponding singlet operator $\tilde \cO_* (x)$ in the replica theory will mix with other operators. The first kind of operators that it can mix with are operators of the same form $\sum _A \cO  _{i,A} = \tilde \cO _i$. These just correspond to operators $\cO  _i(x)$ in the pure theory that $\cO_* (x)$ can mix with. But, there are additional singlet operators in the replicated theory.  From products of two replicated operators, we can form the singlet $\tilde  \cO  _{ij}(x) = \sum _{A \neq B} \cO _{i,A}(x) \cO  _{j,B}(x)  $.\footnote{Note that after turning on disorder there can be short-distance singularities in the product of $\cO_{i,A}$ and $\cO_{j,B}$ corresponding to other singlet operators, and we assume that these have been subtracted.} Similarly, singlets can be formed from products of three operators, and so on. Since operators can mix only with operators of lower dimension, and there is a finite number of operators with dimensions $[\cO  _i] \le [\cO_* ]$ for the first kind of operators, $[\cO _i] + [\cO  _j] \le [\cO_* ]$ for the second kind, and so on ($[\cO ]$ is the dimension of $\cO $), there is a finite number of operators with which $\tilde \cO_* (x)$ can mix.

We may wonder what is the meaning of this mixing in the language of the disordered theory, since while $\tilde \cO _i$ are the replica analog of the local operators $\cO _i$ in the disordered theory, there are no local operators in the disordered theory corresponding to the replica operators with a higher number of replicas such as $\tilde \cO _{ij} $. Recalling that the integral of $\sum _{A \neq B} \cO _{A} \cO  _{B}$ corresponds to the Gaussian term in the disorder distribution, it is natural to associate these operators with the disorder distribution.
Using \autoref{section:Ph_non_Gaussian}, if we denote the coupling multiplying the operator $\cO _i$ in the disordered action by $g^i$, we see that $\tilde \cO _i$ in the replica corresponds to having a constant part for $g^i$, while $\tilde \cO _{ij}$ corresponds in the disordered theory to correlations between non-constant $g^i$ and $g^j$.
So the constant coupling for $\tilde \cO_{ij}$ corresponds to some moment of the disorder distribution.
Thus, in the language of the disordered system we have a mixing between all the coupling constants and all the distribution parameters (moments).
This can also be seen using the approach of the running distribution and inhomogeneous couplings described in \autoref{subsection:local_RG_approach}. 

While this accounts for these operators mixings from the point of view of the corresponding mixing of couplings, we also saw in \autoref{subsection:local_RG_approach} directly what is the origin of the mixing of operators. This came from a mixing of local operators with coefficients depending explicitly on the disorder field.

To see the implications of such mixings for disorder-averaged correlation functions, assume for simplicity that there is a single operator $\tilde \cO _{ij} $ with which $\tilde  \cO $ can mix. 
The relevant equation in the replicated theory that will lead to the disordered connected correlation function of $\cO $ now becomes
\begin{equation}
\begin{split}
& M \pder{}{M} \langle \sum_{A_1} \cO _{A_1}(x_1)\cdots \sum _{A_k} \cO _{A_k}(x_k)\rangle + \beta _{\bar \lambda _i} \pder{}{\bar \lambda _i}  \langle \sum_{A_1} \cO _{A_1}(x_1)\cdots \sum _{A_k} \cO _{A_k}(x_k)\rangle + \\
& + k \gamma _{\tilde \cO} \langle \sum_{A_1} \cO _{A_1}(x_1)\cdots \sum _{A_k} \cO _{A_k}(x_k)\rangle + \\
& \left[ \gamma _{\tilde \cO ,\tilde \cO _{ij} } \langle \sum _{A \neq B} \cO _{i,A} \cO _{j,B}(x_1) \sum _{A_2} \cO _{A_2}(x_2) \cdots \sum _{A_k} \cO _{A_k}(x_k)\rangle + \cdots \right] = 0 .
\end{split}
\end{equation}
Taking the derivative with respect to $n$ and the $n \to 0$ limit, using
\begin{equation}
\sum _{A \neq B} \langle \cO _{i,A}(x_1) \cO _{j,B} (x_1) \sum _{A_2} \cO _{A_2} (x_2) \cdots \rangle = n(n-1) \langle \cO _{i,1}(x_1) \cO _{j,2}(x_1) \sum _{A_2} \cO _{A_2}(x_2) \cdots \rangle 
\end{equation}
and  \eqref{eq:RGnq_product_connected_coorelator_relation_replica}, leads to
\begin{equation}
\begin{split}
& M \pder{}{M} \mean{\langle \cO (x_1) \cdots \cO (x_k) \rangle_{conn} } + \at[\beta _{\bar \lambda _i} ]{n=0} \pder{}{\bar \lambda _i} \mean{\langle \cO (x_1) \cdots \cO (x_k) \rangle_{conn} } \, +\\
& \qquad + k \at[\gamma _{\tilde \cO } ]{n=0} \mean{\langle \cO (x_1) \cdots \cO (x_k) \rangle_{conn} } \, - \\
& \qquad - \at[\gamma _{\tilde \cO, \tilde\cO _{ij} } ] {n=0} \left[  \sum _{\substack{\text{Partitions of } \{2,\dots ,k\} \\ \text{into } S_1,S_2}}  \mean{\langle \cO_i (x_1) \prod _{l \in S_1} \cO (x_l)\rangle_{conn} \langle \cO_j (x_1) \prod _{l \in S_2} \cO (x_l)\rangle_{conn} } + \right. \\
& \qquad\qquad\qquad\qquad\qquad +   (x_1 \leftrightarrow x_2) + \cdots + (x_1 \leftrightarrow x_k)  \Bigg] = 0 .
\end{split}
\end{equation}

Restoring the disorder strength and returning to the notation in which the beta and gamma functions in the disordered theory stand for the appropriate ones in the replicated theory at $n=0$, the full GCS equation is
\begin{equation} \label{eq:RGnq_RG_eq_conn_more_general}
\begin{split}
& \left(M \pder{}{M} + \beta _{\qd} \pder{}{\qd} + \beta _{\lambda _i}  \pder{}{\lambda _i}  + k \gamma _{ \cO } \right) \mean{\langle \cO (x_1) \cdots \cO (x_k) \rangle_{conn} } \, - \\
& \qquad- \gamma _{ \cO ,\cO _{ij} } \left[  \sum _{\substack{\text{Partitions of } \{2,\dots ,k\}\\ \text{into } S_1,S_2}}  \mean{\langle \cO_i (x_1) \prod _{l \in S_1} \cO (x_l)\rangle_{conn} \langle \cO_j (x_1) \prod _{l \in S_2} \cO (x_l)\rangle_{conn} } + \right. \\
& \qquad \qquad\qquad\qquad+   (x_1 \leftrightarrow x_2) + \cdots + (x_1 \leftrightarrow x_k)  \Bigg] = 0 .
\end{split}
\end{equation}
These are non-trivial GCS equations for the connected correlators, which precisely agree with what we found (under additional simplifying assumptions) using the local RG approach \eqref{eq:local_RG_diff_conn}. Note that if $\gamma _{\cO ,\cO _{ij}} $ is vanishing, then $\gamma _{\cO } $ just reduces to what was denoted by $\gamma '$ in the simplest case considered in \eqref{eq:RGnq_RG_eq_conn}.
Our discussion implies that equation \eqref{eq:RGnq_RG_eq_conn} is valid only when there is no mixing of $\tilde \cO $ with multi-replica operators. Mixing with operators with more replicas correspond in the local RG approach to including terms of higher order in $h(x)$, either in the operator mixings or in the disorder distribution (or both).
The corresponding generalization of \eqref{eq:RGnq_RG_eq_conn_more_general} is straightforward.

\subsubsection{Non-connected correlation functions}

Consider next general (non-connected) disordered correlation functions. Here, already for the simplest case in which $\cO _A$ mix only among themselves (that led to \eqref{eq:RGnq_RG_eq_conn}) there is a non-trivial GCS equation.

Look again at correlation functions of $k$ $\cO $'s.
Set $A=B= \cdots =1$ in \eqref{eq:RGnq_CS_replica} to get
\begin{equation}
\begin{split}
& M \pder{}{M} \langle \cO _1(x_1) \cO _1(x_2) \cdots \rangle + \beta  _{\bar \lambda _i} \pder{}{\bar \lambda _i} \langle\cO _1(x_1) \cO _1(x_2) \cdots \rangle + \\
& \qquad+k\gamma ' \langle \cO _1(x_1) \cO _1 (x_2) \cdots \rangle + \left( \gamma '' \sum _A \langle \cO _A (x_1) \cO _1(x_2) \cdots \rangle  + \cdots \right) = 0 .
\end{split}
\end{equation}
Taking $n \to 0$ and using \eqref{eq:RGnq_disconnected_coorelator_relation_replica_generalized}, this gives
\begin{equation}
\begin{split}
M \pder{}{M} \mean{ \langle \cO (x_1) \cdots  \cO (x_k)  \rangle} + \at[\beta _{\bar \lambda _i}]{n=0}  \pder{}{\bar \lambda _i} \mean{ \langle \cO (x_1) \cdots \cO (x_k) \rangle} + k \at[(\gamma '+\gamma '')]{n=0}  \mean{ \langle \cO (x_1) \cdots  \cO (x_k)  \rangle} - \\
- \at[\gamma '']{n=0} \left[ \mean{ \langle \cO (x_1)\rangle \langle \cO (x_2) \cdots \cO (x_k)\rangle} + \mean{ \langle \cO (x_2) \rangle \langle \cO (x_1) \cO (x_3) \cdots \cO (x_k) \rangle} + \cdots \right] = 0 .
\end{split}
\end{equation}
Recall that the disorder strength $\qd$ should be treated as one of the coupling constants. We see that the GCS equation satisfied by these disordered correlation functions is
\begin{equation}\label{finalrgnonconn}
\begin{split}
&\left( M \pder{}{M} + \beta _{\qd} \pder{}{\qd} + \beta _{\lambda _i}  \pder{}{\lambda _i}  + k (\gamma '+\gamma '') \right)  \mean{ \langle \cO (x_1) \cdots  \cO (x_k)  \rangle} - \\
& \qquad - \gamma '' \left[ \mean{ \langle \cO (x_1)\rangle \langle \cO (x_2) \cdots \cO (x_k)\rangle} + \mean{ \langle \cO (x_2) \rangle \langle \cO (x_1) \cO (x_3) \cdots \cO (x_k) \rangle} + \cdots \right] = 0 .
\end{split}
\end{equation}
This is precisely the contribution \eqref{local_mixing} that we found in the local RG approach, due to mixing with the identity operator. Operator mixings with multi-replica operators will lead to an even more complicated equation, along the lines of our discussion above.

Note that the beta and gamma functions that appear in the GCS equations \eqref{eq:RGnq_RG_eq_conn} and \eqref{finalrgnonconn} for the disordered connected and non-connected correlation functions are intrinsic to the disordered theory, and do not depend on the replica trick; indeed we found similar mixings in the local RG approach.

\subsection{Disordered fixed points} \label{subsection:RGnq_fixed_points}

A disordered theory may flow to a fixed point, defined by vanishing disordered beta functions $\beta _{\lambda _i} =\beta _{\qd} =0$. It can flow to a pure fixed point if $\qd$ (and any other disorder-related coupling constants) flows to zero, but otherwise it will be a random fixed point. We will analyze here the properties of such random fixed points if and when they arise, denoting the anomalous dimensions at the fixed point by $\gamma ^*$, and the couplings by $\lambda _i^*,\qd^*$. 

\subsubsection{Connected correlation functions}

Connected correlation functions of low-dimension operators, obeying \eqref{eq:RGnq_RG_eq_conn}, will have standard scaling behavior at such fixed points.
However, for the more general $S_n$-singlet operators $\tilde \cO $ that can mix, the situation is different. We illustrate this by again looking at the simplest case of an $\tilde \cO $ which mixes only with $\tilde \cO _{ij} $, with some $2\times 2$ mixing matrix $\gamma $ (the same computation can be performed directly with averaged correlation functions, but it will be simpler to use the replica approach here). The CS equations for their 2-point functions in the replica theory at fixed points are 
\begin{equation} \label{eq:RGnq_fixed_point_connected_replica_CS}
\begin{split}
& \left( M \pder{}{M} +2\gamma ^*_{11} \right) \langle \tilde \cO (x) \tilde  \cO (0) \rangle + 2 \gamma ^*_{12} \langle \tilde  \cO (x) \tilde  \cO _{ij} (0)\rangle =0, \\
& \left( M \pder{}{M} +\gamma ^*_{11} +\gamma ^*_{22} \right) \langle \tilde \cO (x) \tilde  \cO _{ij} (0)\rangle + \gamma ^*_{21} \langle \tilde  \cO (x) \tilde  \cO (0)\rangle + \gamma ^*_{12} \langle \tilde \cO _{ij} (x) \tilde  \cO _{ij} (0)\rangle = 0, \\
& \left(M \pder{}{M} +2\gamma ^*_{22} \right) \langle \tilde  \cO _{ij} (x) \tilde  \cO _{ij} (0)\rangle + 2\gamma ^*_{21} \langle \tilde  \cO (x) \tilde  \cO _{ij} (0)\rangle = 0 .
\end{split}
\end{equation}
By the same arguments as before, all these correlation functions are proportional to $n$ as $n \to 0$, and therefore when a derivative with respect to $n$ is taken on these equations, in the limit $n \to 0$ all the correlation functions will simply be replaced by their derivatives with respect to $n$. 
Recall also that as $n \to 0$, the correlator $ \pder{}{n}  \langle \tilde \cO (x) \tilde  \cO (0)\rangle$ approaches $\mean{\langle \cO (x) \cO (0)\rangle_{conn} }$, while the other two-point functions do not approach a connected correlation function. 

The first question to ask is whether there are linear combinations of the three 2-point functions in \eqref{eq:RGnq_fixed_point_connected_replica_CS} such that the CS equations can be written as three homogeneous equations of the form $\left(M \pder{}{M} + 2 \gamma^* \right) G(x)=0$. It can be checked that this can be done, unless the matrix $\gamma  ^*= \begin{pmatrix}\gamma ^*_{11} & \gamma ^*_{12} \\ \gamma ^*_{21} & \gamma ^*_{22} \end{pmatrix}$ is not diagonalizable.

If $\gamma^* $ is not diagonalizable in the $n \to 0$ limit, it can only be brought to the Jordan form $\gamma^*  = \begin{pmatrix}\gamma ^*_{11} & \gamma ^*_{12} \\ 0 & \gamma ^*_{11} \end{pmatrix}$. This can only happen if there is an exact degeneracy in the scaling dimensions (as $n \to 0$), which is unlikely to happen except in special cases such as large $N$ theories, that will be discussed in \autoref{section:classical_disorder_large_N}. The consequences of such degeneracies will be addressed there.

In the general case, there will not be such an exact degeneracy in the anomalous dimensions. There is then a basis of three correlation functions which satisfy a usual CS equation, and at the fixed point each of them is given by a power law. However, when substituted back into the original basis which is distinguished by $ \pder{}{n} \langle \tilde  \cO (x) \tilde \cO (0)\rangle$ giving in the $n \to 0$ limit the disordered connected correlation function $\mean{\langle \cO (x) \cO (0)\rangle_{conn} }$, it implies that $\mean{\langle \cO (x) \cO (0)\rangle_{conn} }$ will generically be a sum of three different power laws. Therefore, the connected part of a disordered correlator will not satisfy the usual power law behavior of pure fixed points, but rather will include combinations of power laws, even at a fixed point. Of course the smallest dimension will always dominate the correlation function at long distances.
The generalization to a higher number of mixing operators is the same, resulting in a larger number of powers in a single disordered connected correlation function.

This is different from a pure unitary (reflection-positive) system, in which we can diagonalize the mixing of operators such that for $k$ operators, only $k$ dimensions appear. Contrary to that, in disordered systems, the correlation functions in the replica that give connected disordered correlation functions are distinguished. In the absence of a mixing with multi-replica operators such as $\tilde \cO _{ij} $, we can diagonalize the $\tilde \cO _i$ operators as usual, but once the $\tilde \cO _i$ operators mix with multi-replica operators, the dimensions of all these multi-replica operators will mix into the connected disordered correlation functions.
So not only is the total number of critical exponents larger than the number of local operators in the disordered theory, because of the presence of the disorder-related couplings and the associated replica operators, but all of these exponents can appear also in the connected correlation functions of the local operators.

In addition, in pure fixed points of unitary theories, the anomalous dimensions $\gamma^*$ must all be real. However, this is not necessarily true for random fixed points (even though the replica theories are unitary for positive integer values of $n$). For such fixed points some of the dimensions (and some of the critical exponents) can be complex; however, complex dimensions must always come in complex-conjugate pairs, so they must arise for pairs of operators that are allowed to mix. Moreover, since the dimensions are continuous along the RG flow, when we flow from a unitary theory, complex dimensions can only appear if two operator dimensions become degenerate along the RG flow and then move off into the complex plane. When we have an operator with a complex dimension $\Delta$, its $2$-point function will scale as $|x|^{-2 {\Re(\Delta)}} \sin \left[ 2 \cdot {\Im(\Delta)} \log(\mu |x|) \right]$, and will feature discrete scale-invariance.

In situations of weak disorder that is seen in perturbation theory around a unitary theory, complex dimensions can arise only if the original unitary theory has a pair of operators with degenerate dimensions. Generically this is not the case, but it is true when our unitary theory is free, which is often the starting point for $\epsilon$-expansions of scalar field theories around $4$ dimensions. In such cases, the double-replica operator related to disorder for some ${\cO}$ is degenerate with the operator ${\cO}^2(x)$ (which is well-defined in a free theory) and will mix with it, potentially leading to complex dimensions (and complex critical exponents). Such a situation was observed for several perturbative random fixed points \cite{aharony1975critical, chen1977mean, khmelnitskii1978impurity, Boyanovsky:1982zz, weinrib1983critical}.

Since disordered fixed points are in general not unitary, there is no argument that scale-invariant disordered fixed points should obey conformal invariance (even when it is obeyed by the replica theories for positive integer $n$).
It would be interesting to understand if and when disordered fixed points have a conformal symmetry.

\subsubsection{Non-connected correlation functions}

Disordered fixed points have even more unusual behavior of their non-connected correlation functions.
For the simpler sort of operators for which $\cO _A$ mix only among themselves, the GCS equations for the connected and the non-connected correlation functions were obtained above.
For the 2-point functions, the GCS equations \eqref{eq:RGnq_RG_eq_conn} and \eqref{finalrgnonconn} that we found reduce at fixed points to
\begin{equation} \label{eq:RGnq_fixed_points_2pf}
\begin{split}
& \left(M \pder{}{M}  +2\gamma '^{*} \right)\mean{\langle \cO(x) \cO(0) \rangle_{conn} }=0 \\
& \left(M \pder{}{M} +2\gamma '^{*}\right) \mean{\langle \cO(x) \cO(0) \rangle} + 2\gamma ''^{*} \mean{\langle\cO(x) \cO(0) \rangle_{conn} }=0 .
\end{split}
\end{equation}
When $\gamma''^{*}\neq 0$ these equations cannot be diagonalized such that in each of them only one combination of correlators appears.\footnote{In the replica theory, taking as a basis the correlation functions $\langle \cO _1 \cO _1\rangle$ and $\langle \cO _1 \cO _2\rangle$ where the indices are replica indices, such a diagonalization is possible for any $n \neq 0$.}

The solution of the first equation is the usual one\footnote{An abuse of notation is used in which a vector raised to some power stands for the norm of that vector raised to the same power.}
\begin{equation}
\mean{\langle \cO (x) \cO (0)\rangle_{conn} }= \frac{M^{-2\gamma '^*} }{x^{2\Delta +2\gamma '^*} } g(\lambda _i^*,\qd^*) \propto \frac{M^{-2\gamma '^*} }{x^{2\Delta +2\gamma '^*} }
\end{equation}
($\Delta $ is the dimension of $ \cO $ at the pure CFT and $g(\lambda _i^*,\qd^*)$ is some integration constant).
Now, solving the second equation, we find
\begin{equation} \label{log_mixing}
\mean{\langle \cO (x) \cO (0)\rangle}= \frac{M^{-2\gamma '^*} }{x^{2\Delta +2\gamma '^*} } \left(h(\lambda _i^*,\qd^*) - 2\gamma ''^*g(\lambda _i^*,\qd^*) \log (xM) \right) \propto \frac{M^{-2\gamma '^*} }{x^{2\Delta +2\gamma '^*} } \left(C_0 +C\log(xM) \right)
\end{equation}
(with $h(\lambda _i^*,\qd^*)$ another integration constant, and $C_0,C$ are also constants).

These solutions exhibit two properties. First, the connected correlation function has a power law behavior as in usual pure fixed points.
It is evident now that disordered connected correlation functions containing only this simple sort of operators must behave at random fixed points as in pure theories, because of the usual CS equations \eqref{eq:RGnq_RG_eq_conn}.
Second, the non-connected correlation function contains a log. The appearance of logs in the non-connected correlation functions of disordered fixed points was found in \cite{Cardy:2013rqg,Gurarie:1999yx,Gurarie:2004ce}, and we reproduce it here in our approach; it is similar to what happens in a logarithmic CFT. Generically this means that some correlation functions are not scale-invariant at the fixed point, because of the mixing of the operators under the RG flow. In other words, there is no basis of operators that transforms homogeneously under scaling. Note that the coefficient $C$ of $\log(x)$ in \eqref{log_mixing} is universal, but the constant term (which can be swallowed into $M$) is not.

\subsubsection{Higher moments of correlation functions}

Since disordered fixed points have no characteristic scale, they will not in general be self-averaging in the large volume limit; we expect the probability distributions of local dimensionless observables to be independent of the scale.
In the lack of self-averaging, a given disordered system is an instance in the probability space, and the complete description of the disorder problem is given by the various moments of observables. Until now we mostly concentrated on average values of observables, but we can just as well consider higher moments of those. One such interesting quantity is the variance of a 2-point function (connected or not), for which we need both the average (squared) of the 2-point function $\mean{\langle\cO(x) \cO(0) \rangle}$ and the average of a product of correlators $\mean{\langle \cO(x) \cO(0) \rangle \langle \cO(x) \cO(0) \rangle}$. 

At a long-distance fixed point the moments $\mean{ \langle \cO (x) \cO (0)\rangle^k}$ of the 2-point function will behave as $1/x^{2X_k} $ (up to possible logs as discussed above), and the behavior of the $X_k$ was analyzed in \cite{Ludwig:1989rj}. The $k$'th moment is given in the replica by the 2-point function of $\cO _{A_1} \cdots \cO _{A_k}(x)$ (with the $A_i$ being different replica indices). This operator is not an $S_n$ singlet but rather decomposes into $k+1$ operators $ \cO ^{(k)} _i$ transforming in different irreducible representations $r_i$, $i=0,\dots ,k$, each with its own scaling dimension at a random fixed point. 
Note that the irreducible representations appearing in $k+1$ are those of $k$ plus one additional irreducible representation \cite{Ludwig:1989rj}.
The lowest dimension among the $\cO^{(k)}_i$ for fixed $k$ will dominate and will fix $X_k$. If there is no mixing between operators of different $k$, e.g.\ by symmetry considerations, then the $X_k$ are independent and give a multi-fractal behavior for the distribution of the 2-point function, with infinitely many critical exponents \cite{Ludwig:1989rj}. This is the case for the $q$-state Potts model as seen for small $q-2$ \cite{Ludwig:1989rj} (see also \cite{Dotsenko:1997wf}) and for finite $q-2$ in \cite{Olson:1999sn}. Otherwise, non-perturbatively, we expect in general the operators transforming under the same irreducible representation $r_i$ for different $k$ to mix with each other, and then the lowest dimension among the different $k$ for a fixed $i$ will dominate the contribution of $r_i$ to $X_k$. If the new irreducible representation $r_{k+1} $ added by increasing $k$ by 1 has a lower dimension, we would then get $X_{k+1} <X_k$, and otherwise $X_{k+1}=X_k$. In any case, $\eta _k=X_k/k$ is non-increasing with increased $k$ by general probability theory (since the $k$'th root of the $k$'th moment is non-decreasing with $k$). As a result, the $k$'th cumulant (e.g., the variance for $k=2$) of the 2-point function will scale as $1/x^{2X_k}$ with the same $X_k$ (since at large distances there will not be contributions more dominant than the $k$'th moment).

It is often more natural to consider the distribution of integrated correlation functions, such as $ \frac{1}{\rm Volume}  \int d^d x d^d y \, e^{iq(x-y)} \langle \cO (x) \cO (y)\rangle$.
As we will discuss below, in the computation of the variance of such integrated 2-point functions of operators of dimension $\Delta < d/2$, the four operators in $\mean{\langle \cO(x) \cO(y) \rangle \langle \cO(x') \cO(y') \rangle}$ will be at generic points, so it will be related to the dimension of $\cO(x)$ rather than to 
the other independent operators in the replicated theory.

Suppose that we are still in the simplest case in which $\cO _A$ mix only among themselves. We can obtain the GCS equation that such an average of product of correlators will obey. To get a closed system of equations, we should write in turn the GCS equations of the correlators that appear in the inhomogeneous parts of each GCS equation (as for instance in \eqref{finalrgnonconn}), giving a system of GCS equations. This is analogous to what was done above for the average of the connected and non-connected 2-point functions, and it can be done in the usual basis $\langle \cO _A \cO _B \cO _C \cO _D\rangle$ in the replica trick (with $n=0$). We may then ask whether logs appear in certain correlation functions at disordered fixed points. When bringing the system of equations to the Jordan form, the highest power of log that will appear in some of the correlation functions is the size of the largest Jordan block minus one. 

When performing this exercise for the case at hand, we find that (at separated points) $\mean{\langle \cO (x_1) \cO (x_2)\rangle_{conn} \langle \cO (x_3) \cO (x_4)\rangle_{conn} }$ satisfies a usual CS equation (with no inhomogeneous terms) and therefore contains no logs. It means that it is given at fixed points by
\begin{equation}
\begin{split}
\mean{\langle \cO (x_1) \cO (x_2)\rangle_{conn} \langle \cO (x_3) \cO (x_4)\rangle_{conn} } =  \frac{g(x_{ij} )}{|x_1-x_2|^{2\Delta^* } |x_3-x_4|^{2\Delta^* } }
\end{split}
\end{equation}
where $g(x_{ij} )$ is a dimensionless function of $x_{ij} =x_i-x_j$, and $\Delta ^* = \Delta +\gamma '^*$. However, $\mean{\langle \cO (x_1) \cO (x_2)\rangle \langle \cO (x_3) \cO (x_4)\rangle }$ does not satisfy a usual CS equation, and in fact at a fixed point it takes the form
\begin{equation} \label{logthree}
\begin{split}
& \mean{\langle \cO (x_1) \cO (x_2)\rangle \langle \cO (x_3) \cO (x_4)\rangle } =  \frac{1}{|x_1-x_2|^{2\Delta^* } |x_3-x_4|^{2\Delta^* } } \cdot \\
&\qquad \cdot   \left( C'_1(x_{ij})+C'_2(x_{ij}) \gamma ''^* \log(Mx)+ C'_3(x_{ij}) (\gamma ''^*)^2 \log^2(Mx) + C'_4(x_{ij}) (\gamma ''^*)^3 \log^3 (Mx) \right)
\end{split}
\end{equation}
where $x$ is any chosen one out of the $x_{ij}$. The highest power of log has a non-zero coefficient if $\gamma ''^* \neq 0$, and the corresponding term in \eqref{logthree} turns out to be proportional to $\log^3(x) \mean{\langle \cO (x_1) \cO (x_2) \cO (x_3) \cO (x_4)\rangle_{conn} }$.
The implications for the integrated correlators will be discussed below.

\subsection{Relation to statistical mechanics}

The results above can be used to compute properties of disordered fixed points, which describe second order phase transitions in disordered materials. Various fixed points of this type were analyzed long ago in the statistical mechanics literature, by the $\epsilon$ expansion and other methods, and it was found that indeed the disorder distribution flows and that it leads to additional critical exponents (see, e.g. \cite{Grinstein:1976zz,andelman1984scale}).

In pure fixed points, when one changes the temperature to go slightly away from a phase transition, all operators allowed by the symmetry are generated. The lowest one $\cE(x)$ controls the specific heat exponent $\alpha$, and higher dimension operators give measurable corrections to the leading scaling behavior. The value of $\alpha$ is related to the dimension of $\cE  (x)$ by 
\begin{equation}
\Delta _{\cE } =d \,\frac{\alpha -1}{\alpha -2}, \qquad\qquad \alpha = \frac{d-2\Delta_{\cE}}{d-\Delta_{\cE}} = (d-2\Delta_{\cE}) \nu.
\end{equation}

At random fixed points the same is true, except that there are extra contributions from the couplings controlling the disorder distribution, or equivalently from multi-replica operators in the replica picture. These mix with the standard couplings/operators, and the phase transition is characterized by the full spectrum of dimensions, containing these additional contributions beyond the ones related to local operators.  Typically the first subleading correction to objects like the specific heat will come from the first disorder-related operator $\Psi$ which in the replica picture includes $\sum_{A \neq B} \cO_A \cO_B$, and its contribution is characterized by a critical crossover exponent $\phi$ defined by
\begin{equation}
\phi = \frac{d-\Delta_{\Psi}}{d-\Delta_{\cE}} = (d-\Delta_{\Psi}) \nu.
\end{equation}
For weak disorder $\phi$ is close to $\alpha$, but in general it could be smaller or larger (as was found in\cite{andelman1984scale}). As we are flowing at long distances to a disordered fixed point, $\phi <0$ in the IR (in order to avoid extra fine-tuning). The presence of these extra powers is one difference between pure and random fixed points.

One way to characterize disorder is to consider the distribution of macroscopic variables, such as the susceptibility, in different realizations of the disorder. At a pure fixed point the susceptibility has a fixed value at large volume, while at a random fixed point it will generally have some distribution even at large volume, since random fixed points are not self-averaging \cite{aharony1996absence}; in particular the parameters of the distribution cannot be measured by going to larger systems, but only by manufacturing many systems with different random couplings. The distribution of macroscopic variables is not controlled directly by the distribution of the microscopic couplings that we discussed earlier (which is scheme-dependent), but rather by other properties of the disordered fixed point. For instance, a natural macroscopic object to look at in a finite-volume system is
\begin{equation}
\chi = {1\over {\rm Volume}} \int d^dx d^dy \langle \sigma(x) \sigma(y) \rangle_{conn},
\end{equation}
for some operator $\sigma (x)$; in the Ising model this gives the magnetic susceptibility. At a random fixed point this has some variance $\mean{\chi^2} - \mean{\chi}^2$, whose ratio to $\mean{\chi}^2$ 
is given by
\begin{equation}\label{defrchi}
R_{\chi} \equiv \frac{\int d^d x d^d y d^d z d^d w \mean{\langle \sigma(x) \sigma(y) \rangle_{conn} \langle \sigma(z) \sigma(w) \rangle_{conn}}}{\left( \int d^dx d^dy \mean{ \langle \sigma(x) \sigma(y) \rangle_{conn} }\right)^2} - 1.
\end{equation}
Note that as long as the dimension of $\sigma$ obeys $\Delta_{\sigma} < d/2$, the integrals are dominated by the regime where all the points are far from each other, of order the size of the system, so the short distance singularities where different operators come together (and one sees the effects of the operator $\sum_{A \neq B} \sigma_A \sigma_B$) do not contribute. In the replica approach $R_{\chi}$ involves ratios of 4-point functions to 2-point functions squared, which can be used to characterize the disordered fixed point; in \cite{aharony1996absence} $R_{\chi } $ was used to define the dimensionless disordered coupling $\qd$ at the fixed point, and since $R_{\chi}$ vanishes when the disorder goes to zero, this definition is as good as any other definition (the microscopic value of $\qd$ is scheme-dependent).

Since the correlators in \eqref{defrchi} are scale-invariant at a fixed point, as discussed above, $R_{\chi } $ approaches a constant at a disordered fixed point. This was used in \cite{aharony1996absence} to conclude that there is no self-averaging (for weakly disordered fixed points $R_{\chi}$ is small and there is an approximate self-averaging). Similar behavior is expected also for higher moments of the distribution of $\chi$, that will generically not be Gaussian at the fixed point.

The arguments above are relevant when $\Delta_{\sigma} < d/2$ at the random fixed point. For operators with $\Delta \geq d/2$, the correlation functions appearing in the variance are dominated by short-distances, so the averages of products will scale as a lower power of the volume than the product of the averages. Thus, correlation functions of such operators, and in particular the specific heat, will be self-averaging.

We can also consider non-connected correlation functions instead of the connected ones above; these are relevant for instance for scattering off disordered materials (see e.g. \cite{pelcovits1985structure}). These quantities are even less self-averaging when $\Delta < d/2$. The reason for this is that as we saw above, while in the non-connected 2-point function there is a single $log$ and in the square of it there is a $log^2$, in the average of a product of 2-point functions there is a $log^3$. Therefore for a system of linear size $L$, the normalized variance $R_{\chi}$ of integrated 2-point functions of operators with $\Delta < d/2$ will behave as $\log (ML)$, which approaches infinity (rather than a constant) in the infinite volume limit. This happens for operators for which $\gamma '' \neq 0$, that is, there is a mixing between the different $\{ \cO _A \}$ operators. Generically such a mixing will occur, unless there is a symmetry under which the different $\cO _A$'s transform differently. We discussed such symmetries at the end of \autoref{subsection:RGnq_replicated_theory}. For instance, in the random-bond Ising model (where we couple disorder to a $\mathbb{Z}_2$-invariant operator) there is a replica $(\mathbb{Z} _2)^n$ symmetry and $\gamma''=0$ for any $\mathbb{Z}_2$-charged operator $\sigma$. Thus in this case a logarithm will not appear in the variance of integrated two-point functions of $\sigma$, but it may appear in other correlation functions and in other examples.

\section{Classical disorder in large $N$ theories} \label{section:classical_disorder_large_N}

In this section, which can be skipped if desired, we discuss classical disorder in large $N$ field theories, at leading order in $1/N$, where they correspond to `generalized free fields'. There are three new features compared to our general discussion in \autoref{section:RG_classical_disorder}. First, we can obtain exact results for the correlation functions in the presence of disorder, that will demonstrate the GCS equations we derived.  Second, when the disorder is marginal, there is always a marginal operator also in the original theory before adding the disorder, which complicates the renormalization group flow. And finally, in large $N$ replica theories there are always degeneracies between operators involving a different number of replicas, which lead to logarithms already in the connected correlation functions. All these properties are only valid at leading order in the $1/N$ expansion, though they affect also higher orders in this expansion. Some of these properties are also present in free field theories with disorder, but in that case the degeneracies mentioned above are lifted by the renormalization group flow, while in large $N$ theories they remain.

\subsection{The RG flow in a disordered generalized free field theory} \label{subsection:RGnq_test_CS}

Throughout this section we will use the generalized free field CFT, which gives the leading order in the $1/N$ expansion. It is similar to a free field theory, except that the basic operator has a general scaling dimension.
Disorder in this theory was studied in \cite{Aharony:2015aea}, and we will see how the results found there are consistent with our GCS equations.

In a generalized free theory there are basic operators that do not talk to each other at leading order in the large $N$ limit, so we can concentrate on a single basic `generalized free field' $\cO (x)$. Its correlation functions are given by Wick's theorem with the contraction
\begin{equation} \label{eq:RGnq_GFF_2pf}
\langle \cO (x) \cO (0)\rangle=\frac{1}{x^{2\Delta  _{\cO } }},
\end{equation}
and the only operators we need to keep are products of $\cO$ with itself and with its derivatives, that are well-defined with no short-distance singularities.
This behavior characterizes both vector and matrix large $N$ theories; in matrix large $N$ theories $\cO $ is called a single-trace operator, and its products are called multi-trace operators.

We will couple disorder to $ \cO(x)$, and will be interested in the case where the disorder is marginal according to the Harris criterion, namely $\Delta _{\cO } =d/2$.
Note that the momentum space correlation function \eqref{eq:RGnq_GFF_2pf} diverges logarithmically and a cutoff is needed, so we will work in position space. For this case the deformation by $\cO ^2$ is also marginal at large $N$, and so it is natural to consider
\begin{equation} \label{eq:RGnq_GFF_disordered}
S=S_0+ \frac{\lambda }{2} \int d^d x\, \cO ^2(x)+\int d^dx\, h(x) \cO (x) .
\end{equation}
The second term is a `double-trace deformation' in the case of a large $N$ matrix theory, while the third term is the disorder term. We take the random distribution of $h(x)$ to be the Gaussian one, with width $\qd$. 
For large $N$, only correlation functions including the specific operator $\cO $ in \eqref{eq:RGnq_GFF_disordered} and its products are modified by these interactions.

Correlation functions in this section will be calculated using conformal perturbation theory. We will treat $\lambda $, $\qd$ and $h(x)$ as small parameters (the typical disorder field for a small $\qd$ is indeed small) and expand in them to get Feynman diagrams. When averaging disordered correlation functions, $h(x)$ is eliminated in favor of $\qd$. We will work up to second order in the couplings (that is $\qd^2$, $\qd\lambda $ and $\lambda ^2$).
The integrals that are needed in this section are of the following form
\begin{equation} \label{eq:RGnq_GFF_integrals}
\begin{split}
& \int d^d z \frac{1}{(x-z)^d (z-y)^d} = \frac{2 S_{d-1} \log (\Lambda |x-y|) + C_1}{(x-y)^d}, \\
& \int d^dz d^d w \frac{1}{(x-z)^d (z-w)^d (w-y)^d} = \frac{4 S_{d-1} ^2 \log ^2 (\Lambda |x-y|) + C_2 \log (\Lambda |x-y|) + C_3}{(x-y)^d}, 
\end{split}
\end{equation}
up to terms which vanish when the UV cutoff $\Lambda  \to \infty $, where $S_{d-1} $ is the volume of the $(d-1)$-dimensional sphere. The constants $C_1,C_2,C_3$ can be evaluated, but they are scheme dependent and their exact value  will not be needed. 

The replicated theory corresponding to the generalized free field theory \eqref{eq:RGnq_GFF_disordered} with Gaussian disorder is
\begin{equation}
S_{replica} =\sum _A S_{0,A} + \frac{\lambda }{2} \sum _A \int d^dx\, \cO _A^2 (x)- \frac{\qd}{2} \sum _{A,B} \int d^dx\, \cO _A(x) \cO _B(x) ;
\end{equation}
note that here we included in the last term also the case $A=B$, even though it can be swallowed into $\lambda$.
The Feynman diagrams in this theory are quite simple and consist of insertions along the propagators.
The renormalized 2-point functions in the replicated theory are found to be\footnote{We mostly use the renormalized correlation functions, in which we eliminate the short distance cutoff $\Lambda $ in favor of a finite energy scale $M$. This is achieved by a redefinition of the coupling constants and the fields such that there is no dependence in correlation functions on $\Lambda $ (where terms that vanish as $\Lambda  \to \infty $ are ignored). The bare correlation functions could instead be used just as well.}
\begin{equation} \label{eq:RGnq_GFF_replica_2pf}
\begin{split}
&\langle \cO _A(x) \cO _B(0) \rangle^{replicated}  \cdot x^d  = \delta _{AB} + 2S_{d-1}  (\qd-\delta _{AB} \lambda ) \log (Mx) + \\
&\qquad\qquad +(\qd^2n-2\qd \lambda +\delta _{AB} \lambda ^2) \log (Mx) \cdot  \left( C_2-2C_1 S_{d-1} +4S_{d-1} ^2 \log (Mx) \right) + \cdots.
\end{split}
\end{equation}
Either from the CS equation, or from the relations $\beta = M \pder{\lambda }{M} $ and $\gamma  =Z^{-1}  \frac{M}{2} \pder{Z}{M} $ with fixed bare quantities ($Z$ is the wavefunction renormalization matrix),
the beta and gamma functions that are obtained are
\begin{equation}
\begin{split}
\gamma _{AB} &= \delta _{AB} \left( S_{d-1} \lambda  + \left( S_{d-1}  C_1 - \frac{C_2}{2} \right) \lambda ^2 \right) - S_{d-1}  \qd +\left(C_1 S_{d-1}  - \frac{C_2}{2} \right) (-2\lambda \qd + n \qd^2) + \cdots , \\
\beta _{\lambda } &=2S_{d-1} \lambda ^2 + \cdots , \\
\beta _{\qd} &= 4S_{d-1}  \lambda \qd - 2n S_{d-1}  \qd^2 + \cdots .
\end{split}
\end{equation}
The operator $\cO $ corresponds to the simplest kind of operators considered in \autoref{section:RG_classical_disorder}, for which $\cO_A$ mix only among themselves. Therefore disordered connected correlation functions of $\cO $ satisfy \eqref{eq:RGnq_RG_eq_conn}.

The beta functions of the disordered theory, and the gamma function that enters the connected correlation functions, are thus \cite{Aharony:2015aea}
\begin{equation} \label{eq:RGnq_double_trace_conn_beta_gamma}
\begin{split}
\at[\gamma ' ]{n=0} &= S_{d-1} \lambda  + \left(S_{d-1}  C_1 - \frac{C_2}{2} \right) \lambda ^2 +\cdots ,\\
\at[\beta _{\lambda } ]{n=0} &= 2S_{d-1} \lambda ^2 +\cdots ,\\
\at[\beta _{\qd} ]{n=0} &= 4 S_{d-1}  \lambda  \qd +\cdots .
\end{split}
\end{equation}

A simple correlation function to be considered in the disordered theory is $\mean{\langle \cO (x) \cO (0) \rangle_{conn} }$, which actually satisfies $\mean{\langle \cO (x) \cO (0) \rangle_{conn} }= \langle \cO (x) \cO (0) \rangle_{conn}$ in this case since it happens to be independent of the disorder. 
From \eqref{eq:RGnq_GFF_replica_2pf} follows \cite{Aharony:2015aea}
\begin{equation} \label{eq:RGnq_GFF_OO_conn}
\mean{\langle \cO (x) \cO (0) \rangle_{conn} } \cdot x^d= 1-2\lambda  S_{d-1}  \log (M x)+\lambda ^2 \log (Mx) \left(C_2-2C_1 S_{d-1} +4 S_{d-1} ^2 \log (Mx) \right) +\cdots.
\end{equation}
This correlation function indeed satisfies the GCS equation \eqref{eq:RGnq_RG_eq_conn} with the beta and gamma functions in \eqref{eq:RGnq_double_trace_conn_beta_gamma}.

Note that if one sets $\lambda=0$ and considers the leading $1/N$ correction, as in \cite{Aharony:2015aea}, then the leading correction in \eqref{eq:RGnq_GFF_OO_conn} goes as $\log^2(Mx)$, which does not take the standard form of an anomalous dimension. This comes from a combination of the standard anomalous dimension with the fact that the operators $\sum_A \cO _A^2$ and $\sum_{A\neq B} \cO _A \cO _B$ no longer have the same dimension in the replica theory, giving an extra logarithm from the different running of the corresponding couplings. This extra logarithm is not related to the ones discussed in \autoref{section:RG_classical_disorder} and in the next subsection, which appear also at fixed points of the renormalization group.

\subsection{Composite operators} \label{subsection:RGnq_composite_operators}

In cases where the pure CFT is a free theory or a large $N$ theory, there are composite operators such as the operator $\cO ^2(x)$ considered above. In addition to correlation functions of $\cO $ as were used in the previous subsection, we may consider correlation functions of $\cO ^2$. For them however, we have the mixing that was considered in \autoref{subsection:RGnq_RG_equations}. That is, the replicated operator $\sum _A \cO _A^2$ mixes with the double-replica operator $\sum _{A \neq B} \cO _A \cO _B$. This is a particular case of the discussion that led to \eqref{eq:RGnq_RG_eq_conn_more_general}. The disordered connected correlation functions of the composite operator thus obey
\begin{equation} \label{eq:RGnq_CS_O2_from_replica}
\begin{split}
& \left(M \pder{}{M} + \beta _{\qd} \pder{}{\qd} + \beta _{\lambda _i}  \pder{}{\lambda _i}  + k \gamma _{ \cO ^2} \right) \mean{\langle \cO ^2(x_1) \cdots \cO ^2(x_k) \rangle_{conn} } \, - \\
& - \gamma '_{ \cO ^2} \left[  \sum _{\substack{\text{Partitions of } \{2,\dots ,k\}\\ \text{into } S_1,S_2}}  \mean{\langle \cO (x_1) \prod _{l \in S_1} \cO ^2(x_l)\rangle_{conn} \langle \cO (x_1) \prod _{l \in S_2} \cO ^2(x_l)\rangle_{conn} } + \right. \\
& \qquad\qquad\qquad +   (x_1 \leftrightarrow x_2) + \cdots + (x_1 \leftrightarrow x_k)  \Bigg] = 0 .
\end{split}
\end{equation}

In addition, in large $N$ theories there can be degeneracies in the scaling dimensions of operators, and in particular the two operators above, $\sum \cO _A^2$ and $\sum _{A \neq B} \cO _A \cO _B$, are degenerate in the large $N$ and $n \to 0$ limit.\footnote{As $n \to 0$, $\tilde \cO $ and $\tilde \cO _A$ have the same dimensions \cite{Cardy:2013rqg} (see footnote \ref{footnote:Lifshitz_same_n_to_0_dim}), and at large $N$, $[\tilde \cO ^2]=2[\tilde \cO ]$ and $[\tilde \cO _A^2]=2[\tilde \cO _A]$. As a consequence of these, in the limit $n \to 0$  for large $N$, $\sum_A \cO _A^2$ and $\sum _{A \neq B} \cO _A \cO _B$ have the same dimension.} Therefore, in a large $N$ disordered fixed point, we are generically in a situation in which the anomalous dimension matrix can only be brought to a Jordan form, as in the discussion below \eqref{eq:RGnq_fixed_point_connected_replica_CS}.
In such cases, the three CS equations \eqref{eq:RGnq_fixed_point_connected_replica_CS} can be brought to the form \eqref{eq:RGnq_fixed_points_2pf}, with one additional equation of the form of the second equation there. The solutions to these equations involve logarithms. When brought back to the original basis of correlation functions, we see that in the large $N$ limit also in the disordered \emph{connected} correlation functions there will be logarithms. The phenomenon of the anomalous dimension matrix taking a Jordan form and resulting in logarithms was understood in \cite{Gurarie:1993xq}.

%

To demonstrate this, we will change basis and use the basis of representations of $S_n$. The coupling $\lambda $ is associated with the operator $\sum _A \cO _A^2= \frac{1}{n} \tilde  \cO ^2 + \sum _A \tilde  \cO _A^2$, while $\qd$ is the coefficient of the operator $ \tilde \cO ^2=\sum _{A,B} \cO _A \cO _B$. The replicated theory in the basis of $\tilde \cO $ and $\tilde  \cO _A$ is thus
\begin{equation} \label{eq:RGnq_GFF_replica_action_Sn_basis}
S_{replica} = \sum _A S_{0,A} + \frac{1}{2} \left(\frac{\lambda }{n} -\qd\right) \int d^dx\,  \tilde  \cO ^2(x)+ \frac{\lambda }{2} \int d^dx \, \sum _A \tilde \cO _A^2 .
\end{equation}
Using the relations to $\cO _A$, we get the two point functions in the generalized free field theory
\begin{equation}
\begin{split}
\langle\tilde \cO (x) \tilde \cO (0)\rangle_0 &= \frac{n}{x^d}, \\
\langle\tilde  \cO _A(x) \tilde  \cO (0) \rangle_0 &= 0, \\
\langle \tilde \cO _A(x) \tilde  \cO _B(0)\rangle_0 &= \frac{\delta _{AB} -1/n}{x^d} .
\end{split}
\end{equation}
The second one vanishes as expected by the symmetry.

Since $\tilde \cO $ and $\tilde  \cO _A$ are decoupled, at large $N$ the correlation functions of only $\tilde \cO $'s and correlation functions of only $\sum _A \tilde  \cO _A^2$ are renormalized multiplicatively. The renormalized correlators and gamma functions to second order can be found to be
\begin{equation} \label{eq:RGnq_GFF_composte_example_1}
\begin{split}
& \langle \tilde \cO ^2(x)  \tilde  \cO ^2(0) \rangle \frac{x^{2d} }{2n^2} = \\
&\qquad = 1+4S_{d-1}  (n\qd-\lambda ) \log (Mx) +
2(\lambda -n \qd)^2 \log (Mx) \cdot \left(C_2-2C_1 S_{d-1} +6 S_{d-1} ^2 \log (Mx) \right) +\cdots \\
&\ \gamma _{\tilde \cO ^2} = 2 S_{d-1}  (\lambda -n \qd) + (2C_1 S_{d-1} -C_2)(\lambda -n \qd)^2 + \cdots,
\end{split}
\end{equation}
and
\begin{equation} \label{eq:RGnq_GFF_composte_example_2}
\begin{split}
\langle\sum _A & \tilde \cO _A^2 (x) \sum _B \tilde  \cO _B^2(0)\rangle \frac{x^{2d}}{2(n-1)} = \\
&= 1-4 S_{d-1} \lambda  \log (Mx) + 2\lambda ^2 \log (Mx) \cdot \left(C_2-2C_1 S_{d-1} +6 S_{d-1} ^2 \log (Mx) \right) +\cdots \\
\gamma _{\sum \tilde \cO _A^2} &= 2 S_{d-1} \lambda  + (2C_1 S_{d-1} -C_2) \lambda ^2 +\cdots .
\end{split}
\end{equation}
Of course the same renormalizations of $\lambda ,\qd$ were used as in $\langle\cO _A \cO _B\rangle$ above.

In this example, $\tilde  \cO ^2=\sum _{A,B} \cO _A\cO _B$ and $\sum _A \tilde  \cO _A^2 = \sum _A \cO _A^2- \frac{1}{n} \tilde  \cO ^2$ diagonalize the anomalous dimensions. The operator relevant for the disordered connected correlation functions, $\sum _A \cO _A^2$, is not an eigenvector of the anomalous dimensions. Hence the GCS equation mixes its disordered connected correlation functions with other correlation functions.

A simple example of the GCS equation arises in the theory with $\lambda =0$. A-priori setting $\lambda =0$ should be avoided since this deformation is expected to be generated and should be included. However, at large $N$ taking $\lambda =0$ is consistent. By considering the replicated action \eqref{eq:RGnq_GFF_replica_action_Sn_basis} in the basis $ \tilde \cO $ and $\tilde  \cO _A$,  the $\tilde  \cO $ sector and $\tilde  \cO _A$ sector are decoupled. If we begin with $\lambda =0$, the $\tilde \cO _A$ sector will remain free, and so the double-trace deformation will never be generated.

The GCS equation will be tested on the bare correlators for simplicity, which is like testing the RG flow at the cutoff scale $\Lambda $. Using perturbation theory in the disordered theory and then averaging over the disorder, we get the following simple correlators, which are exact in $\qd$ (there are no higher order corrections) \cite{Aharony:2015aea}:
\begin{equation} \label{eq:RGnq_GFF_composite_correlators}
\begin{split}
\mean{\langle \cO ^2(x) \cO ^2(0) \rangle_{conn} }\cdot x^{2d} &= 2+4\qd\left( 2 S_{d-1}  \log (\Lambda x)+C_1 \right), \\
\mean{\langle\cO (0) \cO ^2(x)\rangle_{conn} \langle\cO (0)\rangle_{conn} } \cdot x^{2d} &= 2\qd \left(2 S_{d-1}  \log (\Lambda x) + C_1 \right) .
\end{split}
\end{equation}
At first sight the log in the 2-point function (at all scales) is confusing.
However, the GCS equation \eqref{eq:RGnq_CS_O2_from_replica} for the 2-point function of $\cO ^2$ is
\begin{equation} \label{eq:RGnq_GFF_composite_2pf_RG_eq}
\begin{split}
& \left( \Lambda  \pder{}{\Lambda } +\beta _{\qd} \pder{}{\qd} +2 \gamma _{\cO^2} \right) \mean{\langle \cO ^2(x) \cO ^2(0)\rangle_{conn} }  -4 \gamma '_{ \cO^2} \mean{\langle\cO (0) \cO ^2(x)\rangle_{conn} \langle\cO (0)\rangle_{conn} } = 0 .
\end{split}
\end{equation}
In the evaluation of $\langle \cO _A \cO _B\rangle$ in the replicated theory with $\lambda =0$, the term with $k$ powers of $\qd$ comes with $n^{k-1} $. Therefore, for $n=0$ the beta function vanishes: $\beta _{\qd} \at{\lambda =n=0}=0$ to all orders in $\qd$. Thus \eqref{eq:RGnq_GFF_composite_correlators} and \eqref{eq:RGnq_GFF_composite_2pf_RG_eq} give
\begin{equation}
\begin{split}
8 S_{d-1} \qd +2 \gamma _{ \cO ^2} \left(2+4\qd (2 S_{d-1}  \log (\Lambda x) + C_1) \right) - 4 \gamma '_{\cO^2}  \cdot 2\qd \left(2 S_{d-1}  \log (\Lambda x) + C_1 \right) = 0.
\end{split}
\end{equation}
We see that for
\begin{equation} \label{eq:RGnq_GFF_exact_gamma_no_double_trace}
\gamma _{ \cO ^2} = \gamma '_{ \cO ^2} = -2 S_{d-1} \qd
\end{equation}
the GCS equation is indeed satisfied by \eqref{eq:RGnq_GFF_composite_correlators} to all orders in $\qd$.

This term is scheme independent and should match the results \eqref{eq:RGnq_GFF_composte_example_1}, \eqref{eq:RGnq_GFF_composte_example_2} found at the beginning of this subsection. There we used another basis of operators. The current basis is related to the previous one by $\sum_A \cO _A^2 = \frac{1}{n} \tilde \cO ^2 + \sum _A \tilde  \cO _A^2$ and $ \sum _{A \neq B} \cO _A \cO _B = \frac{n-1}{n} \tilde  \cO ^2 - \sum _A \tilde  \cO _A^2$, that is, by the transformation matrix
\begin{equation} \label{eq:RGnq_GFF_composite_basis_transformation}
O = \begin{pmatrix}\frac{1}{n} & 1 \\ \frac{n-1}{n} & -1\end{pmatrix} .
\end{equation}
In the previous basis the anomalous dimensions matrix $\gamma $ was diagonal. In the new basis $\cO'_i = O_{ij} \cO _j$, the anomalous dimensions are $\tilde \gamma = O\gamma  O^{-1} $. Using \eqref{eq:RGnq_GFF_composte_example_1}, \eqref{eq:RGnq_GFF_composte_example_2}, \eqref{eq:RGnq_GFF_composite_basis_transformation} above, and then substituting $n=\lambda=0$, we get
\begin{equation}
\tilde \gamma  =  \begin{pmatrix}-2 S_{d-1} \qd & -2 S_{d-1} \qd \\ 2 S_{d-1}  \qd & 2 S_{d-1}  \qd \end{pmatrix}
\end{equation}
(this result is from the calculation to order $\qd^2$). The first line indeed matches the values \eqref{eq:RGnq_GFF_exact_gamma_no_double_trace}.

\section{The renormalization group for quantum disorder} \label{section:RG_quantum_disorder}

The quantum version of quenched disorder, in which the disorder varies in spatial directions but there is also a time direction, is described in a similar way to the classical case, but it involves a few additional complications.

We now think about the quantization of a many body system in the presence of disorder, which might for instance be caused by impurities. The basic many body system will be defined on a $d$-dimensional space. In the path integral quantization of the system, we use fields which are defined on the $d$-dimensional space, as well as one additional time direction. In the absence of the disorder, we assume for simplicity that the system is critical (conformal) and Lorentz-invariant, though the generalization of our results to general systems which can be non-relativistic and non-scale-invariant is straightforward, and this assumption does not affect any of our conclusions. The path integral quantization is based on the action which is a functional of the fields defined on the $(d+1)$-dimensional Minkowski space. Denoting the coordinates of space by boldface letters $\vecx$ (having $d$ components) and time by $t_M$, the disorder field is taken to vary only in space $h=h(\vecx)$. We will also use $x,y,\dots $ for the $d+1$ dimensional spacetime coordinates. As before we denote by $S_0$ the action of the pure system. In the presence of the disorder coupled to the operator $\cO _0$, the action is
\begin{equation}
S = S_0- \int d^d\vecx dt_M \, h(\vecx) \cO _0(\vecx,t_M) .
\end{equation}
We will analytically continue Minkowski space to a Euclidean space using $t_M=-it$, in which the action becomes 
\begin{equation} \label{eq:Lifshitz_action_with_disorder}
S_E = S_{0,E} + \int d^d\vecx dt\, h(\vecx) \cO _0(\vecx,t) .
\end{equation}
In the following we will work exclusively in Euclidean space, and omit the subscripts denoting that.\footnote{In Minkowski space this means that we will only consider time-ordered correlation functions.} The total spacetime dimension will be denoted by
\begin{equation}
\dcft = d+1.
\end{equation}

Exactly as before we will have a probability distribution $P[h(\vecx)]$ for the disorder, and an average of a quantity $X$ with respect to it is denoted by $\mean{X}$. The Gaussian probability distribution \eqref{gaussian} is defined as before. 

In the presence of a particular disorder configuration $h(\vecx)$, the symmetries under space translations and spacetime rotations are broken, but the time translation symmetry remains. After averaging, similarly to the classical case, if $P[h]$ is symmetric under translations and rotations, the space translation and rotation symmetries are restored (we assume they are not spontaneously broken). However, the symmetry under full spacetime rotations is not restored, and instead of this $SO(d+1)$ symmetry, we are left in the averaged correlation functions with an $SO(d) \times \mathbb{Z} _2$ symmetry, where the $\mathbb{Z} _2$ acts as a simultaneous time reversal and reflection of one spatial dimension. As a consequence, at disordered critical points we should not expect in general a scaling symmetry under which $\vecx \to \lambda \vecx$ and $t \to \lambda t$, even if we start from a relativistic pure system. A more generic possibility is that Lifshitz scaling may emerge, under which $\vecx \to \lambda \vecx$ and $t \to \lambda ^z t$, where $z$ is the dynamical exponent.

The Harris criterion for the case of quantum disorder is modified as follows. Denoting again the dimension of $\cO _0$ in the pure theory by $\Delta _0$ and taking a Gaussian disorder distribution, the width $\qd$ has now the dimension $[\qd]=d+2-2\Delta _0$. This means that disorder is relevant for $\Delta _0 < \frac{d+2}{2} $, marginal for $\Delta _0 = \frac{d+2}{2} $ and irrelevant for $\Delta _0 > \frac{d+2}{2} $. Assuming that disorder couples to the lowest dimension operator ${\cal E}(x)$ whose dimension is related to the critical exponent $\nu = 1 / (d+1-\Delta_0)$, this means that we should have $\nu \le 2/d$ for disorder to be marginal or relevant.
As before, the case where we may have perturbative control is when the operator is close to the marginal value $\Delta _0 = \frac{d+2}{2} $.


The analysis of \autoref{subsection:RGnq_replicated_theory} goes through for the quantum case,
but with $P[h]$ independent of time such that $h$ varies only along the spatial directions. The replicated theory is defined just in the same way. In particular, for Gaussian disorder, the replicated action is
\begin{equation} \label{eq:Lifshitz_basic_replica_action}
S_{replica} = \sum _{A=1}^n S_{0,A} -\frac{\qd}{2} \sum _{A,B=1}^n \int d^d\vecx dt dt'\, \cO _{0,A} (\vecx,t) \cO _{0,B}(\vecx,t') .
\end{equation}
Note that in \eqref{eq:Lifshitz_basic_replica_action} we do need to include in the sum the terms with $A=B$, since the two operators are generally separated. In general we will have singularities in the operator we added to the action when $A=B$ and $t' \to t$; this will have interesting consequences below.

In \eqref{eq:Lifshitz_basic_replica_action} we see the main complication in the case of quantum disorder: the replica theory is not local in time. 
Because of this it is far from obvious that we can use Wilsonian RG methods; however, the local RG point of view discussed in \autoref{subsection:local_RG_approach} suggests that we can, and we will argue (and show explicitly in examples) that for $n\to 0$ we can indeed use the renormalization group. For finite values of $n$ the theories \eqref{eq:Lifshitz_basic_replica_action} suffer from IR divergences related to the extra time integral and we will not try to make sense of them here.


There are many interesting works about quantum disorder (see \cite{sachdev2001quantum} for a survey of the subject). We will relate our general analysis to two specific examples.
In \cite{Boyanovsky:1982zz} a weakly coupled $O(m)$ model was investigated, and the breaking of the symmetry between space and time was noticed. Additionally, the renormalization group was studied for this model, and the dynamical exponent $z$ was evaluated perturbatively.
In \cite{Hartnoll:2014cua}, holography was used to study a large $N$ system with exactly marginal quantum disorder. 
It was found that Lifshitz scaling emerges at low energies, and the dynamical exponent $z$ was calculated analytically and numerically as a function of the dimensionless $\qd$.\footnote{Another model where Lifshitz scaling arises from holographic disorder was studied in \cite{Garcia-Garcia:2015crx}.}

The purpose of this section is to study the renormalization group for the case of quantum disorder, emphasizing general properties. There are at least two main goals. The first is to address the general appearance of Lifshitz scaling both qualitatively (showing where it comes from) and quantitatively (evaluating perturbatively the dynamic exponent $z$). Second, we would like to see if there are GCS equations for the case of quantum disorder as well, given that the replica theory now is non-local. We will again mostly be interested in disordered fixed points, 
and will ignore various difficulties (such as Griffiths-McCoy singularities \cite{griffiths1969nonanalytic,mccoy1969incompleteness,mccoy1969theory}) in practical realizations and measurements of such fixed points.

\subsection{Universal properties of quantum disorder}\label{universal}

We start from a pure theory and add disorder to it, starting from the simplest case of a Gaussian disorder distribution (we will discuss the generalization later). In the language of the replica trick this deformation of the pure theory is described by \eqref{eq:Lifshitz_basic_replica_action}.
We would like to see what happens as we flow along the renormalization group. A useful way to do that is to use conformal perturbation theory: we assume that the disorder $\qd$ is small, and expand in it. Denoting the partition function of the pure theory by $Z_0$, the partition function of the deformed theory is
\begin{equation} \label{eq:Lifshitz_CPT_expansion}
\begin{split}
\frac{Z}{Z_0} &= 1 + \frac{\qd}{2}  \sum _{AB} \int d^d\vecx dt dt' \langle\cO _{0,A}(\vecx,t) \cO _{0,B}(\vecx,t') \rangle +\\
&+ \frac{\qd^2}{8} \sum _{AB} \sum _{CD} \int d^d\vecx_1 dt_1 dt'_1 d^d\vecx_2 dt_2 dt'_2 \langle \cO _{0,A}(\vecx_1,t_1) \cO _{0,B}(\vecx_1,t'_1) \cO _{0,C}(\vecx_2,t_2) \cO _{0,D}(\vecx_2,t'_2) \rangle + \cdots .
\end{split}
\end{equation}
The expectation values are evaluated in the pure replica theory. Even if some expectation value vanishes, we do not set it to zero since we are interested also in the results with operator insertions. Whenever operators in this expansion are coincident, we will get divergences. This is dealt with as usual by introducing a cutoff. A simple cutoff is introduced by having a minimal distance $1/\Lambda $ between spacetime points. For instance this is the case when the system is defined on a (space-time) lattice. To implement the renormalization group, we modify the coupling constants such as to cancel these cutoff dependences, ensuring that the physics at long distances remains the same as the cutoff changes.

A new phenomenon that occurs in quantum disorder 
is that the bi-local operator that we added to the action \eqref{eq:Lifshitz_basic_replica_action} needs regularization. The terms with $A=B$ are singular as $t'\to t$, leading to a (generally divergent) mixing of this bi-local operator with integrals of local operators $\sum_A \int d^d \vecx dt\, \cO'_A(\vecx,t)$. We will discuss this mixing in more detail in the next subsection. In general these local operators appear already in the original pure action $S_0$, so including them does not add anything new to the RG analysis. However, the disorder-related operator that we added can mix also with operators that are not Lorentz-invariant but are only $SO(d)$ invariant. These do not appear in the original action assuming that we start from a relativistic theory, and they must also be added now to the action, beginning at order $\qd$. In a general RG flow we would need to add all such operators and they would all flow.

A particularly interesting operator, that mixes with all the bi-local operators related to quantum disorder, is the integral of the time-time component of the energy-momentum tensor $T_{00}(\vecx,t)$, namely the Hamiltonian integrated over time. This is always a marginal operator, so its coefficient is dimensionless, and it can exhibit universal logarithmic divergences in a perturbative expansion. 
Adding this operator will be interpreted below as a stretching of the time direction relative to the spatial directions, but for now let us just analyze the RG flow of the replica theory and show that we have to add this operator to the action; we do this in the case where the disorder is marginal such that we can perform explicit perturbative computations.


In order to make sense of \eqref{eq:Lifshitz_CPT_expansion} at the leading order in $\qd$, note that when $t'$ is close to $t$ in the linear term in $\qd$, we can use the OPE $\cO_0 \times \cO_0 $.  We show in \autoref{section:Lifshitz_T_in_OO} that in this OPE there is a universal term
\begin{equation} \label{eq:Lifshitz_CPT_OOT_OPE}
\cO_0 (x) \cO_0 (0) \supset \frac{c_{\cO \cO T} }{c_T} \frac{x^{\mu } x^{\nu } }{x^{2\Delta_0 -\dcft+2} } T_{\mu \nu } (0) \qquad \text{(in } \dcft \text{ dimensions),}
\end{equation}
where $c_T$ and $c_{\cO \cO T} $ are defined in \eqref{eq:Lifshitz_T_in_OO_TT} and \eqref{eq:Lifshitz_T_in_OO_OOT} (with $\cO=\cO _0 $ there). Applying it to our case, we set $\dcft =d+1$ and $\Delta_0 = (d+2)/2$ corresponding to marginal disorder, and get in \eqref{eq:Lifshitz_CPT_expansion}
\begin{equation}
\frac{\qd}{2} \sum _A \int d^d\vecx dt dt' \frac{c_{\cO \cO T} }{c_T} \frac{1}{|t - t'|} \langle T_{00,A} (\vecx,t) \rangle \sim \qd \frac{c_{\cO \cO T} }{c_T}  \log (\Lambda _t) \sum _A \int d^d\vecx dt \, \langle T_{00,A} (\vecx ,t)\rangle .
\end{equation}
The notation $\Lambda _t$ stands for the short distance cutoff in the time direction, and is meant to emphasize that in a relativistic theory it does not matter in what directions precisely we use a cutoff, but here it may be more natural to use different cutoffs in the space and time directions.

This means that a deformation proportional to $T_{00} $ is generated by the RG flow.\footnote{A similar logarithmic running of the coefficients of the energy-momentum tensor was seen also in non-relativistic theories (see, for example, \cite{Korovin:2013bua}).} We should therefore modify \eqref{eq:Lifshitz_basic_replica_action} to\footnote{The motivation for denoting the new coupling constant by $h_{00}$ is that when coupling a field theory to a curved space, the linear order deformation corresponding to the metric variation $h_{\mu \nu } $ is $h_{\mu \nu } T^{\mu \nu } $. However this is just a notation and no coupling to a curved space will be needed here.}
\begin{equation} \label{eq:Lifshitz_replica_action}
S_{replica} = \sum _A S_{A,0} - \frac{\qd}{2} \sum _{A,B} \int d^d \vecx dt dt'\, \cO _{0,A}(\vecx ,t) \cO _{0,B}(\vecx ,t') + h_{00} \sum _A \int d^d \vecx dt \, T_{00,A} (\vecx ,t) .
\end{equation}
To leading order the flow of the coupling $h_{00} $ is such as to compensate for the logarithmic cutoff dependence that was found,
\begin{equation}
\delta h_{00}  = \frac{\qd c_{\cO \cO T} }{c_T} \log (\Lambda _t) + O(\qd^2).
\end{equation}
The corresponding beta function is
\begin{equation} \label{eq:Lifshitz_CPT_h00_beta}
\beta _{h_{00} } = \frac{\qd c_{\cO \cO T} }{c_T} + O(\qd^2).
\end{equation}
The new term in \eqref{eq:Lifshitz_replica_action} is a single-replica term, so this deformation corresponds to adding $h_{00} \int d^d \vecx dt\, T_{00}(\vecx, t)$ to the original disordered theory. Note that we could also add terms proportional to $T^{\mu}_{\mu}$, in the disordered theory or in the replica theory, but since we start from a CFT this vanishes by the equations of motion, so it can be removed by a field redefinition. In particular, adding $T_{00}$ is the same as adding $T_{ii}$ with an opposite sign.
This is true only around the CFT, since along the RG flow we have only a well defined $T_{00} $.

Even though this result used the replica theory corresponding to a Gaussian disorder, it holds for a general local $P[h]$. Defining $\qd$ through $\mean{h(\vecx )h(\vecy )}=\qd \delta (\vecx -\vecy )$, the replica theory will include the $\qd$ term in \eqref{eq:Lifshitz_basic_replica_action} that was used, as well as additional terms from the higher moments of the distribution, see \autoref{section:Ph_non_Gaussian}. If the Gaussian disorder is marginal then all these additional terms are irrelevant when $\Delta_0 > 1$ (namely $d > 0$), and do not affect \eqref{eq:Lifshitz_CPT_h00_beta}.


The generated $T_{00} $ deformation leads to an anisotropy between space and time; in fact it is equivalent to rescaling time, as will be shown now. For a general theory with action $S$, the variation of the action from the infinitesimal transformation $x'_{\mu } =x_{\mu } +\epsilon _{\mu } (x)$ is $\delta S = - \int d^{\dcft } x \, \partial _{\mu } \epsilon _{\nu } T^{\mu \nu } $, as follows from Noether's theorem.\footnote{Another way to see this is to couple the theory to a curved background space and use the usual convention for the energy momentum tensor (see for instance \cite{Osborn:1993cr})
\begin{equation}
T^{\mu \nu } = - \frac{2}{\sqrt{g} } \frac{\delta S}{\delta g_{\mu \nu } } .
\end{equation}
After restricting to flat space, the variation of the action under $x'_{\mu } =x_{\mu } +\epsilon _{\mu } $ is the same as written above. }
Therefore for an infinitesimal time dilation, $t'=t(1+\epsilon)$, 
\begin{equation} \label{eq:Lifshitz_action_variation_under_time_rescaling}
\delta S = -\epsilon  \int d^d\vecx dt \, T_{00} (x,t) .
\end{equation}
Since at leading order $h_{00} $ is the coefficient of the total energy of the replica theory (or of the original theory), 
we conclude that a small $h_{00} $ deformation is equivalent to rescaling time. The relation between correlation functions in the presence of this $h_{00} $ deformation (denoted in the following equation by adding an $h_{00} $ subscript) and correlation functions without it, is (for scalar operators)
\begin{equation} \label{eq:Lifshitz_correlators_relation_to_time_dilation}
\begin{split}
\langle \cO _{i_1,A_1}(\vecx _1,t_1) \cdots \cO _{i_k,A_k} (\vecx _k,t_k) \rangle_{h_{00} } &= \langle \cO _{i_1,A_1}(\vecx _1,t_1) \cdots \cO _{i_k,A_k} (\vecx _k,t_k) \rangle - \\
&- h_{00}  \sum_{i=1} ^k t_i \pder{}{t_i} \langle \cO _{i_1,A_1}(\vecx _1,t_1) \cdots \cO _{i_k,A_k} (\vecx _k,t_k) \rangle + O(h_{00} ^2).
\end{split}
\end{equation}

So far we discussed the first (linear) order in perturbation theory of \eqref{eq:Lifshitz_replica_action}. Consider next higher orders. Since $h_{00} $ is just equivalent to time rescaling, terms in the perturbative expansion which include $h_{00} $ will not give rise to new contributions which diverge as the cutoff is removed (such as logarithmic terms). This can be checked explicitly to second order.

Thus, the only remaining contribution to the RG flow from \eqref{eq:Lifshitz_replica_action} at second order in the couplings is from the $\qd^2$ term in \eqref{eq:Lifshitz_CPT_expansion}.
Cutoff dependent terms arise when two or more operators become close to each other. This $\qd^2$ term, which multiplies the operator $\cO _A \cO _B \cO _C \cO _D$ (with the spacetime arguments and the `0' subscript left implicit), will contribute to the running of various operators. One contribution comes from the region where $\cO _A$,$\cO _C$ are close to each other, as well as $\cO _B$,$\cO _D$ being close (and the symmetric region obtained by $C \leftrightarrow D$). In these regions we can use the OPE $\cO_0 \times \cO_0 $. Defining $c_{\cO \cO } $ and $c_{\cO \cO \cO } $ as the coefficients in the two and three point functions of $\cO_0 $
\begin{equation}
\begin{split}
\langle \cO_0 (x_1) \cO_0 (x_2) \rangle &= \frac{c_{\cO \cO } }{(x_1-x_2)^{2\Delta_0 } } , \\
\langle \cO_0 (x_1) \cO_0 (x_2) \cO_0 (x_3) \rangle &= \frac{c_{\cO \cO \cO } }{| (x_1-x_2)(x_1-x_3)(x_2-x_3)|^{\Delta_0 } } ,
\end{split}
\end{equation}
there is the following term in the OPE (evaluated by the same strategy used in \autoref{section:Lifshitz_T_in_OO})
\begin{equation}
\cO_0 (x) \cO_0 (0) \supset \frac{1}{x^{\Delta_0 } } \frac{c_{\cO \cO \cO } }{c_{\cO \cO } } \cO_0 (0) .
\end{equation}
Using this OPE in the above regions (with $\Delta_0 =(d+2)/2$) gives a contribution logarithmic in the cutoff to the disorder operator corresponding to $\qd$. It is compensated by taking
\begin{equation} \label{Lifshitz_CPT_beta_g_partial}
\delta \qd = - \frac{\qd^2}{2} S_{d-1} \frac{c_{\cO \cO \cO } ^2}{c_{\cO \cO } ^2} \frac{\pi  \Gamma \left( \frac{d}{4} \right)^2}{\Gamma \left( \frac{d+2}{4} \right)^2} \log (\Lambda _{\vecx})
\end{equation}
(where again $S_{d-1} $ is the volume of the $d-1$ dimensional sphere), giving a universal contribution to the beta function of $\qd$. Unfortunately, this is only one contribution, and we cannot infer from it the beta function of $\qd$ to order $\qd^2$. There is for instance the region where the three operators $\cO _A$,$\cO _B$,$\cO _C$ are close together, and behave as a single $\cO _A$, giving together with the $\cO _D$ an additional correction to the running of $\qd$. This is not the usual OPE of two operators and cannot be evaluated for a general theory. Using the OPE to get the contribution \eqref{Lifshitz_CPT_beta_g_partial} is the same as what is done in a usual local field theory to get the $\qd^2$ term in the beta function of a coupling $\qd$ corresponding to a local operator. The region where three operators are close together is what gives the $\qd^3$ term in the beta function of a local interaction, for which there is no formula in terms of simple CFT data. In the disordered case, this complication appears already at order $\qd^2$. We will see an explicit example of this in \autoref{subsection:Lifshitz_around_free_theory}.

A straightforward generalization of quantum disorder is when the disorder is homogeneous in an arbitrary number $d_t$ of directions, such that classical disorder corresponds to $d_t=0$ while quantum disorder is $d_t=1$.
This is actually useful in many contexts, such as theories of constant random couplings ($d_t=\dcft$), or theories with disorder that is homogeneous in some directions and not others.
The Harris criterion states that coupling disorder to an operator of dimension $\Delta_0 $ will be relevant if $\Delta_0 < \frac{\dcft + d_t}{2} $ (again $\dcft $ includes both the directions in which the disorder varies and those in which it is homogeneous), irrelevant if $\Delta_0 > \frac{\dcft + d_t}{2} $ and marginal for $\Delta _0 = \frac{\dcft +d_t}{2} $. Using the OPE \eqref{eq:Lifshitz_CPT_OOT_OPE} as before, there will be generated $T_{\alpha \alpha} $ deformations for $\alpha =0,\dots ,d_t-1$ along the directions on which the disorder does not depend:
\begin{equation} \label{wlog}
\begin{split}
& \frac{\qd}{2}  \frac{c_{\cO \cO T} }{c_T} \sum _A \int d^{\dcft -d_t} \vecx d^{d_t} t d^{d_t} t' \sum _{\alpha ,\beta  =0} ^{d_t-1} \frac{(t'-t)_{\alpha } (t'-t)_{\beta } }{|t'-t|^{2\Delta_0 -\dcft +2} } T_{\alpha \beta ,A} (\vecx ,t) \sim \\
& \qquad \sim \frac{\qd}{2}  \frac{c_{\cO \cO T} }{c_T} \frac{S_{d_t-1} }{d_t} \log (\Lambda _t) \sum _A \int d^{\dcft -d_t} \vecx d^{d_t} t \sum _{\alpha =0} ^{d_t-1} T_{\alpha \alpha ,A} (\vecx ,t) .
\end{split}
\end{equation}
By analogy with the quantum case, the directions in which the disorder is homogeneous were denoted by $t$.
The disorder was taken again to be marginal or very close to marginal (in the latter case the log in \eqref{wlog} is the first term in an expansion in powers of log).
The deformation
\begin{equation} \label{eq:Lifshitz_generated_T00_general_dt}
\delta S_{replica} = \sum _A h_{00}  \sum _{\alpha  =0} ^{d_t-1} \int d^{\dcft -d_t}\vecx  d^{d_t} t \, T_{\alpha \alpha  ,A} (\vecx ,t)
\end{equation}
is thus generated with
\begin{equation} \label{eq:Lifshitz_CPT_h00_beta_general_d_t}
\beta _{h_{00 } } = \frac{\qd c_{\cO \cO T} }{2 c_T} \frac{S_{d_t-1} }{d_t} +\cdots = \frac{\qd c_{\cO \cO T} }{2 c_T} \frac{\pi^{d_t/2}}{ \Gamma((d_t+2)/2)} + \cdots.
\end{equation}
This is again equivalent to a rescaling of the directions in which the disorder is homogeneous (and the first term on the second line of equation \eqref{eq:Lifshitz_correlators_relation_to_time_dilation} is modified to a sum over all the $d_t$ directions).

\subsection{Operator mixings and generated interactions in the theory} \label{subsection:Lifshitz_replica_interactions}

In quantum disorder the non-locality of the replica theory in time naively allows many more possibilities for operator mixings and interactions compared to the classical disorder case, and we would like to see which ones actually arise. We will show that as expected the only new interactions that arise are those that can be interpreted as corrections to the disorder distribution.


The replica theory suggests, as we will argue in a moment, that operator mixings in quantum disorder should be non-local in the time direction. Such a mixing seems very surprising since the disordered theories are local, so their RG evolution should not include any non-local effects. However, the non-local effects arise after averaging over the disorder, because the disorder distribution correlates disorder at the same position at different times. For example, if one repeats the analysis around \eqref{local_mixing} of the local RG flow, one finds that in the same setup described there, the correlation function $\mean{\langle \cO_i(\vecx,t) \cdots \rangle}$ mixes with
\begin{equation} \label{eq:Lifshitz_local_RG_average_mixing}
-\qd \left( \mean{\langle \bigg(\int dt'\, \cO_0(\vecx,t')\bigg) \cO_j(\vecx,t) \cdots \rangle} - \mean{\langle \int dt'\,  \cO_0(\vecx,t') \rangle \langle \cO_j(\vecx,t) \cdots \rangle} \right).
\end{equation}
Unlike in classical disorder, the operators $\cO_0$ and $\cO_j$ are now at different times, so there is no short-distance singularity, and \eqref{eq:Lifshitz_local_RG_average_mixing} is well-defined. For connected correlation functions this implies that $\mean{\langle \cO_i(\vecx,t) \cdots \rangle_{conn}}$ mixes with $\mean{\langle (\int dt'\,\cO_0(\vecx,t')) \cO_j(\vecx,t) \cdots \rangle_{conn}}$, without any additional contributions.

Going back to the replica theory, the non-locality of the replica action in time is reflected in singularities that are independent of some of the time differences between local operators.
Considering a correlation function including operators $\cO_{0,A}(\vecx,t)$ and $\cO_{0,B}(\vecy,t')$, already at leading order in perturbation theory in $\qd$ we obtain a UV divergence from the interaction term in \eqref{eq:Lifshitz_basic_replica_action}, whose form is independent of $t$ or $t'$. The origin of this divergence comes from one of the operators in the disorder interaction $\qd$ approaching an external operator. In this region we can use the OPE, and conclude that the UV divergence can be cancelled by a mixing of the operator $\cO _{0,A}(\vecx ,t)$ with the operator $\sum _B \int dt'\, \cO _{0,B}(\vecx ,t')$ with a divergent coefficient.

More generally, in conformal perturbation theory, an operator $\cO'_A(\vecx,t)$ will mix with any operators that appear with singular coefficients when we bring it together with the interaction vertices in \eqref{eq:Lifshitz_CPT_expansion}. The general form of such operators that $\cO '_A(\vecx ,t)$ can mix with is
\begin{equation}\label{eq:Lifshitz_local_operator_in_x}
\cO''_A(\vecx,t) \sum_{A_1} \int dt_1\, \cO^{(1)}_{A_1}(\vecx,t_1) \cdots \sum_{A_k} \int dt_k\, \cO^{(k)}_{A_k}(\vecx,t_k).
\end{equation}
These general mixings allow us to cancel any singularities where two operators come together.
Our previous example involved the special case where $\cO''$ was the identity operator.
Note that in the action any non-locality is related to an extra integral over time, and this always comes with a sum over the replica index, restricting the possible mixings to the form \eqref{eq:Lifshitz_local_operator_in_x}. In other words, every independent integration over time comes with a separate $S_n$ symmetry rotating the replica indices of the operators with that time variable. 

If we look at integrated operators, that can appear in the action, the same argument implies that operators of the form
\begin{equation}\label{mixings2}
\int d^d\vecx \sum_{A_1} \int dt_1\, \cO^{(1)}_{A_1}(\vecx,t_1) \cdots \sum_{A_k} \int dt_k\, \cO^{(k)}_{A_k}(\vecx,t_k)
\end{equation}
only mix among themselves, so that if we start from such an operator in the action (as above), only such operators will be generated by the renormalization group flow. Note that even though the $k$ operators in \eqref{mixings2} are generically at different times, their product at the same spatial point still needs to be regularized, as we saw in the example of the previous subsection. In particular the scaling dimension of such a product of operators will not simply be the sum of the dimensions of its constituents; we will see an explicit example of this in \autoref{subsection:Lifshitz_around_free_theory}. Note also that there can be contributions when one of the operators $\cO^{(j)}$ is equal to the identity operator. These contributions will be proportional to the volume of the time direction so that they are IR-divergent, but they are also proportional to $n$ so they will vanish as $n\to 0$.

Next, let us verify that the replica interactions that are generated are precisely the ones related to the standard couplings and to the disorder distribution.
We can identify those using the approach to the RG using the flow of inhomogeneous couplings (\autoref{subsection:local_RG_approach}).
First,  we have the various homogeneous couplings that are generated along the flow, and give the usual local replica terms $\int d^d\vecx dt \, \sum _A \cO _{i,A} (\vecx ,t)$. In addition to that, different couplings become inhomogeneous in the spatial directions under the RG, and the disorder distribution flows as well. Using \autoref{section:Ph_non_Gaussian}, the quadratic correlations between the different inhomogeneous couplings give in the replica $ \int d^d\vecx dt dt' \, \sum _{A,B} \cO _{i,A} (\vecx ,t) \cO _{j,B} (\vecx ,t')$, and higher moments in the distribution (generated even if they are not there in the ultra-violet) give higher multi-replica interactions $\int d^d\vecx dt_1\cdots dt_k \, \sum _{A_1,\dots ,A_k} \cO _{i_1,A_1}(\vecx ,t_1) \cdots \cO _{i_k,A_k} (\vecx ,t_k)$. These are precisely the terms that we encountered above.

\begin{figure}[h]
\centering
\includegraphics[width=0.7\textwidth]{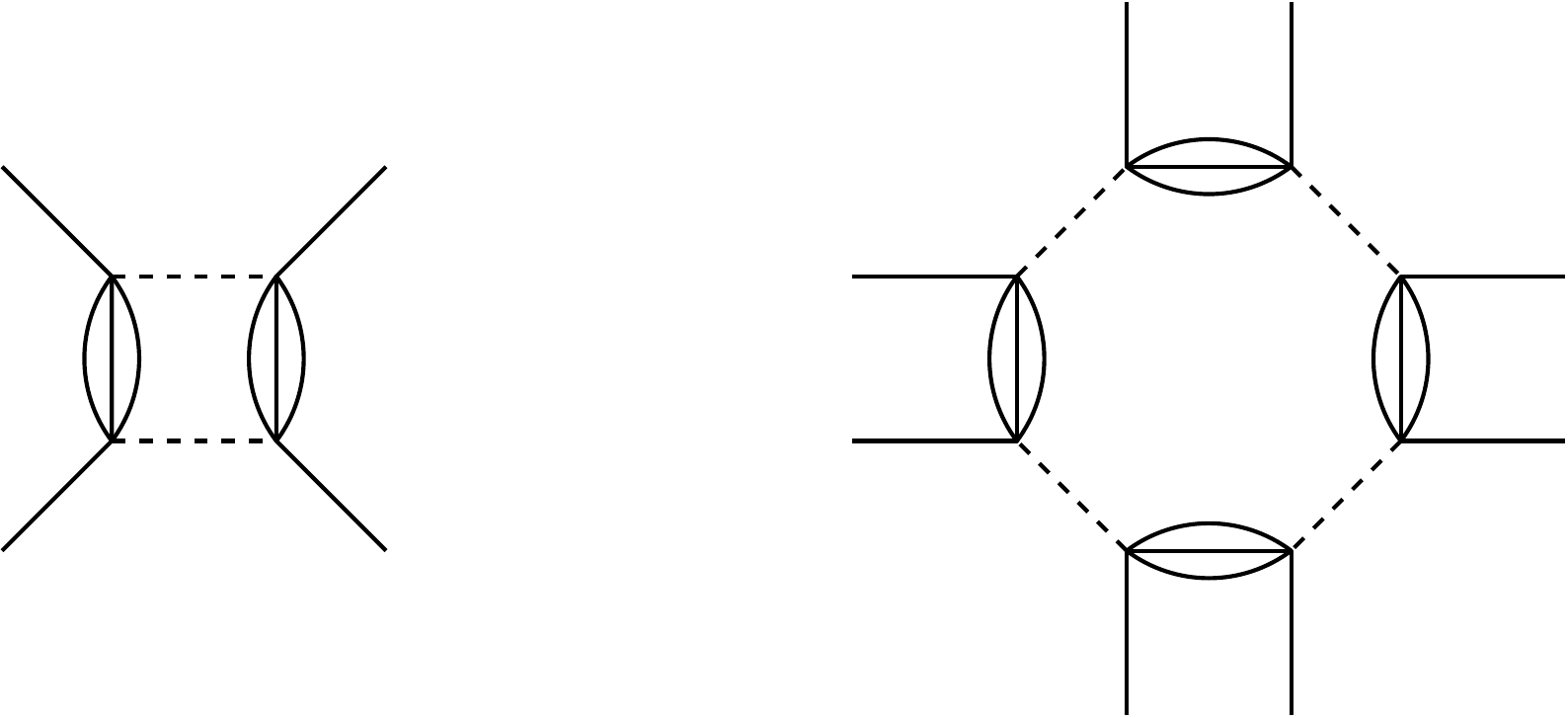}
\caption{Examples of diagrams that generate non-local terms in the replica action.}
\label{fig:Lifshitz_non_local_diagrams}
\end{figure}

As an example where the disorder distribution is modified, consider a scalar theory with disorder coupled to $\cO_0 = \varphi ^4$ in $\dcft =3$ dimensions, which is marginal by the Harris criterion. The disorder corresponds to the replica coupling
\begin{equation} \label{eq:Lifshitz_generated_interactions_phi4}
\qd \int d^d\vecx dt dt' \sum _{A,B} \varphi _A^4(\vecx,t) \varphi _B^4(\vecx,t') .
\end{equation}
The following coupling is consistent with the symmetries
\begin{equation}
\int d^d\vecx dt dt' \sum _{A,B} \varphi _A^2(\vecx,t) \varphi _B^2(\vecx,t'),
\end{equation}
and indeed it is generated (with a divergent coefficient) already at order $\qd^2$, for instance by the diagram appearing on the left-hand side of \autoref{fig:Lifshitz_non_local_diagrams} (the dashed line represents the non-local interaction \eqref{eq:Lifshitz_generated_interactions_phi4} that can connect different replicas; for the precise Feynman rules in a similar theory see \autoref{subsection:Lifshitz_around_free_theory}).
This interaction corresponds to Gaussian disorder coupled to the operator $\varphi ^2$ in the original local disordered theory. Higher moments for this disorder will also be generated (not only Gaussian); for example the coupling
\begin{equation}
\int d^d\vecx dt_1dt_2dt_3dt_4 \sum _{A,B,C,D} \varphi _A^2(\vecx,t_1) \varphi _B^2(\vecx,t_2) \varphi _C^2(\vecx,t_3) \varphi _D^2(\vecx,t_4)
\end{equation}
is generated through the diagram on the right hand side of \autoref{fig:Lifshitz_non_local_diagrams}. There are analogous diagrams generating all higher order moments as well.
Note that when the dimension of all operators (consistent with the symmetries) in the original theory is larger than one, interactions with higher moments have higher dimensions and will be suppressed at low energies. However, when there is an operator of dimension one, $\int dt \,{\cal O}(\vecx,t)$ is dimensionless, so all higher moments will be equally important at low energies (this is what happens in our example). If there is an operator of dimension less than one, these higher moments are expected to dominate. We will assume for simplicity that these situations do not arise; in such a case it is reasonable to use a basis for the operators which is polynomial in the original local operators, despite the non-locality in time. In other cases, like the disordered quantum Ising chain (see chapter 21.6 of \cite{sachdev2001quantum} for a review and references) it is more suitable to study the flow of the full disorder distribution.

Note that if one only takes into account the overall $S_n$ replica symmetry, then there are additional operators that are allowed, even though they have no mapping to the disorder distribution.
One example is the local operator $\delta S_1 = \tilde \qd \int d^d\vecx dt \, \sum _{A,B} \cO _{i,A} (\vecx ,t) \cO _{j,B} (\vecx ,t)$ which would correspond to classical disorder, and is not expected to be generated since the couplings in our disordered theory depend only on space and cannot develop a dependence on time. Another example is
\begin{equation} \label{eq:Lifshitz_non_generation_of_g_prime}
\delta S_2 = - \qd' \sum _{A} \int d^d\vecx dt dt'\, \cO _A(\vecx,t) \cO _A(\vecx,t'),
\end{equation}
which seems to correspond to a non-local interaction in the disordered theory. However, all operators of these types cannot be generated, because every time integration comes with a sum over the replica index, and every sum over the replica index comes with a time integration. In other words, all interactions between different replicas are independent of the time difference between the replicas, and all non-local interactions in time are independent of the replica indices.

Since this argument that time integrations always come with sums over replica indices and vice versa will be used several times, let us mention that this can also be seen diagrammatically in perturbation theory. All diagrams, such as those in \autoref{fig:Lifshitz_non_local_diagrams}, will always consist of $k$ sub-diagrams which are connected among themselves by dashed lines. In such diagrams, both the time coordinates and the replica indices are uncorrelated between the different sub-diagrams, as can be seen by the types of interactions that we have. In momentum space they come with $k$ separate delta functions for energies. As a result, any generated interaction (or operator mixing) will maintain this property.

\subsection{Generalized Callan-Symanzik equations} \label{subsection:Lifshitz_RG_equations}

Quantum disorder differs from the classical case in several aspects, modifying the GCS equations and their implications. The first difference is the anisotropy between space and time. The main other difference is that the mixings of local operators are not the same as those in the classical case. On the one hand, the sorts of mixing we saw for classical disorder are not present in quantum disorder. First, there is no mixing of the single-replica operators with local multi-replica operators (such as the mixing we had between $\tilde \cO $ and $\tilde \cO _{ij} $). The reason for this is that $k$-replica operators always appear with $k$ independent time coordinates; alternatively, the absence of such a mixing can be seen by a diagrammatic argument as above. Similarly, local operators from different replicas cannot mix as in  \eqref{eq:RGnq_mixing_O_A}, and so in this sense $\gamma ''$ is zero. On the other hand, as mentioned above, single-replica operators can mix with integrated multi-replica operators \eqref{eq:Lifshitz_local_operator_in_x}. We shall see how this mixing affects the disorder-averaged correlation functions.

As in classical disorder, the couplings related to the disorder distribution flow and mix with the standard couplings that are there already in the original theory (i.e.\ both types of couplings appear in the beta functions). 
Indeed, the mixing of the couplings is consistent with the argued mixing of the integrated operators. Alternatively we can associate all the couplings with ``local'' operators of the form \eqref{eq:Lifshitz_local_operator_in_x} which mix with each other.

\subsubsection{Connected correlation functions}

For connected correlation functions of local operators, we should consider the replica operators of the form $\sum _A \cO '_A(\vecx ,t)$. These mix with the operators \eqref{eq:Lifshitz_local_operator_in_x} summed over $A$. When considering connected correlation functions, we can simply forget about the mixing with operators \eqref{eq:Lifshitz_local_operator_in_x} in which $\cO ''$ is the identity operator, since their sum over $A$ will introduce an extra factor of $n$ giving no contribution as $n \to 0$. 

Let us start by considering operators $\cO $ that do not have this sort of mixing (in particular this is the case for the lowest dimensional operators assuming that all operator dimensions are greater than $1$).
The derivation of \autoref{subsection:RGnq_RG_equations} goes through once we include all the couplings in the replica theory.
As before these include $\qd$ (and possibly the other parameters of the joint disorder probability distribution) and the various homogeneous couplings of the original theory, which we denote by $\lambda _i$.
The homogeneous couplings should include all operators consistent with the symmetries, and in particular $h_{00} $.
As in \autoref{section:RG_classical_disorder}, we get
\begin{equation}
\begin{split}
& \left( M \pder{}{M} + \beta _{\qd} \pder{}{\qd} + \beta _{\lambda _i} \pder{}{\lambda _i} + \beta _{h_{00} } \pder{}{h_{00} } + k \gamma ' \right) \mean{\langle \cO (x_1) \cdots \cO (x_k) \rangle_{conn} } =0 .
\end{split}
\end{equation}

We can also use the equivalent language in which $h_{00} $ is replaced by time rescaling to write these equations in a different form. Using \eqref{eq:Lifshitz_correlators_relation_to_time_dilation} we obtain for scalar operators
\begin{equation}
\pder{}{h_{00} } \langle \cO _{A_1} (x_1) \cdots \cO _{A_k}(x_k) \rangle_{h_{00} } = - \sum _{i=1}^k t_i \pder{}{t_i} \langle \cO _{A_1} (x_1) \cdots \cO _{A_k} (x_k) \rangle + O(h_{00} ) .
\end{equation}
The GCS equations then become for scalar operators\footnote{Note that the effect of the higher orders in $h_{00} $ is that the infinitesimal transformation is exponentiated, resulting in rescaling the times $t_i$, and changing variables in the equations to the rescaled times. }
\begin{equation}\label{eq:Lifshitz_GCS_connected_simple}
\begin{split}
& \left( M \pder{}{M} + \beta _{\qd} \pder{}{\qd} + \beta _{\lambda _i} \pder{}{\lambda _i} + \gamma  _t \sum _i t_i \pder{}{t_i } + k \gamma ' \right) \mean{\langle \cO (x_1) \cdots \cO (x_k) \rangle_{conn} } =0 ,
\end{split}
\end{equation}
with the `anomalous dimension of time' defined (using \eqref{eq:Lifshitz_CPT_h00_beta}) by
\begin{equation} \label{eq:Lifshitz_RG_gamma_t}
\gamma _t \equiv - \beta _{h_{00} } = - \frac{\qd c_{\cO \cO T } }{c_T}  + \cdots .
\end{equation}

We can now move on to the more general case where the local operators mix with the non-local operators \eqref{eq:Lifshitz_local_operator_in_x}. For simplicity, let us assume that $\sum _A \cO _A(\vecx ,t)$ mixes only with $\sum _A \cO _{i,A} (\vecx ,t) \cdot \sum _B \cO _{j,B} (\vecx ,E=0)$, where we abbreviated $\cO (\vecx ,E=0)= \int dt'\, \cO (\vecx ,t')$. The situation here differs from the classical case in two respects. First, the mixing occurs with an operator which is non-local in time. This will be reflected in a similar non-locality in time in the GCS equation; this non-locality in time is not some artifact of the replicated theory, but originates in the correlations between the values of the disorder at large time separation, as we saw at the beginning of \autoref{subsection:Lifshitz_replica_interactions}. Second, the sum includes not only $A \neq B$, but also $A=B$. This implies that the disorder-averaged connected correlation functions mix only among themselves (as opposed to the classical case). By following the usual derivation using \eqref{eq:RGnq_connected_coorelator_relation_replica}, we get (for a scalar $\cO$)
\begin{equation} \label{eq:Lifshitz_GCS_connected_general}
\begin{split}
& \left(M \pder{}{M} +\beta _{\qd} \pder{}{\qd} +\beta _{\lambda _i} \pder{}{\lambda _i} + \gamma _t \sum _i t_i \pder{}{t_i}  +k \gamma _{\cO } \right) \mean{\langle \cO (\vecx_1,t_1) \cdots \cO (\vecx_k,t_k)\rangle_{conn}  } + \\
& \qquad \gamma _{\cO ,\cO _{ij} } \Bigg[\mean{ \langle \cO _i (\vecx _1,t_1) \cO _j(\vecx _1,E=0)  \cO (\vecx _2,t_2) \cdots \cO (\vecx _k,t_k)\rangle_{conn} } +\\ 
& \qquad  \qquad  + (x_1 \leftrightarrow x_2) + \cdots + (x_1 \leftrightarrow x_k)  \Big] = 0 .
\end{split}
\end{equation}
This is exactly of the form of mixing found below \eqref{eq:Lifshitz_local_RG_average_mixing} using the local RG approach.
Note that equation \eqref{eq:Lifshitz_GCS_connected_general} and the other GCS equations given below will be presented in the form where we do not have $h_{00} $, and are valid for scalar operators $\cO $ (otherwise small straightforward modifications are needed, which originate in the modification of \eqref{eq:Lifshitz_correlators_relation_to_time_dilation}). We can always use the form with $h_{00} $, valid for generic operators, by replacing
\begin{equation}
\gamma _t \sum _i t_i \pder{}{t_i} \to \beta _{h_{00} } \pder{}{h_{00} } .
\end{equation}

\subsubsection{Fixed points and the dynamical exponent $z$} \label{subsubsection:Lifshitz_fixed_points}

At quantum disordered fixed points, the equations above generically lead to Lifshitz behavior. Naively at a fixed point all beta functions, including $\beta_{h_{00}}$, should vanish. However, allowing it to take a constant non-zero value $(-\gamma_t^*)$ still gives a fixed point, just with different scaling. To see this let us solve the simplest GCS equation \eqref{eq:Lifshitz_GCS_connected_simple} for the connected 2-point function at a fixed point,
\begin{equation} \label{eq:Lifshitz_2pf_CS_eq}
\left(M \pder{}{M}  +   \gamma _t^* t \pder{}{t} + 2 \gamma '^* \right) \mean{\langle \cO (x) \cO (0)\rangle_{conn} }=0 .
\end{equation}
The solution is
\begin{equation} \label{eq:Lifshitz_fixed_point_2pf}
\mean{\langle \cO (x) \cO (0) \rangle_{conn} } = \frac{M^{-2 \gamma '^*} }{\vecx^{2\Delta +2 \gamma '^*} } F \left( \frac{t}{M^{\gamma _t^*} \vecx ^{1+\gamma _t^*} }, \lambda _i^*,\qd^* \right),
\end{equation}
with the function $F$ undetermined from the GCS equation. This 2-point function is invariant under the rescaling $\vecx \to \lambda \vecx $, $t \to \lambda ^z t$, $\cO \to \lambda^{-\Delta^*} \cO$, with dimension $\Delta^* =\Delta+\gamma '^*$ and
\begin{equation} \label{eq:Lifshitz_z_exponent_relation_to_beta_h00}
z = 1+\gamma _t^* .
\end{equation}
Since $\beta_{h_{00}}$ is non-zero already at leading order in the disorder, we see that fixed points generically have such a generalized scale-invariance. Such fixed points do not describe theories that have hyperscaling violation at low energies.

Clearly the entire theory, and not only such a correlation function, is invariant under this Lifshitz scaling, as can be seen by the RG flow. Rescaling the RG scale $M$ by a factor $b$ is the same as rescaling space and time $\vecx \to \vecx /b$, $t \to t/b$. At a fixed point all beta functions vanish, except for a constant $\beta _{h_{00} } $ which we allowed, so that under an infinitesimal RG step $h_{00} \to h_{00} -\beta _{h_{00} } \log(b)$. As explained around \eqref{eq:Lifshitz_action_variation_under_time_rescaling}, this is equivalent to keeping $h_{00} $ fixed (together with all the other couplings) and rescaling $t \to t(1+\beta _{h_{00} } \log(b) ) \sim t b^{\beta _{h_{00} } } $. Therefore we find that the theory is invariant under the scaling $\vecx \to \vecx /b$, $t \to t / b^{1-\beta _{h_{00} } } $, which is the Lifshitz scaling with the exponent $z$ just found. If the pure theory was Lifshitz invariant with exponent $z_{pure} $ rather than the relativistic $z_{pure} =1$, the RG transformation we would start from is $\vecx \to \vecx /b$, $t \to t/b^{z_{pure} } $ since under this RG transformation the pure theory is a fixed point with the corresponding scaling dimensions. Then again $\beta _{h_{00} } $ gives the anomalous scaling exponent, so that $z_{random} =z_{pure} + \gamma ^*_t$. (Equivalently, in solving \eqref{eq:Lifshitz_2pf_CS_eq}, we would have to use dimensional arguments based on the scaling behavior of the UV fixed point, for which $[M]=1$, $[\vecx]=-1$, $[t]=-z_{pure} $, $[\cO ]=\Delta $, so that in \eqref{eq:Lifshitz_fixed_point_2pf} the power of $\vecx$ inside $F$ becomes $z_{pure} +\gamma_t^*$.)

Equation \eqref{eq:Lifshitz_z_exponent_relation_to_beta_h00} holds also in pure non-relativistic systems (see, for example, \cite{Korovin:2013bua}). However, the usual argument for it in such systems uses the appearance of the beta function in the expression for the trace of the energy-momentum tensor, while in disordered theories only the components of this tensor associated to time translations exist. Our arguments show that only the existence of the Hamiltonian is required for \eqref{eq:Lifshitz_z_exponent_relation_to_beta_h00}.

For classical disorder, as reviewed above, the response of the IR fixed point to the disorder deformation corresponded to a critical exponent $\phi$ that was independent of its response to homogeneous deformations (related to the critical exponents $\nu$ or $\alpha$). This was related to the fact that the two corresponded to two different local operators in the replica theory, with unrelated dimensions in the IR. Naively in the quantum disorder case this would not be true, since one would say that the dimension of the replica operator corresponding to the disorder $\sum_{A,B} \int d^d\vecx dt dt' \, \cO_{0,A}(\vecx,t) \cO_{0,B}(\vecx,t')$, involving two operators at separate points, is simply related to the dimension of $\cO_0$, so that there is no independent critical exponent.\footnote{This claim appears, for instance, in \cite{sachdev2001quantum}.} However, our discussion above shows that this is not true, and this operator does have an independent anomalous dimension, corresponding to an independent critical exponent $\phi$. We will see an example of this below. It would be interesting to test this experimentally at quantum critical points.
In particular, we can define the exponent $\phi $ in analogy to the classical case through the dimension $\Delta _{\Psi} $ of the disorder-related operator which includes $ \sum _{A,B} \int dt dt' \, \cO _{0,A}(\vecx,t) \cO _{0,B}(\vecx,t') $, by $\Delta _{\Psi} = d - \frac{\phi }{\nu } $. Then in the IR disordered fixed point we expect $\phi <0$.
Using the incorrect relation mentioned above between the dimension of the disorder operator and the dimension of $\cO_0 $, this leads to $\nu  > 2/d$. However, independently of this, it is argued that this inequality is still true\cite{Chayes:1986ju}.

For a perturbative fixed point, we find at leading order using \eqref{eq:Lifshitz_RG_gamma_t} and the Ward identity \eqref{eq:Lifshitz_WI_for_cOOT} 
\begin{equation} \label{eq:Lifshitz_z_exp1}
z \approx  1-\qd \frac{c_{\cO \cO T} }{c_T}   = 1 + \qd \frac{c_{\cO \cO } }{c_T}  \frac{\dcft \Delta_0 }{(\dcft -1)} \frac{\Gamma (\dcft /2)}{2\pi ^{\dcft /2} }  ,
\end{equation}
where $c_{\cO \cO}$ is the coefficient of the two-point function of $\cO_0(x)$.\footnote{We can choose any normalization we want for $\cO_0 (x)$, but the combination $\qd c_{\cO \cO } $ is independent of this.}
In this computation the disorder was chosen to be marginal or very close to marginal in order to allow for a perturbative fixed point, and thus to leading order
\begin{equation} \label{eq:Lifshitz_z_exp2}
z\approx 1+\frac{\qd}{2}  \frac{c_{\cO \cO } }{c_T} \frac{\dcft (\dcft +1)}{\dcft -1 } \frac{\Gamma (\dcft /2)}{2\pi ^{\dcft /2} }   .
\end{equation}
This is a universal formula for the dynamical exponent $z$, valid for any weak scalar disorder. We will see examples below.


As in \autoref{section:RG_classical_disorder}, the anomalous dimensions of operators that mix with each other can be complex. However, since the energy is conserved, $T_{00} $ is well-defined and does not mix with any other operators, and therefore $\gamma _t^*$ (and thus $z$) will always be real.


In quantum disorder, the GCS equations generally mix connected correlation functions among themselves (as we saw in \eqref{eq:Lifshitz_GCS_connected_general}). But still, we do not expect simple scaling behavior for correlation functions of
the operators that have non-trivial mixing with multi-replica operators. Consider for instance a 2-point function $\mean{\langle \cO(\vecx ,t)\cO (0)\rangle_{conn} }$ satisfying \eqref{eq:Lifshitz_GCS_connected_general}, which then mixes with $\mean{\langle \cO _i(\vecx ,t)\cO _j(\vecx ,E=0) \cO (0)\rangle_{conn} }$. The latter, however, cannot be treated as a connected correlation function of 2 operators (in particular, it is not the same as $\mean{\langle \left( \cO _i(\vecx ,t) \cO _j(\vecx ,E=0) \right) \cO (0)\rangle_{conn} }$ where the 2 operators in the parenthesis are treated as a single operator for the purpose of the connectedness), and therefore we cannot perform a diagonalizing transformation among the local operators (including those multiplied by zero-energy operators) that will bring a connected correlation function to a simple scaling behavior.

The generalization to several directions $d_t>0$ on which the disorder does not depend (as discussed in \autoref{universal}) includes summing over all those directions in the $t \pder{}{t} $ terms in the GCS equation \eqref{eq:Lifshitz_GCS_connected_simple} (and the other GCS equations), and $\gamma _t=-\beta _{h_{00} } $ given by \eqref{eq:Lifshitz_CPT_h00_beta_general_d_t}. For $0 <d_t < \dcft $, there is an $SO(\dcft -d_t) \times SO(d_t)$ symmetry in averaged correlation functions, and the solution of the connected 2-point function is still given by \eqref{eq:Lifshitz_fixed_point_2pf} where now $t$ there stands for $|t| = \sqrt{\sum _{\alpha =0} ^{d_t-1} x_{\alpha } ^2} $, the norm in the $d_t$ directions (and $\vecx $ still stands there for the norm in the other $\dcft -d_t$ directions). The Lifshitz scaling is $x_{\mu } \to \lambda  x_{\mu } $ for $\mu =d_t,\dots ,\dcft -1$ and $x_{\alpha } \to \lambda ^z x_{\alpha } $ for $\alpha =0,\dots ,d_t-1$, with $z=1+\gamma _t^*$. For weak disorder (using a marginal or very close to marginal disorder, with the appropriate $\Delta_0 $ for a general $d_t$)
\begin{equation}
\begin{split}
z & \approx 1 -\qd \frac{ c_{\cO \cO T} }{c_T} \frac{S_{d_t-1} }{2d_t} = 1+ \qd \frac{c_{\cO \cO } }{c_T} \frac{\dcft \Delta_0 }{\dcft -1} \frac{\Gamma  (\dcft /2)}{2\pi ^{\dcft /2} } \frac{S_{d_t-1} }{2d_t} \approx \\
& \approx 1 + \frac{\qd}{4} \frac{c_{\cO \cO } }{c_T} \frac{\dcft (\dcft +d_t)}{d_t (\dcft -1)} \frac{\Gamma (\dcft /2)}{\pi ^{(\dcft -d_t)/2} \Gamma (d_t/2)} .
\end{split}
\end{equation}

\subsubsection{Non-connected correlation functions} \label{subsubsection:Lifshitz_non_connected_correlators}

Let us begin with operators $\cO _A(\vecx ,t)$ that do not mix with other operators \eqref{eq:Lifshitz_local_operator_in_x}. For such operators, we find the simple equation
\begin{equation} \label{Lifshitz_GCS_non_connected_simple}
\left( M \pder{}{M} + \beta _{\qd} \pder{}{\qd} + \beta _{\lambda _i}  \pder{}{\lambda _i}  + \gamma _t \sum _i t_i \pder{}{t_i} + k \gamma ' \right)  \mean{ \langle \cO (x_1) \cdots  \cO (x_k)  \rangle} = 0.
\end{equation}
We could think that because there is no mixing among the $\{\cO _A\}$, there is a significant difference for this simplest sort of operators compared to classical disorder, where we found a non-trivial GCS equation mixing different correlation functions. However, this is not precisely the case as we will see in a moment, since the operators satisfying \eqref{Lifshitz_GCS_non_connected_simple} are analogous to the cases in classical disorder where we had a $G^n$ symmetry under which the $\{\cO _A\}$ transformed differently, and resulted in $\gamma ''=0$ in \eqref{eq:RGnq_mixing_O_A}.

Consider a more general operator $\cO _A(\vecx ,t)$ and assume again for simplicity that it mixes only with $\cO _{i,A} (\vecx ,t) \cdot \sum _B \cO _{j,B} (\vecx ,E=0)$ (the generalization to more general mixings follows along similar lines). Writing the CS equation in the replicated theory for correlation functions of $\cO _{A=1} $, taking the $n \to 0$ limit and using \eqref{eq:RGnq_disconnected_coorelator_relation_replica_generalized} we find
\begin{equation} \label{eq:Lifshitz_GCS_non_connected_more_general}
\begin{split}
& \left(M \pder{}{M} + \beta _{\qd} \pder{}{\qd} + \beta _{\lambda _i} \pder{}{\lambda _i} + \gamma _t \sum _i t_i \pder{}{t_i} + k \gamma _{\cO } \right) \mean{\langle \cO (\vecx _1,t_1) \cdots \cO (\vecx _k,t_k)\rangle } + \\
& + \gamma _{\cO ,\cO _{ij} } \Bigg[ \bigg( \mean{\langle \cO _i(\vecx _1,t_1) \cO _j(\vecx _1,E=0) \cO (\vecx _2,t_2) \cdots \cO (\vecx _k,t_k) \rangle } - \\
& \qquad \qquad - \mean{\langle \cO _j(\vecx _1,E=0) \rangle \langle \cO _i(\vecx _1,t_1) \cO (\vecx _2,t_2) \cdots \cO (\vecx _k,t_k)\rangle } \bigg)  + \\
& \qquad \qquad + (x_1 \leftrightarrow x_2) + \cdots +(x_1 \leftrightarrow x_k) \Bigg] = 0 .
\end{split}
\end{equation}
Again this is precisely what we saw in \eqref{eq:Lifshitz_local_RG_average_mixing}.

A difference from considering connected correlation functions is that while there we could forget about a mixing of local operators with operators \eqref{eq:Lifshitz_local_operator_in_x} in which $\cO ''=1$, this is no longer true for non-connected correlation functions.
This mixing corresponds to a mixing with the identity operator in the local RG approach, contributing only to disorder-averaged non-connected correlation functions.
In particular, we may have such a mixing with $\cO _j = \cO $\, that is, a mixing of $\cO _A(\vecx,t)$ with $\sum _B \cO _B(\vecx ,E=0)$. This gives us a quantum disorder analog of the $\gamma ''$ mixing in classical disorder. Note that the latter time-integrated operator mixes into the former operator, but clearly it does not happen in the other direction (because of the time dependence), and therefore this mixing does not give rise to new anomalous dimensions of local operators, and the dimension of the local operator is the same as the dimension of the integrated operator (plus $z$) as could be expected by the definition of a scaling dimension.\footnote{The disorder operator had an independent dimension since it is constructed by a product of operators, and as usual such operators get anomalous dimensions.} This is the same as in classical disorder where $\gamma ''$ did not modify scaling dimensions.\footnote{\label{footnote:Lifshitz_same_n_to_0_dim}We mentioned that the operators that should have well defined dimensions are $\tilde \cO $ and $\tilde \cO _A$ which form $S_n$ irreducible representations. Both in classical and quantum disorder, a mixing of $\cO _A$ and $\sum _B \cO _B$ given by $\gamma ''$ is independent of $A$ and therefore can contribute only to the dimension of $\tilde \cO $. But this contribution will come with an explicit factor of $n$ which goes to zero as $n \to 0$. Therefore $\tilde \cO $ and $\tilde \cO _A$ have the same dimension.} It does, though, change the GCS equation. 
Using equation \eqref{eq:Lifshitz_GCS_non_connected_more_general} with $\cO _i = 1$, $\cO _j = \cO $, and changing notation $\gamma _{\cO } \to \gamma '$, $\gamma _{\cO ,\cO _{ij} } \to \gamma ''$, we find the analog of \eqref{finalrgnonconn}
\begin{equation}  \label{eq:Lifshitz_GCS_non_connected_gamma_pp_analog}
\begin{split}
& \left(M \pder{}{M} + \beta _{\qd} \pder{}{\qd} + \beta _{\lambda _i} \pder{}{\lambda _i} + \gamma _t \sum _i t_i \pder{}{t_i} + k \gamma ' \right) \mean{\langle \cO (\vecx _1,t_1) \cdots \cO (\vecx _k,t_k)\rangle } + \\
& + \gamma'' \Bigg[ \bigg( \mean{\langle \cO (\vecx _1,E=0) \cO (\vecx _2,t_2) \cdots \cO (\vecx _k,t_k) \rangle } - 
\mean{\langle \cO (\vecx _1,E=0) \rangle \langle \cO (\vecx _2,t_2) \cdots \cO (\vecx _k,t_k)\rangle } \bigg)  + \\
& \qquad \qquad + (x_1 \leftrightarrow x_2) + \cdots +(x_1 \leftrightarrow x_k) \Bigg] = 0 .
\end{split}
\end{equation}

Such a mixing is independent of the time coordinate of the local operator and is therefore a mixing only of $ \frac{1}{\beta} \cO_A (\vecx ,E=0)$ (where $\beta$ is the volume of the time direction) with $\sum _B \cO _{B} (\vecx ,E=0)$.
Going to energy space, this GCS equation becomes
\begin{equation}
\begin{split}
& \left(M \pder{}{M} + \beta _{\qd} \pder{}{\qd} + \beta _{\lambda _i} \pder{}{\lambda _i} - \gamma _t \left( \sum _{i=1} ^k E_i \pder{}{E_i}+k-1\right) + k \gamma ' \right) \mean{\langle \cO (\vecx _1,E_1) \cdots \cO (\vecx _k,E_k)\rangle } + \\
& \qquad + \gamma '' \Bigg[ 2\pi \delta (E_1) \bigg( \mean{\langle \cO (\vecx _1,E_1) \cdots \cO (\vecx _k,E_k)\rangle } - \mean{\langle \cO (\vecx _1,E_1)\rangle\langle \cO (\vecx _2,E_2) \cdots \cO (\vecx _k,E_k)\rangle} \bigg) +\\
& \qquad \qquad + (x_1 \leftrightarrow x_2) + \cdots +(x_1 \leftrightarrow x_k) \Bigg] = 0 ,
\end{split}
\end{equation}
where each averaged correlator (and averaged product of correlators) is defined after factoring out an overall energy conserving delta function as usual, and in this equation $\sum _i E_i=0$ (alternatively we can eliminate using energy conservation one of the energies so that we are left with $k-1$ energies, and the sum in the first line of this equation is restricted to the $k-1$ appropriate terms).
We see that in non-connected correlation functions there will be $\delta (E)$ terms, as happens also in different physical situations. As mentioned, this mixing does not affect connected correlators, and therefore there are no $\delta (E)$ terms in connected correlation functions, which are related to thermodynamic quantities.

Contrary to the $\gamma ''$ mixing in classical disorder, there are no logs in disorder-averaged correlation functions of local operators at quantum disordered fixed points. The reason for this is that such logs require degeneracies and non-diagonalizability of operator dimensions. However, in quantum disorder there is no mixing between the local $\cO _A(\vecx ,t)$ and $\sum _B \cO _B(\vecx ,t)$.
Note that $\gamma ''$ in \eqref{eq:Lifshitz_GCS_non_connected_gamma_pp_analog} is dimensionful, and we have no dimensionful quantities at disordered fixed points.

This is all at zero temperature (or more physically at large distances, but much smaller than $\beta = 1/T$). At finite temperature $T$, the Euclidean time dimension is compact. At large distances we can use a Kaluza-Klein reduction on the time direction. We then get that the quantum disorder problem in $d+1$ dimensions, reduces at large distances to classical disorder in $d$ dimensions. Each operator $\cO _i$ gives rise to a tower of operators $\cO _i(\vecx ,E)$ which are local in $\vecx $ (the classical disorder coordinates). 
As before, along the RG, independent replica indices come with independent time integrations, and thus the disorder is coupled only to zero-energy operators. 
Similarly, while we expect a mixing among the $\{\cO _A\}$ for any $\cO $ in classical disorder (and not only for the operators to which disorder was coupled), here we get such a mixing only for the zero-energy operators (this is the mixing of $\cO _A(\vecx ,E)$ with $\sum _B \cO _B(\vecx ,E)$). Note that anyway only the zero-energy modes are relevant for long distance physics.

\subsubsection{Example of the generalized Callan-Symanzik equation}

Let us check the GCS equation \eqref{eq:Lifshitz_GCS_connected_simple} on the 2-point function of $\cO_0(x)$ to which disorder was coupled up to order $\qd$. Using perturbation theory in $\qd$ in the replica theory as before, we have
\begin{equation}
\begin{split}
& \langle \cO_{0,A} (\vecx,t) \cO_{0,B} (0)\rangle _{\qd} = \langle \cO _{0,A}(\vecx,t) \cO _{0,B}(0)\rangle + \\
&\qquad\qquad+ \frac{\qd}{2}  \int d^d\vecx_1 dt_1 dt'_1 \sum _{C,D} \langle \cO _{0,A}(\vecx,t) \cO _{0,B}(0) \cO _{0,C}(\vecx_1,t_1) \cO _{0,D}(\vecx_1,t'_1)\rangle + \cdots . 
\end{split}
\end{equation}
The correlation functions on the right hand side are again evaluated in the un-deformed CFT.

For marginal disorder the term of order $\qd^0$ is
\begin{equation}\label{eq:twop}
\langle \cO _{0,A}(\vecx,t) \cO _{0,B}(0)\rangle = \frac{c_{\cO \cO } \delta _{AB} }{(\vecx^2+t^2)^{\frac{d+2}{2} } } .
\end{equation}

Next consider the term of order $\qd^1$ for $A=B$. The contribution from $C=D \neq A$ is just the order $\qd^0$ contribution times a correction to the vacuum energy. We are left with the correction from $C=D=A=B$ which is
\begin{equation}
\frac{\qd}{2}  \int d^d\vecx_1 dt_1 dt'_1 \langle \cO _{0,A}(\vecx,t) \cO _{0,A}(0) \cO _{0,A}(\vecx_1,t_1) \cO _{0,A}(\vecx_1,t'_1) \rangle
\end{equation}
(no sum over $A$).
This is not universal, but we can compute one universal term in the OPE expansion of the 4-point function.
Using the $T_{00} $ term in the OPE $\cO_0 \times \cO_0 $ we get the term
\begin{equation}
\begin{split}
& \frac{\qd}{2}  \int d^d\vecx_1 dt_1 dt'_1\frac{c_{\cO \cO T} }{c_T} \frac{1}{|t_1-t'_1|} \langle \cO _{0,A}(\vecx,t) \cO _{0,A}(0) T_{00,A} (\vecx_1,t_1) \rangle \sim \\
& \qquad\qquad \sim \qd \frac{c_{\cO \cO T} }{c_T}  \log (\Lambda _t) \int d^d\vecx_1 dt_1\langle \cO _{0,A}(\vecx,t) \cO _{0,A}(0) T_{00,A} (\vecx_1,t_1) \rangle .
\end{split}
\end{equation}
This and \eqref{eq:twop} are independent of $n$, so they contribute to $\mean{\langle \cO_0(\vecx,t) \cO_0(0) \rangle}$.

The first contribution to the CS equation \eqref{eq:Lifshitz_GCS_connected_simple} to this order comes from $\gamma _t \sum t_i \pder{}{t_i} $ acting on the $O(\qd^0)$ term. On the two point function this is
\begin{equation}
\begin{split}
\gamma _t t \pder{}{t} \frac{c_{\cO \cO } }{(\vecx^2+t^2)^{(d+2)/2} } = -\gamma _t (d+2) c_{\cO \cO } \frac{t^2}{(\vecx^2+t^2)^{(d+4)/2} } .
\end{split}
\end{equation}
The second contribution comes from the log divergent term of order $\qd$ (as we already noted in \autoref{subsection:RGnq_composite_operators} we can use $\Lambda $ instead of $M$ in the CS equation)
\begin{equation}
\begin{split}
& \Lambda _t \pder{}{\Lambda _t} \left[ \qd \frac{c_{\cO \cO T} }{c_T}  \log (\Lambda _t) \int d^d\vecx_1dt_1 \langle\cO _{0,A}(\vecx,t)\cO _{0,A}(0)T_{00,A} (\vecx_1,t_1)\rangle \right] = \\
&\qquad\qquad = \qd \frac{c_{\cO \cO T} }{c_T} \int d^d\vecx_1dt_1 \langle\cO _{0,A}(\vecx,t)\cO _{0,A}(0)T_{00,A} (\vecx_1,t_1)\rangle .
\end{split}
\end{equation}
We evaluate this integral explicitly in \autoref{section:Lifshitz_OOT_integral}. Setting in \eqref{eq:eseven} $\Delta =(d+2)/2$, we get
\begin{equation}
\qd \frac{c_{\cO \cO T} }{c_T} (d+2) c_{\cO \cO } \frac{-t^2}{(\vecx^2+t^2)^{(d+4)/2} } .
\end{equation}
In the GCS equation we claimed that $\gamma _t$ is given by \eqref{eq:Lifshitz_RG_gamma_t},
and indeed with this value the two terms cancel and the GCS equation is satisfied. Note that the GCS equation implies the non-trivial statement that any logarithmic divergences arising from the additional terms that we did not explicitly compute above must be proportional to the leading order two-point function.

\subsubsection{Large $N$ quantum disorder}

As in \autoref{section:classical_disorder_large_N}, large $N$ provides an interesting test of the GCS equations. Let us consider the quantum version of the generalized free field theory representing a large $N$ limit, with the replicated theory given by
\begin{equation}
S_{replica} = \sum _A S_{0,A} - \frac{\qd}{2} \sum _{A,B} \int d^d\vecx dt dt' \, \cO _A(\vecx ,t) \cO _B(\vecx ,t') .
\end{equation}

In most of the examples we consider, the disorder is taken to be marginal (or close to marginal), but in general we do not have to restrict to the marginal case. In fact, here it will be more interesting to consider the relevant case in which the dimension of $\cO $ is taken to be $\Delta = \frac{d+1}{2} $. The double trace deformation is marginal, but as in \autoref{section:classical_disorder_large_N} we can set it to zero. The 2-point functions of $\cO _A$ are given by
\begin{equation} \label{largenqd}
\begin{split}
& \langle \cO _A(\vecx ,t) \cO _B(0) \rangle = \frac{\delta _{AB} }{\left( \vecx ^2 +t^2\right)^{(d+1)/2} } + \qd \int d^d\vecz  dt_1 dt_2 \frac{1}{\left( (\vecz -\vecx )^2+(t_1-t)^2\right)^{(d+1)/2} } \frac{1}{\left(\vecz ^2 + t_2^2\right)^{(d+1)/2} }  = \\
& \qquad = \frac{\delta _{AB} }{\left( \vecx ^2 +t^2\right)^{(d+1)/2} }+ \qd \frac{\pi \Gamma  \left( \frac{d}{2} \right)^2}{\Gamma \left( \frac{d+1}{2} \right)^2} \frac{2 S_{d-1} \log(\Lambda  \vecx )+C_1}{\vecx ^d} ,
\end{split}
\end{equation}
where we used \eqref{eq:RGnq_GFF_integrals}. There are no higher order corrections in $\qd$ for $n \to 0$. Equation \eqref{largenqd} can be renormalized using $\left(\cO _A\right)_R = \cO _A + c \cdot \qd \int dt' \, \sum _{B} \cO _B(\vecx ,t')$ for an appropriately chosen $c$, but we may also apply the GCS equations directly with $M \to \Lambda $. We may test the non-connected GCS equation \eqref{eq:Lifshitz_GCS_non_connected_gamma_pp_analog}, which for the 2-point function is
\begin{equation}
\begin{split}
\left( \Lambda  \pder{}{\Lambda } +\beta _{\qd} \pder{}{\qd} +\gamma _t t \pder{}{t} +2\gamma '\right) \mean{\langle \cO (\vecx ,t) \cO  (0) \rangle} + 2\gamma '' \mean{\langle \cO (\vecx ,E=0) \cO (0)\rangle_{conn} }= 0.
\end{split}
\end{equation}
In large $N$ the central charge $c_T$ is large and thus we expect $\gamma _t=0$. This GCS equation is indeed satisfied, to all orders in $\qd$, with the correlators given by
\begin{equation}
\begin{split}
& \mean{\langle \cO (\vecx ,t) \cO (0)\rangle}=  \frac{1 }{\left( \vecx ^2 +t^2\right)^{(d+1)/2} }+ \qd \frac{\pi \Gamma  \left( \frac{d}{2} \right)^2}{\Gamma \left( \frac{d+1}{2} \right)^2} \frac{2 S_{d-1} \log(\Lambda  \vecx )+C_1}{\vecx ^d},\\
& \mean{\langle \cO (\vecx ,E=0) \cO (0)\rangle_{conn} }= \frac{\sqrt{\pi } \Gamma \left(\frac{d}{2} \right)}{\Gamma \left( \frac{d+1}{2} \right)} \frac{1}{\vecx ^d} .
\end{split}
\end{equation}
The anomalous dimensions are $\gamma '=0$, $\gamma '' = -S_{d-1} \qd \frac{\sqrt{\pi } \Gamma (d/2)}{\Gamma \left((d+1)/2\right)} $, and indeed $\gamma _t=0$.

\subsection{Example 1 : perturbation theory in scalar field theories} \label{subsection:Lifshitz_around_free_theory}

As our first example, let us analyze scalar field theories with quantum disorder coupled to $\varphi^2(x)$, and
compare our analysis above to a computation performed by Boyanovsky and Cardy \cite{Boyanovsky:1982zz}. They considered an $O(m)$ symmetric model of $m$ real scalar fields $\varphi _i$, with disorder coupled to $\varphi ^2 = \sum _{i=1} ^m \varphi _i^2$.
The dimension of spacetime is taken to be $\dcft =4-\epsilon $, and the number of time dimensions (on which the disorder does not depend) is taken to be $\epsilon _d \ll 1$, with a Gaussian distribution $\mean{h(x)h(y)}= \qd \delta ^{(\dcft -\epsilon _d)} (\vecx-\vecy)  $.
This is a particular case of the generalization that we had of $d_t$ dimensions along which the disorder is constant, with $d_t=\epsilon _d$. 
The disorder coupling $\qd $ is dimensionless for $\epsilon=\epsilon_d=0$, and one can perform an expansion in $\epsilon$ and $\epsilon_d$. This is the reason why $d_t=\epsilon _d$ is taken to be small, while eventually one is interested in the quantum disorder case $d_t=1$.
Let us keep $\dcft $ and $d_t$ general, but keeping the disorder marginal or very close to being marginal such that we can use perturbation theory (giving the constraint $d_t \approx \dcft -4$ since $\Delta = \dcft -2$ for $\varphi ^2$), and compare this to the case of \cite{Boyanovsky:1982zz}.
The replicated action used in \cite{Boyanovsky:1982zz} is
\begin{equation}\label{bcaction}
\begin{split}
S_{replica} &= \sum _A \int d^{\dcft -d_t} \vecx d^{d_t} t \left[ \frac{1}{2} (\nabla _{\perp} \varphi _A)^2  + \frac{\alpha }{2} (\nabla _{\parallel} \varphi _A)^2 + \frac{m^2}{2}  \varphi _A^2  \right] - \\
&\qquad- \frac{\qd }{2} \int d^{\dcft } x d^{\dcft } x' \delta ^{(\dcft - d_t)} (\vecx-\vecx') \sum _{A,B} \varphi _A^2(x) \varphi _B^2(x') .
\end{split}
\end{equation}
Here $\nabla _{\perp} $ are the derivatives along the spatial $\dcft - d_t$ directions, and $\nabla _{\parallel} $ is along the $d_t$ temporal directions. This model was analyzed in \cite{Boyanovsky:1982zz} up to two loops in a double expansion in $\epsilon ,\epsilon _d$ (in which case a local $\varphi ^4$ interaction should also be included), and they then inspected the renormalization group flow and its fixed points. They observed the need to introduce a parameter $\alpha $ which parameterizes the anisotropy of the model, and that this $\alpha $ flows under the renormalization group. 

The correction to the action that we found earlier in this section matches exactly this modification of the temporal part of the kinetic term, for the case where the CFT is a free scalar theory.
Indeed, in a free scalar theory, the improved energy-momentum (EM) tensor takes the form (see \cite{Osborn:1993cr})
\begin{equation}
T_{\mu \nu } (x) = \partial _{\mu } \varphi  \cdot \partial _{\nu } \varphi  - \frac{1}{4(\dcft -1)} \left[(\dcft -2)\partial _{\mu } \partial _{\nu } +\delta _{\mu \nu } \partial ^2\right] \varphi ^2 .
\end{equation}
The term we added to the action in \eqref{eq:Lifshitz_generated_T00_general_dt} becomes
\begin{equation}
h_{00} \sum _A \sum _{\alpha =0} ^{d_t-1} \int d^{\dcft } x \, T_{\alpha \alpha ,A} (x) = h_{00} \sum _A \int d^{\dcft } x ( \nabla _{\parallel} \varphi _A)^2,
\end{equation}
which is exactly the $\alpha $ term in \eqref{bcaction} with $h_{00} \leftrightarrow \frac{\alpha -1}{2} $.
The running of $\alpha $ using \eqref{eq:Lifshitz_CPT_h00_beta_general_d_t} is thus expected to be
\begin{equation}
\beta _{\alpha } \approx  \frac{\qd c_{\cO \cO T} }{c_T} \frac{S_{d_t-1} }{d_t} .
\end{equation}
What is left is to find the relevant OPE coefficients $c_T$ and $c_{\cO \cO T}$ for the free theory. With the usual free propagator
\begin{equation} \label{eq:Lifshitz_scalar_2pf}
\langle \varphi_i (x) \varphi_j (0) \rangle = \frac{\delta_{ij}}{(\dcft -2) S_{\dcft -1} } \frac{1}{x^{\dcft -2} },
\end{equation}
it is checked easily that the two-point function of the EM tensor is of the form \eqref{eq:Lifshitz_T_in_OO_TT} with
\begin{equation} \label{eq:Lifshitz_cT_free_theory}
c_T = \frac{m \dcft }{(\dcft -1) S_{\dcft -1} ^2}
\end{equation}
(see also \cite{Osborn:1993cr}).
Next, a calculation of the correlation function $\langle \varphi ^2(x_1) \varphi ^2(x_2) T_{\mu \nu } (x_3)\rangle$ gives the expected form \eqref{eq:Lifshitz_T_in_OO_OOT} with\footnote{
This result can also be checked using the Ward identity \eqref{eq:Lifshitz_WI_for_cOOT} (the energy-momentum tensor was chosen with the appropriate normalization). In our case
\begin{equation}
c_{\cO \cO } =\frac{2m}{(\dcft -2)^2 S_{\dcft -1} ^2} 
\end{equation}
and $\Delta  = \dcft -2$.}
\begin{equation}
c_{\cO \cO T} = - \frac{2 m \dcft }{(\dcft -1)(\dcft -2)S_{\dcft -1} ^3} .
\end{equation}
Thus we find
\begin{equation}
\beta _{\alpha } \approx - \frac{2\qd}{(\dcft -2)S_{\dcft -1} } \frac{S_{d_1-1} }{d_t} = - \frac{2\qd}{(\dcft -2) d_t} \pi ^{(d_t-\dcft )/2} \frac{\Gamma (\dcft /2)}{\Gamma (d_t/2)} .
\end{equation}
Using the marginality of the disorder $d_t\approx \dcft -4$ we get
\begin{equation} \label{eq:Lifshitz_phi2_example_Lifshitz}
\beta _{\alpha } \approx - \frac{\qd}{2 \pi ^2} .
\end{equation}
Interestingly, this is independent of $\dcft $; however, this is a special property of the case $\cO_0 = \varphi ^2$ (as can be seen by the diagrams evaluated below). We can now compare it to \cite{Boyanovsky:1982zz}. The correct relation between $\qd$ and the corresponding rescaled $\delta $ that appears in their equation (3.26) is $\qd=8\pi ^2 \delta $ as we will verify below (this is not precisely the normalization written below equation (3.26) there). With this normalization we find an agreement between \eqref{eq:Lifshitz_phi2_example_Lifshitz} and their equation for $\xi _{\alpha } $ in (3.26b) (using the fact that $\alpha =1$ to leading order).

\subsubsection{Quantum disorder for $\varphi^2$ in four space dimensions}

\begin{fmffile}{Feyndiagrams}
\fmfcmd{%
vardef cross_bar (expr p, len, ang) =
((-len/2,0)--(len/2,0))
rotated (ang + angle direction length(p)/2 of p)
shifted point length(p)/2 of p
enddef;
style_def crossed expr p =
cdraw p;
ccutdraw cross_bar (p, 3mm, 45);
ccutdraw cross_bar (p, 3mm, -45)
enddef;}

The quantum disorder case corresponds to $d_t=1$. For this case we can still use perturbation theory when disorder couples to $\cO_0 =\varphi ^2$, if we take the marginal case of $\dcft =5$.
In this subsection we present a detailed analysis of this case, to confirm our general expectations discussed above. In particular we want to verify that the anomalous dimension of the non-local-in-time operator related to the disorder is not just twice the anomalous dimension of $\cO_0$.

\begin{figure}[h]
\centering
\includegraphics[width=0.7\textwidth]{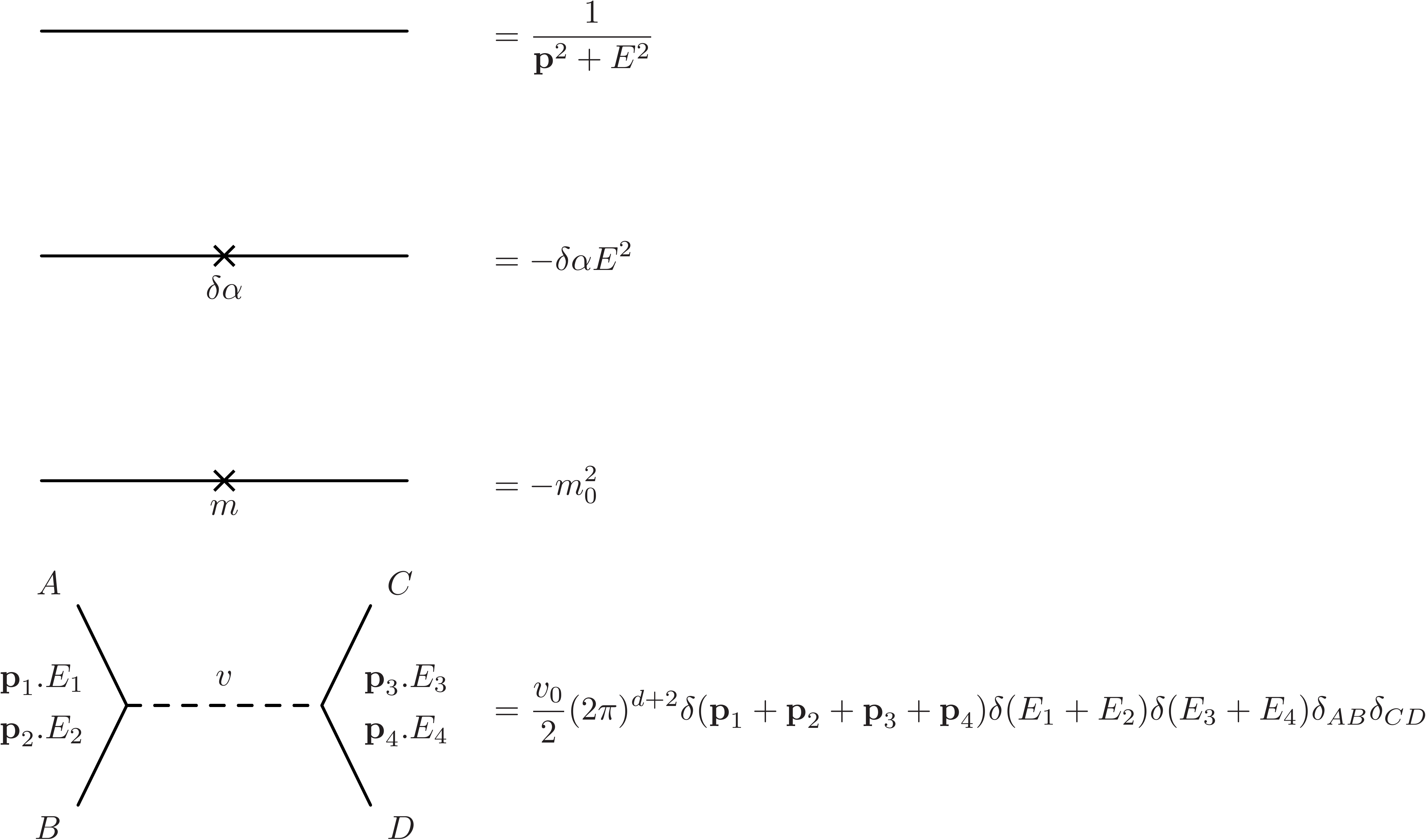}
\caption{Feynman rules for the $\dcft =5$ theory with disorder coupled to $\cO_0 = \varphi ^2$.}
\label{fig:Lifshitz_5d_Feynman_rules}
\end{figure}

In $5d$ the homogeneous $\varphi^4$ coupling is irrelevant so we do not have to include it. The replicated action 
is then \eqref{bcaction} which we rewrite as
\begin{equation} \label{eq:Lifshitz_5d_model_action}
S= \sum _A \int d^d\vecx dt \left[\frac{1}{2} \sum _{i=1} ^d (\partial _i \varphi _A)^2 + \frac{\alpha }{2} (\partial _t \varphi _A)^2 + \frac{m_0^2}{2} \varphi _A^2\right]  - \frac{\qd_0}{2}  \sum _{A,B} \int d^d\vecx dt dt' \, \varphi _A^2(\vecx,t) \varphi _B^2(\vecx,t').
\end{equation}
From now on, we use a single scalar for simplicity, but in fact the renormalization group coefficients we will compute hold also for $m$ real scalars (they are independent of $m$); the reason is that in this case, the $O(m)$ model is obtained simply by taking the $S_n$ indices $A$ to stand for a pair of $S_n$ and $O(m)$ indices $A \to (A,i)$, and since we take $n \to 0$, there is no dependence on $m$.
Subscripts were added to the couplings in \eqref{eq:Lifshitz_5d_model_action} to indicate that these are the bare couplings. When doing renormalization, we will write $\qd_0$ in terms of the renormalized couplings. We treat $m_0^2$ and $\delta \alpha  = \alpha -1$ as counter-terms which are included in the perturbation theory. Also, an overall wavefunction renormalization should be considered ($\varphi$ in \eqref{eq:Lifshitz_5d_model_action} is the bare field). The Feynman rules that are obtained, including the non-local coupling $\qd$, are shown in \autoref{fig:Lifshitz_5d_Feynman_rules}; their normalization is the one above, which differs by symmetry factors from the standard normalization. The dashed lines signify the non-local operators (similarly to \cite{Boyanovsky:1982zz}).

Dimensional regularization will be used as $d=4-\epsilon $. $m^2=0$ will not be needed.  Additionally, only one-particle-irreducible (1PI) diagrams are considered, since they are sufficient to expose the entire renormalization structure of the theory.
We will write the expressions for the various diagrams in dimensional regularization, and omit their evaluation since they reduce eventually to the same expressions as in the well-known $\varphi ^4$ scalar theory in four spacetime dimensions. We are interested in the renormalization group functions up to second order in $\qd$ and therefore we will need the full evaluation of the diagrams to first order, and only the cutoff dependent terms in second order (omitting as usual the terms that go to zero as the cutoff is removed).

Let us start with the two point function of $\varphi $. For any fixed $A$ there is a symmetry $\varphi _A \to -\varphi _A$, $\varphi _B \to \varphi _B$ ($B \neq A$), so $\langle\varphi _A \varphi _B \rangle =0$ for $A \neq B$. We therefore consider $\langle \varphi _A \varphi _A\rangle$ with fixed replica index $A$. We will omit the external leg propagators of the 1PI diagrams and so the zeroth order diagram is (the momenta are split to spatial momenta $\vecp$ and energy $E$)
\begin{equation}
\begin{gathered}
  \begin{fmfgraph*}(110,60)
    \fmfleft{i1}
    \fmfright{o1}
    \fmf{plain}{i1,v1}
    \fmf{plain}{v1,o1}
     \marrow{a}{up}{top}{$p_1,E_1$}{i1,v1}
     \marrow{b}{up}{top}{$p_2,E_2$}{o1,v1}
  \end{fmfgraph*}
\end{gathered} = \vecp_1^2+E_1^2 .
\end{equation}
We omit in all of the diagrams contributing to this correlation function the overall momentum conservation $(2\pi )^{d+1} \delta (\vecp_1+\vecp_2)\delta (E_1+E_2)$.

To first order we have only a single diagram

\begin{equation} \label{firstdiag}
\begin{split}
& \begin{gathered}
  \begin{fmfgraph*}(110,60)
    \fmfleft{i1}
    \fmfright{o1}
    \fmf{plain,tension=2}{i1,v1}
    \fmf{dashes,label=$\qd$}{v1,v2}
    \fmf{plain,tension=2}{v2,o1}
    \fmf{plain,left}{v1,v2}
  \end{fmfgraph*}
\end{gathered} = 4\qd_0 \int \frac{d^d\vecp}{(2\pi )^d} \frac{1}{\vecp^2+E_1^2} = - \frac{\qd_0}{4\pi ^2} E_1^2 \left( \frac{2}{\epsilon } -\gamma +1+\log \left( \frac{4\pi }{E_1^2}\right) \right) .
\end{split}
\end{equation}
Note that we omit the arbitrary dimensionful quantity that needs to be introduced in order to fix the units in dimensional regularization, and will restore it later on. Since there is no cutoff dependent $\vecp_1^2$ term, there is no need for a wavefunction renormalization to first order. In order to have no dependence on the cutoff, \eqref{firstdiag} should be compensated using the diagram
\begin{equation} \label{eq:Lifshitz_phi2_example_alpha_CT}
\begin{gathered}
  \begin{fmfgraph*}(110,60)
    \fmfleft{i1}
    \fmfright{o1}
    \fmf{crossed,label=$\delta \alpha$}{i1,o1}
  \end{fmfgraph*}
  \end{gathered}  = -\delta \alpha  E_1^2,
\end{equation}
fixing $\delta \alpha $ to first order to be
\begin{equation} \label{eq:Lifshitz_phi2_example_alpha_CT_value}
\delta \alpha  = -\frac{\qd}{2\pi ^2 \epsilon } +O(\qd^2) .
\end{equation}
$1/\epsilon  $ is essentially $\log (\Lambda)$ and so this $\delta \alpha $ is the same as the one found before (it gives rise to the beta function \eqref{eq:Lifshitz_phi2_example_Lifshitz}).

To second order (in which $\qd _0^2$ can be replaced by $\qd^2$), we have first of all the following diagram which is fixed from the previous evaluation
\begin{equation} \label{eq:Lifshitz_phi2_example_2_loop_exp1}
\begin{split}
& \begin{gathered}
  \begin{fmfgraph*}(110,60)
    \fmfleft{i1}
    \fmfright{o1}
    \fmf{plain,tension=2}{i1,v1}
    \fmf{dashes,label=$\qd$}{v1,v2}
    \fmf{plain,tension=2}{v2,o1}
    \fmf{crossed,left,label=$\delta \alpha $}{v1,v2}
  \end{fmfgraph*}
  \end{gathered}  =\\
  &= -4\qd_0 (\delta \alpha) E_1^2 \int \frac{d^d\vecp}{(2\pi )^d} \frac{1}{(\vecp^2+E_1^2)^2} = \frac{\qd^2 E_1^2}{8\pi ^4 \epsilon } \left( \frac{2}{\epsilon } -\gamma +\log \left( \frac{4\pi }{E_1^2} \right) \right) + \text{ finite } + O(\qd^3).
\end{split}
\end{equation}
As in the first order diagram, the renormalization of the coupling $\qd_0 = \qd +  G_1 \qd^2$ gives the contribution
\begin{equation} \label{eq:Lifshitz_phi2_example_2_loop_exp2}
-G_1 \frac{\qd^2}{4\pi ^2} E_1^2 \left(\frac{2}{\epsilon } -\gamma +1+\log \left( \frac{4\pi }{E_1^2} \right) \right),
\end{equation}
where here and below we write only the cutoff dependent part to second order (and use the $\sim$ sign to indicate that). There are also two new diagrams at second order:
\begin{equation} \label{eq:Lifshitz_phi2_example_2_loop_exp3}
\begin{split}
& \begin{gathered}
  \begin{fmfgraph*}(130,60)
    \fmfleft{d1,i1}
    \fmfright{d2,o1}
    \fmf{plain,tension=2}{i1,v1}
    \fmf{dashes,label=$\qd$}{v1,v2}
    \fmf{plain,tension=2}{v2,o1}
\fmffreeze 
\fmf{phantom,tension=4}{d1,v3}
\fmf{phantom,tension=4}{v4,d2}
\fmf{dashes,label=$\qd$}{v3,v4}
    \fmf{plain,tension=3}{v1,v3}
    \fmf{plain,tension=3}{v2,v4}
\fmf{plain,right}{v3,v4}
  \end{fmfgraph*}
  \end{gathered}  =  16 \qd^2 \int \frac{d^d\vecp}{(2\pi )^d} \frac{1}{(\vecp^2+E_1^2)^2} \int \frac{d^d\vecp'}{(2\pi )^d} \frac{1}{\vecp'^2+E_1^2} \sim \\
&  \qquad\qquad\qquad\qquad\qquad\qquad \sim - \frac{\qd^2}{4 \pi ^4} E_1^2 \left( \frac{1}{\epsilon ^2} - \frac{\gamma }{\epsilon } +\frac{1}{2\epsilon } +\frac{1}{\epsilon } \log \left( \frac{4\pi }{E_1^2} \right) \right) ,
\end{split}
\end{equation}
and
\begin{equation} \label{eq:Lifshitz_phi2_example_2_loop_exp4}
\begin{split}
& \begin{gathered}
  \begin{fmfgraph*}(130,60)
    \fmfleft{i1}
    \fmfright{o1}
\fmftop{d1}
\fmfbottom{d2}
    \fmf{plain}{i1,v1}
    \fmf{plain}{v1,v2}
    \fmf{plain}{v2,v3}
    \fmf{plain}{v3,v4}
    \fmf{plain}{v4,o1}
\fmffreeze
\fmf{phantom}{v2,d1}
\fmf{phantom}{v4,d2}
    \fmf{dashes,label=$\qd$,left}{v1,v3}
    \fmf{dashes,label=$\qd$,right,label.side=left}{v2,v4}
  \end{fmfgraph*}
  \end{gathered}  =  16 \qd^2 \int \frac{d^d\vecp}{(2\pi )^d} \frac{d^d\vecp'}{(2\pi )^d} \frac{1}{\vecp^2+E_1^2} \frac{1}{\vecp'^2+E_1^2} \frac{1}{(\vecp_1-\vecp+\vecp')^2+E_1^2} \sim \\
& \qquad\qquad\qquad\qquad\qquad\qquad \sim - \frac{\qd^2}{8\pi ^4} E_1^2 \left[ \frac{3}{\epsilon ^2} + \frac{9}{2\epsilon } - \frac{3\gamma }{\epsilon } +\frac{3}{\epsilon } \log \left(\frac{4\pi }{E_1^2} \right)\right] - \frac{\qd^2}{32\pi ^4} \frac{\vecp_1^2}{\epsilon } .
\end{split}
\end{equation}
There are also the following two diagrams
\begin{equation}
\begin{split}
 \begin{gathered}
  \begin{fmfgraph*}(130,60)
    \fmfleft{d1,i1}
    \fmfright{d2,o1}
    \fmf{plain,tension=2}{i1,v1}
    \fmf{plain}{v1,v2}
    \fmf{plain,tension=2}{v2,o1}
\fmffreeze 
\fmf{phantom,tension=15}{d1,v3}
\fmf{phantom,tension=15}{v4,d2}
    \fmf{dashes,tension=3}{v1,v3}
    \fmf{dashes,tension=3}{v2,v4}
\fmf{plain,right,tension=4}{v3,v4,v3}
  \end{fmfgraph*}
  \end{gathered} ,
\begin{gathered}
  \begin{fmfgraph*}(130,60)
  \fmfleft{i1,i2}
\fmftop{d1}
\fmfbottom{d2}
\fmf{plain,tension=3}{i1,v1}
\fmf{plain,tension=3}{v1,i2}
\fmf{dashes}{v1,v2}
\fmf{plain}{v3,v2,v4}
\fmf{phantom,tension=3}{d1,v3}
\fmf{dashes}{v3,v4}
\fmf{phantom,tension=3}{v4,d2}
\fmf{plain,left,tension=0.2}{v3,v4}
    \end{fmfgraph*}
  \end{gathered}  
\end{split}
\end{equation}
having a $\delta (0)$ of energy, proportional to the volume of the time direction. This vanishes in dimensional regularization, but in any case these diagrams are proportional to $n$ and thus vanish as $n \to 0$ in any regulator. In the language we used before they are related to mixing with the integrated identity operator, and all diagrams of this type will vanish as $n\to 0$.

As was mentioned, for any fixed $A$ there is a symmetry $\varphi _A \to -\varphi _A$ with the rest unchanged. $\varphi _A$ and $\varphi _B$ ($B \neq A$) transform differently under it, and thus they renormalize multiplicatively and do not mix (this is actually true even without this symmetry as we argued before). Since to second order we do have a cutoff dependent $\vecp_1^2$ term, we introduce a wavefunction renormalization
\begin{equation}
(\varphi _A)_R = \varphi _{A} \left( 1+ \frac{\qd^2}{32\pi ^4\epsilon } +\cdots \right)^{1/2} .
\end{equation}
This wavefunction renormalization multiplying the zeroth order diagram, should be combined with the expressions \eqref{eq:Lifshitz_phi2_example_2_loop_exp1}--\eqref{eq:Lifshitz_phi2_example_2_loop_exp4}, and with the $\delta \alpha $ counter-term to second order, to give all the contributions to this order in perturbation theory. Cancellation of the cutoff-dependent $\log (E_1)$ terms fixes the coupling renormalization $G_1 = - \frac{2}{\pi ^2\epsilon } $ and so
\begin{equation} \label{eq:Lifshitz_5d_model_g_ren}
\qd_0 = \qd - \frac{2}{\pi ^2\epsilon } \qd^2 + \cdots .
\end{equation}
The remaining dependence is compensated by the second order term in
\begin{equation}
\delta \alpha  = - \frac{\qd}{2\pi ^2\epsilon } + \frac{\qd^2}{4\pi ^4 \epsilon } \left( \frac{5}{2\epsilon } - \frac{5}{8} \right) + \cdots .
\end{equation}
We can now find the beta function of $\qd$. In order to do that in dimensional regularization we should restore dimensions by introducing the energy scale $\mu $. Defining the dimensionless coupling corresponding to $\qd$ by $\bar \qd$, they are related through $\qd = \bar \qd \mu ^{\epsilon } $. The beta function is given by $\mu  \pder{\bar \qd}{\mu } $ when $\qd_0$ is kept fixed, giving
\begin{equation}\label{gbeta}
\beta _{\qd} = - \frac{2}{\pi ^2} \qd^2 + \cdots .
\end{equation}

We can test the consistency of our results by considering the fully connected 1PI four-point functions of $\varphi $. Consider first the correlation function $\langle \varphi _A(\vecp_1,E_1) \varphi _A(\vecp_2,E_2) \varphi _B(\vecp_3,E_3) \varphi _B(\vecp_4,E_4)\rangle$ with $A \neq B$. Define
\begin{equation}
\delta _{12,34} = (2\pi )^{d+2} \delta (\vecp_1+\vecp_2+\vecp_3+\vecp_4) \delta (E_1+E_2)\delta (E_3+E_4)
\end{equation}
with similar definitions for different choices of indices.
In considering this correlation function, we evaluate only the diagrams proportional to $\delta _{12,34} $ and we leave this factor implicit. The leading order diagram is
\begin{equation}
\begin{split}
& \begin{gathered}
  \begin{fmfgraph*}(130,60)
\fmfleft{i1,i2}
\fmfright{o1,o2}
\fmf{plain}{i1,v1}
\fmf{plain}{i2,v1}
\fmf{dashes,label=$\qd$}{v1,v2}
\fmf{plain}{v2,o1}
\fmf{plain}{v2,o2}
\fmflabel{$2$}{i1}
\fmflabel{$1$}{i2}
\fmflabel{$4$}{o1}
\fmflabel{$3$}{o2}
  \end{fmfgraph*}
  \end{gathered}  = 4\qd_0 .
\end{split}
\end{equation}

To second order 
we have a cutoff dependent contribution from
\begin{equation}\label{otherdiagram}
\begin{split}
& \begin{gathered}
  \begin{fmfgraph*}(130,60)
\fmfleft{i1,i2}
\fmfright{o1,o2}
\fmf{plain}{i1,v1}
\fmf{plain}{i2,v1}
\fmf{dashes,label=$\qd$}{v1,v2}
\fmf{plain}{o1,v3,v2,v4,o2}
\fmf{dashes,tension=0,label=$\qd$}{v3,v4}
\fmflabel{$2$}{i1}
\fmflabel{$1$}{i2}
\fmflabel{$4$}{o1}
\fmflabel{$3$}{o2}
  \end{fmfgraph*}
  \end{gathered}  = 16 \qd^2 \int \frac{d^d\vecp}{(2\pi )^d} \frac{1}{\vecp^2 + E_3^2} \frac{1}{(\vecp_1+\vecp_2+\vecp)^2+E_3^2} \sim \qd^2 \frac{2}{\pi ^2 \epsilon },
\end{split}
\end{equation}
as well as the same divergent contribution for the diagram reflected horizontally (that is interchanging $1,2$ with $3,4$), and from
\begin{equation} \label{eq:Lifshitz_diagram_corresponding_to_OPE_region}
\begin{split}
& \begin{gathered}
  \begin{fmfgraph*}(130,60)
\fmfleft{i1,i2}
\fmfright{o1,o2}
\fmf{plain,tension=3}{i1,v1}
\fmf{dashes,label=$\qd$}{v1,v2}
\fmf{plain,tension=3}{v2,o1}
\fmf{plain,tension=3}{i2,v3}
\fmf{dashes,label=$\qd$}{v3,v4}
\fmf{plain,tension=3}{v4,o2}
\fmf{plain}{v1,v3}
\fmf{plain}{v2,v4}
\fmflabel{$2$}{i1}
\fmflabel{$1$}{i2}
\fmflabel{$4$}{o1}
\fmflabel{$3$}{o2}
  \end{fmfgraph*}
  \end{gathered}  +(3 \leftrightarrow 4)=\\
  & \\
  &\qquad\qquad =  16 \qd^2 \int \frac{d^d\vecp}{(2\pi )^d} \frac{1}{\vecp^2+E_1^2} \frac{1}{(\vecp_1+\vecp_3-\vecp)^2+E_3^2} + (3 \leftrightarrow 4) \sim \frac{4 \qd^2}{\pi ^2\epsilon } .
\end{split}
\end{equation}

There is also the diagram
\begin{equation}
\begin{split}
& \begin{gathered}
  \begin{fmfgraph*}(130,60)
\fmfleft{i1,i2}
\fmfright{o1,o2}
\fmf{plain,tension=3}{i1,v1}
\fmf{plain,tension=3}{i2,v1}
\fmf{dashes}{v1,v2}
\fmf{dashes}{v3,v4}
\fmf{plain,tension=3}{v4,o1}
\fmf{plain,tension=3}{v4,o2}
\fmf{plain,left,tension=0.5}{v2,v3,v2}
\fmflabel{$2$}{i1}
\fmflabel{$1$}{i2}
\fmflabel{$4$}{o1}
\fmflabel{$3$}{o2}
  \end{fmfgraph*}
  \end{gathered} 
\end{split}
\end{equation}
having again a $\delta _E(0)$, and vanishing for $n \to 0$.

We saw that there is no wavefunction renormalization to first order in $\qd$. Therefore, ignoring the last diagram we get that the cutoff dependence should be cancelled by
\begin{equation}
\qd_0=\qd-\frac{2}{\pi ^2\epsilon } \qd^2 +\cdots ,
\end{equation}
consistent with our previous result \eqref{eq:Lifshitz_5d_model_g_ren}.\footnote{In all of the diagrams in this section, since disorder is coupled to $\varphi ^2$ and we had no local interactions, all the energies were fixed and did not appear in loop integrations. As a result, there would actually be no dependence on $d_t$ if it was taken to be arbitrary. Thus the $\beta _{\qd}$ result will also be the same for the case of $d_t=\epsilon _d$ as in \cite{Boyanovsky:1982zz}. Comparing it to the equation for $\beta _{\delta } $ in (3.26a) of \cite{Boyanovsky:1982zz}, we find the normalization relation between our $\qd$ and their $\delta $ is $\qd=8\pi ^2\delta $.}

The analysis of the four-point function $\langle \varphi _A \varphi _A \varphi _A \varphi _A \rangle$ with fixed $A$ is the same, just with various permutations of the external legs, so again it is consistent with the previous computation of the beta function.
There is an alternative way to formulate this agreement. If we were using the result for the coupling renormalization as obtained here from the 4-point function, then in the calculation of the 2-point function it would imply that the $\frac{1}{\epsilon } \log (E_1)$ terms cancel, as they should for renormalizability.


Let us now compare the running of $\qd$ to the general conformal perturbation theory analysis of section \ref{universal}. The contribution to $\beta _{\qd}$ that we get by substituting the appropriate OPE coefficients of the case at hand in \eqref{Lifshitz_CPT_beta_g_partial} is $- \frac{1}{\pi ^2} \qd^2$. This gives half of the value we found in \eqref{gbeta}. Indeed, the logarithmic divergence in the diagram \eqref{eq:Lifshitz_diagram_corresponding_to_OPE_region}, which gives half of the contribution to the beta function, is dominated by the OPE region where the operators come together in pairs, and we see that this contribution comes from the $\cO_0 $ term in the OPE. On the other hand, the other diagrams \eqref{otherdiagram} are not dominated by this region, and as discussed above acquire contributions also when three of the operators are close together. This is an illustration of our general discussion in section \ref{universal}.

\end{fmffile}

\subsubsection{Mixing of operators}

\begin{figure}[h]
\centering
\includegraphics[width=0.3\textwidth]{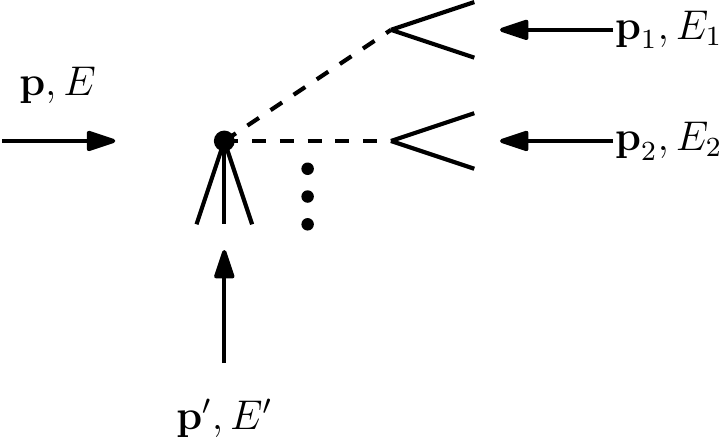}
\caption{A general external state of the form \eqref{eq:Lifshitz_local_operator_in_x}. In this figure we have $\cO ''=\varphi ^3$ and there are several operators integrated over time, which are depicted as $\cO^ {(1)}=\cO ^{(2)} = \cdots =\varphi ^2$.}
\label{fig:Lifshitz_local_in_x_external_vertex}
\end{figure}

We claimed that in general an operator $\cO _A'(\vecx,t)$ can mix with various operators of the form \eqref{eq:Lifshitz_local_operator_in_x}, 
and such operators mix also among themselves.
We would like to show that this effect indeed happens by using the example of this subsection. 
In addition, we argue that taking into account mixing of this form, the correlation functions of operators \eqref{eq:Lifshitz_local_operator_in_x} are renormalizable for $n \to 0$ (and for this it is important that we take the sum over the $A_1,\dots,A_k $ indices in all operators integrated over time; as usual time integration comes with a sum over replicas). We can represent external states of the form \eqref{eq:Lifshitz_local_operator_in_x}, with momentum $(\vecp,E)$, by vertices in Feynman diagrams as shown in \autoref{fig:Lifshitz_local_in_x_external_vertex}. Such an external vertex imposes
\begin{equation}
\vecp +\vecp '+\vecp_1+\vecp_2+\cdots =0,\qquad\qquad E_1=E_2=\cdots =0, \qquad\qquad E+E'=0.
\end{equation}

Now let us test the naive argument that is used in the literature, which says that the disorder operator which multiplies $\qd$ in \eqref{eq:Lifshitz_basic_replica_action} has a dimension that is fixed by the dimension of $\cO_0 (\vecx,t)$.
The dimension of the disorder coupling can be read from the beta function \eqref{gbeta}, and up to order $\qd$ it is
\begin{equation}
[\qd] = - \pder{\beta _{\qd}}{\qd} = \frac{4\qd}{\pi ^2} +\cdots .
\end{equation}

\begin{figure*}[t!]
    \centering
    \begin{subfigure}[t]{0.3\textwidth}
          \centering
          \includegraphics[height=0.6\textwidth]{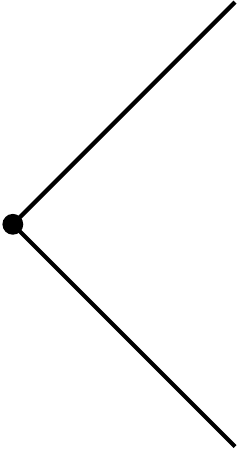}
          \caption{}
          \label{fig:Lifshitz_5d_phi2_dim_1}
    \end{subfigure}
    \begin{subfigure}[t]{0.3\textwidth}
          \centering
          \includegraphics[height=0.6\textwidth]{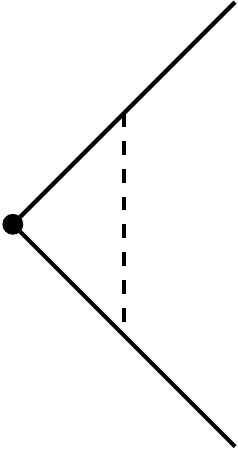}
          \caption{}
          \label{fig:Lifshitz_5d_phi2_dim_2}
    \end{subfigure}
    \begin{subfigure}[t]{0.3\textwidth}
          \centering
          \includegraphics[height=0.6\textwidth]{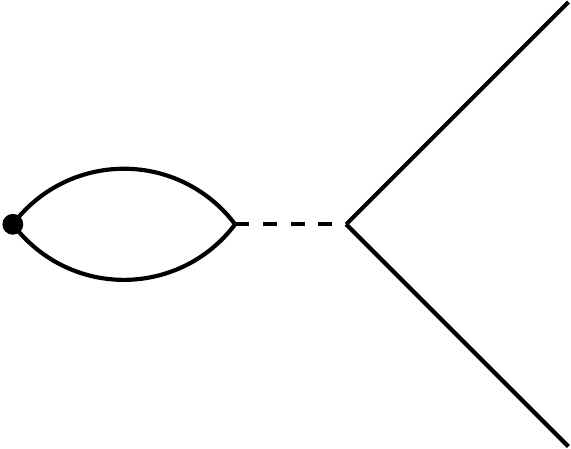}
          \caption{}
          \label{fig:Lifshitz_5d_phi2_dim_3}
    \end{subfigure}
    \caption{Diagrams contributing (except the last one) to the anomalous dimension of the $\varphi (\vecx,t)^2$ operator.}
    \label{fig:Lifshitz_5d_phi2_dim}
\end{figure*}

In order to compare this to the dimension of $\varphi ^2$ we consider the correlation function $\langle \frac{1}{2}  \varphi _A^2(\vecp,E) \varphi _A(\vecp_1,E_1) \varphi _A(\vecp_2,E_2) \rangle$ (with fixed $A$). The diagrams up to order $\qd$ that contribute to the anomalous dimension appear in \autoref{fig:Lifshitz_5d_phi2_dim_1} and \autoref{fig:Lifshitz_5d_phi2_dim_2}. We dropped the diagrams giving a divergence which is simply cancelled by the 1-loop $T_{00} $ running as in \eqref{firstdiag}--\eqref{eq:Lifshitz_phi2_example_alpha_CT_value}. There is also the diagram of \autoref{fig:Lifshitz_5d_phi2_dim_3}, but it has no logarithmic divergence and no contribution to the anomalous dimension (in dimensional regularization it is simply finite). Omitting the usual energy-momentum conservation delta functions, the diagram of \autoref{fig:Lifshitz_5d_phi2_dim_1} gives $1$, while the divergent part of the diagram in \autoref{fig:Lifshitz_5d_phi2_dim_2} is $\frac{\qd}{2\pi ^2\epsilon } $. This implies that the dimension of $\varphi ^2 $ is (adding the anomalous dimension to the classical dimension)
\begin{equation}
\Delta _{\varphi ^2} = 3-\frac{\qd}{2\pi ^2} +\cdots .
\end{equation}
If we use this naively in the disorder operator, we find that the dimension of the integrated disorder operator in \eqref{eq:Lifshitz_basic_replica_action} is $-\frac{\qd}{\pi ^2} $ which does not equal to $-[\qd]$. This shows explicitly that assuming that the dimension of integrated operators separated in time is simply the sum of their dimensions is wrong.

\begin{figure*}[t!]
    \centering
    \begin{subfigure}[t]{0.3\textwidth}
          \centering
          \includegraphics[height=0.6\textwidth]{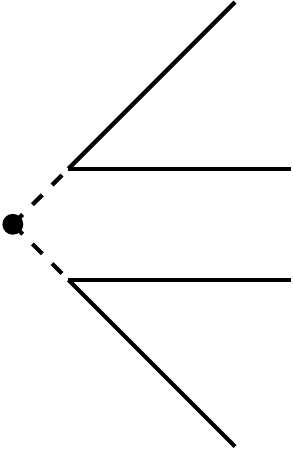}
          \caption{}
          \label{fig:Lifshitz_5d_disorder_op_dim_1}
    \end{subfigure}
    \begin{subfigure}[t]{0.3\textwidth}
          \centering
          \includegraphics[height=0.6\textwidth]{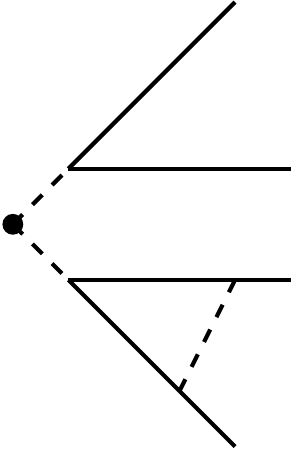}
          \caption{}
          \label{fig:Lifshitz_5d_disorder_op_dim_2}
    \end{subfigure}
    \begin{subfigure}[t]{0.3\textwidth}
          \centering
          \includegraphics[height=0.6\textwidth]{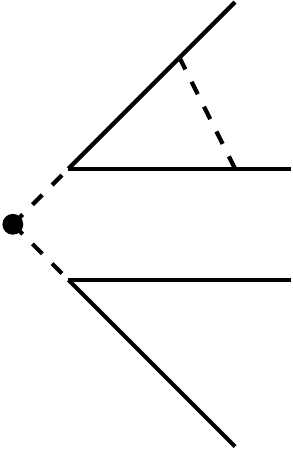}
          \caption{}
          \label{fig:Lifshitz_5d_disorder_op_dim_3}
    \end{subfigure}
    \begin{subfigure}[t]{0.3\textwidth}
          \centering
          \includegraphics[height=0.6\textwidth]{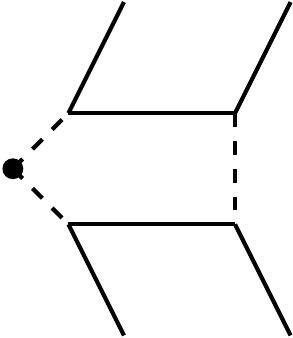}
          \caption{}
          \label{fig:Lifshitz_5d_disorder_op_dim_4}
    \end{subfigure}
    \begin{subfigure}[t]{0.3\textwidth}
          \centering
          \includegraphics[height=0.6\textwidth]{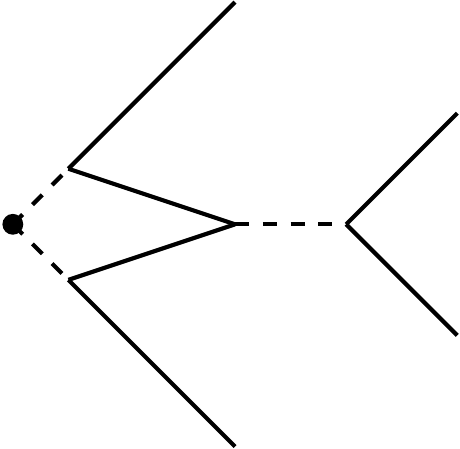}
          \caption{}
          \label{fig:Lifshitz_5d_disorder_op_dim_5}
    \end{subfigure}
    \begin{subfigure}[t]{0.3\textwidth}
          \centering
          \includegraphics[height=0.6\textwidth]{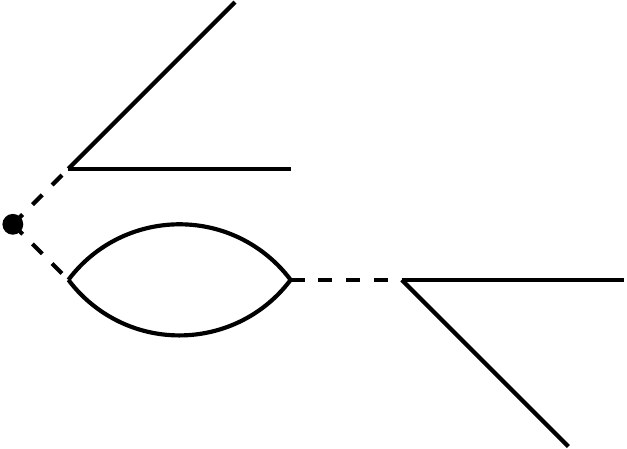}
          \caption{}
          \label{fig:Lifshitz_5d_disorder_op_dim_6}
    \end{subfigure}
    \caption{Diagrams contributing (except the last one) to the anomalous dimension of the (non-local in time) disorder operator.}
\end{figure*}

Instead, we should consider the local (in space) operator related to the disorder
\begin{equation}
\Psi(\vecx) \equiv \sum_{A,B} \int dt_1\, \varphi_A(\vecx,t_1)^2  \int dt_2\, \varphi_B(\vecx,t_2)^2
\end{equation}
as an independent operator which needs to be separately renormalized. We can find its dimension essentially in the same way as we do for local operators in a local field theory. The disorder operator $\Psi(\vecx )$ is of the form \eqref{eq:Lifshitz_local_operator_in_x} with $\cO ''$ being the identity operator, and $\cO ^{(1)}=\cO ^{(2)}= \varphi ^2$. It is convenient to consider the correlation function
\begin{equation}
\begin{split}
\langle \frac{1}{4} \Psi(\vecp )
\sum _{A_1} \varphi _{A_1} (\vecp _1,t_1) \sum _{A_2} \varphi _{A_2} (\vecp _2,t_2) \sum _{A_3} \varphi _{A_3} (\vecp _3,t_3) \sum _{A_4} \varphi _{A_4} (\vecp _4,t_4) \rangle,
\end{split}
\end{equation}
and represent the disorder operator $\Psi(\vecx )$ as in \autoref{fig:Lifshitz_local_in_x_external_vertex}. Let us look at the contributions that give $(2\pi )^{d+2} \delta (\vecp +\vecp _1 +\vecp _2 +\vecp _3+\vecp _4)\delta (E_1+E_2)\delta (E_3+E_4)$ and omit this factor (the other ones are the same). At  tree level we have the diagram of \autoref{fig:Lifshitz_5d_disorder_op_dim_1}. We recognize the corrections of \autoref{fig:Lifshitz_5d_disorder_op_dim_2} and \autoref{fig:Lifshitz_5d_disorder_op_dim_3} as coming from the renormalization of each of the $\varphi ^2$ factors in $\Psi(\vecx )$ (compare to the diagram in \autoref{fig:Lifshitz_5d_phi2_dim_2}). Assuming that the dimension of the disorder operator $\Psi(\vecx )$ is fixed by the dimension of $\varphi ^2$ amounts to taking into account only these diagrams. However, there are additional corrections, such as \autoref{fig:Lifshitz_5d_disorder_op_dim_4}. There is also the diagram of \autoref{fig:Lifshitz_5d_disorder_op_dim_5}, which contributes only for $A=B$ in the sum in $\Psi$, but it gives another sum over replica indices from the other side of the disorder interaction, so it should be taken into account as well. There is an additional diagram shown in \autoref{fig:Lifshitz_5d_disorder_op_dim_6} which has a $\delta (0)$ of energy, but as before this has an extra factor of $n$ with respect to the previous diagrams (the power of $n$ is the number of solid lines in correlators of this form) and vanishes as $n \to 0$. 
Writing only the divergent part of the order $\qd$ diagrams, the correlation function of interest is (up to order $\qd$)
\begin{equation}
2n^2 \left( 1 + \frac{\qd}{2\pi ^2 \epsilon }  + \frac{\qd}{2\pi ^2 \epsilon } + \frac{2\qd}{\pi ^2 \epsilon }+ \frac{\qd}{\pi ^2 \epsilon } \right) .
\end{equation}

As in local field theories we may then define a renormalized disorder operator
\begin{equation}
\begin{split}
\left( \Psi(\vecx ) \right)_R =  \left( 1-\frac{4\qd}{\pi ^2\epsilon }  + \cdots \right) \Psi(\vecx ) + \cdots 
\end{split}
\end{equation}
where we have additional corrections from mixing with other operators (such as the single-replica operator $ \sum _A \int dt_1\, \varphi _A(\vecx,t_1)\partial _{t_1} ^2 \varphi _A(\vecx,t_1)$).
However, it can be seen easily (at least to this order in $\qd$) that these additional operators do not get contributions from the disorder operator $\Psi(\vecx )$, and therefore the mixing matrix is triangular. This means that the anomalous dimension of the disorder operator is still given by the value on the diagonal (even though the eigenvector that corresponds to this value is no longer the trivial one).
Thus, the anomalous dimension to order $\qd$ (note that at this order there is no wavefunction renormalization of $\varphi $) is
\begin{equation}
\gamma  = - \frac{4\qd}{\pi ^2} + \cdots 
\end{equation}
and the dimension is
\begin{equation}
\left[ \Psi(\vecx ) \right]  = \left[ \sum _{A,B} \int dt_1\, \varphi _A^2(\vecx,t_1) \int dt_2\, \varphi _B^2(\vecx,t_2) \right] = 4 - \frac{4\qd}{\pi ^2} +\cdots .
\end{equation}
The dimension of the space integral over this is exactly $(-[\qd])$ as it should be.

\subsubsection{A perturbative random fixed point in $d=4+\epsilon$}

Using the result of \eqref{gbeta} we see that at $d=4+\epsilon $ there is a random fixed point at
\begin{equation}
\qd =\frac{\pi ^2 \epsilon }{2} +O(\epsilon ^2).
\end{equation}
In this fixed point, we can evaluate the dynamical exponent $z$. To leading order we can use the result from the $\dcft =5$ computation (e.g using \eqref{eq:Lifshitz_phi2_example_Lifshitz}, \eqref{eq:Lifshitz_RG_gamma_t} and \eqref{eq:Lifshitz_z_exponent_relation_to_beta_h00}) which gives
\begin{equation}
z = 1+ \frac{\epsilon }{8} +O(\epsilon ^2) .
\end{equation}
The anomalous dimension of the coupling $\qd$ is $4\qd/\pi ^2 + O(\qd^2) = 2\epsilon +O(\epsilon ^2)$ and therefore
\begin{equation}
[\qd ]= \epsilon +O(\epsilon ^2) .
\end{equation}
This is positive and it means that this is a UV-stable fixed point (rather than IR stable). The dimension of the operator $\varphi ^2$ in the random fixed point is
\begin{equation}
\Delta _{\varphi ^2} = 3+\epsilon - \frac{\qd}{2\pi ^2} +O(\qd^2) = 3+\frac{3\epsilon }{4} +O(\epsilon ^2) .
\end{equation}

\subsection{Example 2 : the holographic model of Hartnoll and Santos} \label{subsection:Lifshitz_HS}

An interesting class of theories in which the effects of disorder can be explicitly studied is large $N$ theories. The generalized free field theory of \autoref{section:classical_disorder_large_N} is the simplest case where we keep only the leading order in $1/N$, but other cases can also be studied. When the large $N$ theories are weakly coupled, they can be studied using large $N$ perturbative field theory techniques. Some strongly coupled large $N$ theories can be described by classical gravitational theories in a space of one higher dimension, using 
the AdS/CFT correspondence \cite{Maldacena:1997re, Gubser:1998bc, Witten:1998qj}. 

A specific model for this, with quenched quantum disorder, was studied in \cite{Hartnoll:2014cua,Hartnoll:2015faa,Hartnoll:2015rza}. The holographic theory there was taken to have the metric $g_{\mu \nu } $ and a scalar field $\Phi $ with the action
\begin{equation}\label{classact}
 I = \frac{\kappa}{L^{\dcft -1} } \int_{AdS} d^{\dcft +1}x \sqrt{g} \left( R + \frac{\dcft (\dcft -1)}{L^2} +2 (D\Phi )^2 + 4 m^2 \Phi ^2 \right),
\end{equation}
where $\kappa$ is a constant and $L$ is the curvature radius of anti-de Sitter (AdS) space.
This corresponds to a toy CFT which contains just the energy-momentum tensor (dual to $g_{\mu \nu}$) and a scalar operator $\cO$ dual to $\Phi$, whose dimension is related to the mass of $\Phi$. The normalizations in \eqref{classact} are such that we keep the leading non-trivial order in $1/N$ in the computation of all connected correlation functions; $\kappa$ scales as a power of $N$, such that quantum corrections to \eqref{classact} correspond to higher orders in $1/N$. In \cite{Hartnoll:2014cua,Hartnoll:2015faa} Gaussian disorder was coupled to $\cO$, and the mass of $\Phi $ was chosen so that the disorder is marginal.

In the gravitational description, the source $h(x)$ for $\cO$ appears through the boundary conditions for $\Phi$ at the boundary of AdS space, and one can solve the classical equations perturbatively in the source and then average over the disorder. In \cite{Hartnoll:2014cua,Hartnoll:2015faa} this was done analytically up to second order in the disorder, and numerically for general values of the disorder.
The low-energy behavior is governed in the holographic description by the behavior in the interior of AdS space, and 
the theory was found to flow to a disordered infra-red fixed point, with an average metric in the IR that has a Lifshitz form with a dynamical exponent $z$, depending on $\qd$.

In order to compare their results with ours, we first compute the two and three-point functions following from \eqref{classact}. By the usual rules of the AdS/CFT correspondence we obtain
\begin{equation} \label{eq:Lifshitz_holographic_cs}
\begin{split}
c_T & = \frac{2\kappa}{\pi ^{\dcft /2} } \frac{\dcft +1}{\dcft -1} \frac{\Gamma (\dcft +1)}{\Gamma (\dcft /2)},\\
c_{\cO \cO T} &= - 4\kappa \frac{\dcft \Delta  (2\Delta -\dcft )}{2\pi ^{\dcft } (\dcft -1)} \frac{\Gamma (\Delta ) \Gamma (\dcft /2)}{\Gamma \left( \Delta  - \frac{\dcft }{2} \right)},\\
c_{\cO \cO } &= 4\kappa \frac{(2\Delta -\dcft )\Gamma (\Delta )}{\pi ^{\dcft /2} \Gamma \left(\Delta  - \frac{\dcft }{2} \right)} .
\end{split}
\end{equation}

Substituting this in \eqref{eq:Lifshitz_z_exp2} (together with $\Delta  = (\dcft +1)/2$), our computation gives at leading order
\begin{equation}\label{forz}
z \approx 1+\frac{\qd}{2} \cdot  \frac{\dcft  \cdot \Gamma \left( \frac{\dcft +1}{2} \right) \Gamma \left( \frac{\dcft }{2} \right)^2}{\pi ^{ \frac{\dcft +1}{2} }\Gamma \left( \dcft +1\right) } = 1+  \frac{\pi ^{\dcft /2} }{2\pi } \Gamma \left( \frac{\dcft }{2} \right)  \left( \frac{\qd}{(2\pi )^{\dcft -1} } \right) .
\end{equation}

In order to compare with \cite{Hartnoll:2014cua} we need to match their normalization of the disorder distribution to ours. 
We find that the sources that they introduce obey (when their IR cutoff is removed)
\begin{equation}
\mean{ h(\vecx ) h(\vecy ) } = (2\pi )^{d} \bar V^2 \delta ^{d}(\vecx -\vecy )  \qquad \Rightarrow \qquad \qd = (2\pi )^d \bar V^2 .
\end{equation}
Thus, translating \eqref{forz} to their notation gives
\begin{equation}
z = 1 + \frac{\pi ^{\dcft /2} }{2\pi } \Gamma \left( \frac{\dcft }{2} \right) \bar V^2 + \cdots,
\end{equation}
which is exactly the second order in $\bar V$ term found in \cite{Hartnoll:2014cua}.

These theories were further studied in \cite{Hartnoll:2015rza}, in which a case with relevant disorder (a different choice of $m^2$ in \eqref{classact}) was also numerically analyzed. Both for strong marginal disorder and for relevant disorder, the low-energy behavior showed signs of discrete scale invariance related to complex anomalous dimensions. This is presumably related to our discussion in \autoref{subsection:RGnq_fixed_points} where we argued that along the renormalization group flow of disordered field theories operators can become degenerate and their dimensions can become complex. Note that in this example this cannot happen for weak marginal disorder, since there is no degeneracy of dimensions in the pure theory corresponding to \eqref{classact}, consistent with the results of \cite{Hartnoll:2015rza}.

\section*{Acknowledgements}

We would like to thank A.~Aharony, E.~Altman, M.~Berkooz, S.~Hartnoll, M.~Hogervorst, Z.~Komargodski, Y.~Korovin, J.~Santos, S.~Yankielowicz and P.~Young for useful discussions.
This work was supported in part  by the I-CORE program of the Planning and Budgeting Committee and the Israel Science Foundation (grant number 1937/12), by an Israel Science Foundation center for excellence grant, by the Minerva foundation with funding from the Federal German Ministry for Education and Research, and by the ISF within the ISF-UGC joint research program framework (grant no.\ 1200/14). OA is the Samuel Sebba Professorial Chair of Pure and Applied Physics.  

\appendices

\section{Perturbative local renormalization group} \label{section:local_RG_demo}

For homogeneous (constant) couplings, it can be seen in a perturbative expansion that theories are RG invariant in the Wilsonian sense, and one can obtain a universal formula for the quadratic term in the beta function (see for instance \cite{Cardy:1996xt}). Here we generalize this to inhomogeneous couplings, and see at the leading orders how the Wilsonian local RG works. The starting point is a Euclidean theory with action $S=S_0 + \int d^dx \, g_i(x) \cO_i (x)$ with $g_i(x)$ the inhomogeneous couplings. The cutoff will be taken to be a minimal distance $a$.
The corresponding dimensionless couplings are $u_i=g_i a^{d-\Delta _i} $ (where $\Delta _i$ is the dimension of $\cO _i$). 
The partition function in perturbation theory is (repeated indices are summed over)
\begin{equation}
\begin{split}
\frac{Z}{Z_0} = 1 -a^{\Delta _i-d}  \int d^dx\, u_i(x) \langle \cO _i(x)\rangle + \frac{1}{2}a^{\Delta _i+\Delta _j-2d} \int_{|x_1-x_2| >a}  d^dx_1 d^dx_2 \, u_i(x_1)u_j(x_2)\langle \cO _i(x_1) \cO _j(x_2)\rangle + \cdots .
\end{split}
\end{equation}
Now let us change the cutoff scale $a \to a'=a(1+\epsilon )$ for infinitesimal $\epsilon $ and see how the theory can remain invariant. The first order change is from the explicit powers of $a$, e.g in the first order term the variation is $-(\Delta _i-d) \epsilon  a^{\Delta _i-d} \int d^dx \, u_i(x) \langle \cO _i(x)\rangle$. These can be compensated by $u_i \to u_i +\epsilon (d-\Delta _i)u_i$. The next variation comes from the $a$ dependence of the integral in the second order. Having a cutoff $a$ in this term can be replaced by $a'$ plus the variation
\begin{equation}
\begin{split}
\frac{1}{2} a^{\Delta _i+\Delta _j-2d} \int _{a<|x_1-x_2| \le a(1+\epsilon )} d^dx_1 d^dx_2 \, u_i(x_1)u_j(x_2) \langle \cO_i (x_1) \cO_j (x_2) \rangle .
\end{split}
\end{equation}
For these close-by operators the OPE can be used (we will assume in this appendix that the operators are normalized so that $c_{\cO \cO } =1$ in \eqref{eq:Lifshitz_appendix_cOO_def}). For a scalar operator $\cO _k$ in the OPE we get
\begin{equation}
\begin{split}
& \frac{1}{2} a^{\Delta _i+\Delta _j-2d} c_{ijk}  \int _{a<|x|\le a(1+\epsilon )} d^dx d^dx_2 \, u_i(x_2+x) u_j(x_2) |x|^{\Delta _k-\Delta _i-\Delta _j} \langle \cO _k (x_2)\rangle \approx \\
& \approx \frac{1}{2} a^{\Delta _k-2d} c_{ijk} \int d^dx_2 \, \langle \cO _k(x_2)\rangle u_j(x_2) \int _{a<|x|\le a(1+\epsilon )} d^dx \, \left(u_i(x_2) + x^{\mu } \partial _{\mu } u_i(x_2) + \frac{1}{2} x^{\mu } x^{\nu } \partial _{\mu } \partial _{\nu } u_i(x_2) + \cdots \right) \approx \\
& \approx \frac{1}{2} a^{\Delta _k-d} c_{ijk} \int d^dx_2 \langle \cO _k(x_2)\rangle u_j(x_2) \epsilon S_{d-1} \left(u_i(x_2) + \frac{a^2}{2d} \nabla ^2 u_i(x_2) + \cdots \right) .
\end{split}
\end{equation}
The expression in brackets is the expansion of the angular integral $S_{d-1} ^{-1} \int _{|x|=a} d\Omega \, u_i(x_2+x)$. This variation can also be compensated by changing the functions $u_k(x)$. We see invariance under local RG when the inhomogeneous couplings run as
\begin{equation} \label{eq:local_RG_demo_couplings_flow}
\begin{split}
a \frac{du_k(x)}{da} &= (d-\Delta _k)u_k + \frac{S_{d-1} }{2} \sum _{ij} c_{ijk} u_j(x) \left( u_i(x) + \frac{a^2}{2d} \nabla ^2 u_i(x) + \cdots \right) + \cdots  ,\\
a \frac{dg_k(x)}{da} &= \frac{S_{d-1} }{2} \sum _{ij} c_{ijk} a^{d+\Delta _k-\Delta _i-\Delta _j} g_j(x) \left( g_i(x) + \frac{a^2}{2d} \nabla ^2 g_i(x) + \cdots \right) + \cdots .
\end{split}
\end{equation}
Generically this flow leads to inhomogeneous couplings for all the operators allowed by the symmetries, even if one starts from only one inhomogeneous coupling. We can use \eqref{eq:local_RG_demo_couplings_flow} to obtain the flow of the disorder probability distribution, but we will not do this here.

\section{Derivation of \eqref{eq:RGnq_product_connected_coorelator_relation_replica}} \label{section:derivation_of_product_connected_correlator}

We will prove \eqref{eq:RGnq_product_connected_coorelator_relation_replica} by induction on $I$. For $I=1$ the equation  is \eqref{eq:RGnq_connected_coorelator_relation_replica2} that was proven in the text. The equation for $(I-1)$ is 
\begin{equation} \label{eq:app_product_connected_coorelator_relation_ind}
\begin{split}
& \lim _{n \to 0} \langle \cO _{i_1,1}(x_1) \cdots \cO _{i_{I-1},{I-1}}(x_{I-1}) \sum _{{A_1}=1}^n \cO_{j_1,{A_1}}(y_1) \sum _{{A_2}=1}^n \cO _{j_2,{A_2}}(y_2) \cdots  \sum _{{A_k}=1}^n \cO _{j_k,{A_k}}(y_k) \rangle ^{replicated} = \\
& \qquad = \sum _{\substack{\text{Partitions of }\{1,\dots,k\} \\ \text{into } S_1,\dots ,S_{I-1} }} \mean{\langle \cO _{i_1} (x_1) \prod _{l \in S_1} \cO _{j_l} (y_l)  \rangle_{conn}  \cdots \langle\cO _{i_{I-1}} (x_{I-1}) \prod _{l \in S_{I-1}} \cO _{j_l} (y_l)\rangle _{conn} }.
\end{split}
\end{equation}

Using the $S_n$ symmetry we can write the left-hand side as
\begin{equation} \label{eq:app_product_connected_coorelator_relation_lhs}
\begin{split}
& \lim_{n\to 0} \left[ \left( \langle \cO _{i_1,1}(x_1) {\cO}_{j_1,1}(y_1) \cdots \cO _{i_{I-1},{I-1}}(x_{I-1}) 
\sum _{A_2=1}^n \cO _{j_2,A_2}(y_2) \cdots  \sum _{A_k=1}^n \cO _{j_k,A_k}(y_k) \rangle ^{replicated} + (I-2)\ {\rm perms} \right) + \right. \\
& \left. \qquad (n-I+1) \langle \cO _{i_1,1}(x_1) \cdots \cO _{i_{I-1},{I-1}}(x_{I-1}) 
\cO_{j_1,I}(y_1) \sum _{A_2=1}^n \cO _{j_2,A_2}(y_2) \cdots  \sum _{A_k=1}^n \cO _{j_k,A_k}(y_k) \rangle ^{replicated} \right].
\end{split}
\end{equation}
The bottom line is the one we want to compute. We can compute each term on the top line using the induction step; for instance the first term there is equal to
\begin{equation}\label{eq:app_product_connected_coorelator_relation_firstterm}
\sum _{\substack{\text{Partitions of } \{2,\dots,k\} \\ \text{into } S_1,\dots ,S_{I-1} }} \mean{\langle (\cO _{i_1} (x_1) \cO_{j_1}(y_1)) \prod _{l \in S_1} \cO _{j_l} (y_l)  \rangle_{conn}  \cdots \langle\cO _{i_{I-1}} (x_{I-1}) \prod _{l \in S_{I-1}} \cO _{j_l} (y_l)\rangle _{conn}},
\end{equation}
where $(\cO_{i_1}(x_1) \cO_{j_1}(y_1))$ is considered as a single operator, such that
\begin{equation}\label{eq:app_product_connected_coorelator_relation_joint}
\begin{split}
\langle (\cO _{i_1} (x_1) \cO_{j_1}(y_1)) & \prod_{l\in S} \cO _{j_l} (y_l)  \rangle_{conn} =
\langle \cO _{i_1} (x_1) \cO_{j_1}(y_1) \prod_{l\in S} \cO _{j_l} (y_l)  \rangle_{conn} + \\
& \sum_{{\rm partitions\ of\ }S{\rm \ to\ }{\hat S}_1,{\hat S}_2} \langle \cO _{i_1} (x_1) \prod_{l\in {\hat S}_1} \cO _{j_l} (y_l)  \rangle_{conn}\langle \cO_{j_1}(y_1) \prod_{l\in {\hat S}_2} \cO _{j_l} (y_l)  \rangle_{conn}.
\end{split}
\end{equation}

Plugging this into \eqref{eq:app_product_connected_coorelator_relation_firstterm}, and plugging \eqref{eq:app_product_connected_coorelator_relation_firstterm} into  \eqref{eq:app_product_connected_coorelator_relation_lhs}, the terms coming from the first line of \eqref{eq:app_product_connected_coorelator_relation_joint} (adding together all the permutations) are exactly the same as the ones appearing on the right-hand side of \eqref{eq:app_product_connected_coorelator_relation_ind}, while the terms coming from the second line give
\begin{equation}\label{eq:app_product_connected_coorelator_relation_afive}
(I-1)
\sum _{\substack{\text{Partitions of }\{2,\dots,k\} \\ \text{into } S_1,\dots ,S_{I} }} \mean{\langle \cO _{i_1} (x_1) \prod _{l \in S_1} \cO _{j_l} (y_l)  \rangle_{conn}  \cdots \langle\cO _{i_{I-1}} (x_{I-1}) \prod _{l \in S_{I-1}} \cO _{j_l} (y_l)\rangle _{conn} \langle\cO _{j_1} (y_1) \prod _{l \in S_{I}} \cO _{j_l} (y_l)\rangle _{conn}}\,,
\end{equation}
and we find that \eqref{eq:app_product_connected_coorelator_relation_afive} plus the $n\to 0$ limit of the second line of \eqref{eq:app_product_connected_coorelator_relation_lhs} gives zero, which is precisely what we wanted to prove.

\section{Non-Gaussian disorder} \label{section:Ph_non_Gaussian}

In explicit calculations the disorder distribution $P[h]$ is usually taken to be Gaussian. Let us see how the analysis in the replica trick is modified for a non-Gaussian distribution.
For both classical and quantum disorder we will still assume that the disorder is local in space (that is, there are no correlations between the disorder field at different points in space), but the distribution at each point is not necessarily Gaussian.

Recall the cumulant expansion. Suppose we have random variables $h _i$. The moment generating function is
\begin{equation}
\mean{e^{h_i \cO _i} }= e^{K(\cO_i)},
\end{equation}
where $K(\cO_i)$ is the cumulant generating function. The first few terms in the cumulant generating function are
\begin{equation}\label{cumulants}
\begin{split}
K &= \mean{h_i}\cO _i + \frac{1}{2} (\mean{h_i h_j} - \mean{h_i} \cdot \mean{h_j}) \cO _i \cO _j + \\
&+ \frac{1}{3!} \left( \mean{h_i h_j h_k} - \mean{h_i h_j} \cdot \mean{h_k} - \mean{h_i h_k }\cdot \mean{h_j }- \mean{h_j h_k }\cdot \mean{h_i}+2 \mean{h_i}\cdot \mean{h_j}\cdot \mean{h_k}\right) \cO _i \cO _j \cO _k + \\
&+ \frac{1}{4!} \left( \mean{h_i h_j h_k h_l} - \mean{h_i h_j h_k}\cdot \mean{h_l} -\mean{h_i h_j h_l}\cdot \mean{h_k} -\mean{h_i h_l h_k}\cdot \mean{h_j} -\mean{h_l h_j h_k}\cdot \mean{h_i} \right. - \\
& - \mean{h_i h_j }\cdot \mean{h_k h_l} - \mean{h_i h_k }\cdot \mean{h_j h_l} - \mean{h_i h_l }\cdot \mean{h_j h_k}  + \\
& + 2\mean{h_i h_j}\cdot \mean{h_k}\cdot \mean{h_l} + 2\mean{h_i h_k}\cdot \mean{h_j}\cdot \mean{h_l}+ 2\mean{h_i h_l}\cdot \mean{h_k}\cdot \mean{h_j}+ 2\mean{h_j h_k}\cdot \mean{h_i}\cdot \mean{h_l}+ 2\mean{h_j h_l}\cdot \mean{h_i}\cdot \mean{h_k}+ 2\mean{h_k h_l}\cdot \mean{h_i}\cdot \mean{h_j}- \\
& - \left.  6 \mean{h_i} \cdot \mean{h_j} \cdot \mean{h_k}\cdot \mean{h_l} \right) \cO _i \cO _j \cO _k \cO _l + \cdots .
\end{split}
\end{equation}
The coefficients are the same as those in the expressions for connected correlation functions in terms of the general ones.

\textbf{Classical disorder} (\autoref{section:RG_classical_disorder}).
After doing the replica trick we had the following dependence on the disorder
\begin{equation}
e^{ - \int d^dx\, h(x) \sum _A \cO _A(x)}
\end{equation}
and we would like to integrate it $\int Dh\, P[h]$ to get the replica action from \eqref{eq:RGnq_W_n_def}. Suppose that
\begin{equation} \label{eq:Ph_non_Gaussian_moments}
\begin{split}
& \mean{h(x)}=0 \\
& \mean{h(x) h(y)} = \kappa _2 \delta (x-y) \\
& \mean{h(x)h(y)h(z)} = \kappa_3 \delta (x-y) \delta (x-z) \\
& \mean{h(x)h(y)h(z)h(w)}= \kappa_4 \delta (x-y) \delta (x-z) \delta (x-w) + \\
& \qquad + \kappa^2_2 \delta (x-y)\delta (z-w)+\kappa^2_2 \delta (x-z)\delta (y-w) +\kappa^2_2 \delta (x-w) \delta (y-z)  .
\end{split}
\end{equation}
We get that the disorder gives the following contribution to the replicated action
\begin{equation}
\begin{split}
\delta S_{replica} & = - \frac{\kappa_2}{2} \int d^dx  \sum _{A,B} \cO _A(x) \cO _B(x) + \\
&+  \frac{\kappa_3}{3!} \int d^dx \sum _{A,B,C} \cO _A(x) \cO _B(x) \cO _C (x) - \\
&- \frac{\kappa_4}{4!}  \int d^dx \sum _{A,B,C,D} \cO _A(x) \cO _B(x) \cO _C(x) \cO _D(x) + \cdots .
\end{split}
\end{equation}
If the disorder is chosen to be marginal (that is, saturating the Harris criterion), all the higher cumulants $\kappa_3,\kappa_4,\dots $ of the disorder distribution give rise to irrelevant terms and thus in principle can be dropped if the disorder is regarded as a small perturbation. This justifies using the Gaussian distribution in such a case. In general the $n$'th cumulant of the local disorder distribution maps to a coupling involving $n$ replicas in the replica theory. The generalization to disorder coupled to more than one operator is straightforward.

\textbf{Quantum disorder} (\autoref{section:RG_quantum_disorder}). 
In the quantum disorder case, the disorder appears in the replica trick as
\begin{equation}
e^{- \int d^d\vecx dt\, h(\vecx) \sum _A \cO_A (\vecx,t)} .
\end{equation}
We should then again integrate $\int Dh \, P[h]$. With the same notations as in \eqref{eq:Ph_non_Gaussian_moments}, the effect of the disorder on the replicated action is now
\begin{equation}
\begin{split}
\delta S_{replica} & = - \frac{\kappa_2}{2} \int d^d\vecx dt_1 dt_2 \sum _{A,B} \cO _A(\vecx,t_1) \cO _B(\vecx,t_2) + \\
&+  \frac{\kappa_3}{3!} \int d^d\vecx dt_1 dt_2 dt_3 \sum _{A,B,C} \cO _A(\vecx,t_1) \cO _B(\vecx,t_2) \cO _C (\vecx,t_3) - \\
&- \frac{\kappa_4}{4!}  \int d^d\vecx dt_1dt_2dt_3dt_4 \sum _{A,B,C,D} \cO _A(\vecx,t_1) \cO _B(\vecx,t_2) \cO _C(\vecx,t_3) \cO _D(\vecx,t_4) + \cdots .
\end{split}
\end{equation}
For marginal disorder $\Delta = \frac{d+2}{2} $, once again all the higher $\kappa_3,\kappa_4,\dots $ terms are irrelevant. For other $\Delta $, the number of such terms that are relevant or marginal is finite, unless $\Delta \le 1$.

Note that naively the expansion \eqref{cumulants} contains also terms that are non-local in $\vecx$. However, these all cancel; the replica action must be local in $\vecx$ whenever $P[h]$ involves independent disorder distributions at different points, since the integral over $h(\vecx)$ in \eqref{eq:RGnq_W_n_def} splits in this case into separate integrals at every point $\vecx$.

\section{The energy-momentum tensor in the $\cO \times \cO $ OPE} \label{section:Lifshitz_T_in_OO}

We work with a general Euclidean CFT in $\dcft$ dimensions, in which we have a scalar primary operator $\cO (x)$. To get the OPE coefficient of the EM tensor we use (see e.g \cite{Osborn:1993cr})
\begin{equation} \label{eq:Lifshitz_T_in_OO_TT}
\langle T_{\mu \nu } (x) T_{\rho \sigma } (0) \rangle = \frac{c_T}{x^{2\dcft} } \left( \frac{1}{2} (I_{\mu \rho } I_{\nu \sigma } +I_{\mu \sigma } I_{\nu \rho } )- \frac{\delta _{\mu \nu } \delta _{\rho \sigma } }{\dcft} \right), 
\text{ where } I_{\mu \nu } (x) \equiv \delta _{\mu \nu } -2 \frac{x_{\mu } x_{\nu } }{x^2},
\end{equation}
and
\begin{equation} \label{eq:Lifshitz_T_in_OO_OOT}
\langle \cO (x_1) \cO (x_2) T_{\mu \nu } (x_3)\rangle= c_{\cO \cO T} \frac{V_{\mu } V_{\nu } - \frac{1}{\dcft} \delta _{\mu \nu } V_{\alpha } V_{\alpha } }{x_{13} ^{\dcft-2} x_{23} ^{\dcft-2} x_{12} ^{2\Delta -\dcft+2} }  , \text{ where } V^{\mu } \equiv \frac{x^{\mu } _{13} }{x_{13} ^2} - \frac{x_{23} ^{\mu } }{x_{23} ^2}  \text{ and } x_{ij} \equiv x_i-x_j .
\end{equation}

Using invariance under Lorentz, scaling and translations, we have in the OPE
\begin{equation}
\cO (x_1) \cO (x_2) \supset \alpha  \frac{x_{12} ^{\mu } x_{12} ^{\nu } }{x_{12} ^{2\Delta -\dcft+2} } T_{\mu \nu } (x_2) + \text{ descendants of }T 
\end{equation}
for some constant $\alpha$. We do not write a $\delta ^{\mu \nu } T_{\mu \nu } $ term since the EM tensor is traceless. Therefore we get in the three-point function
\begin{equation}
\begin{split}
\langle \cO (x_1) \cO (x_2) T_{\rho \sigma } (x_3) \rangle &= \alpha  \frac{x_{12} ^{\mu } x_{12} ^{\nu } }{x_{12} ^{2\Delta -\dcft+2} } \langle T_{\mu \nu } (x_2) T_{\rho \sigma } (x_3)\rangle + \text{ higher order in } x_{12} \\
&= \alpha  \frac{x_{12} ^{\mu } x_{12} ^{\nu } }{x_{12} ^{2\Delta -\dcft+2} } \frac{c_T}{x_{23} ^{2\dcft} } \left( \frac{1}{2} I_{\mu \rho } I_{\nu \sigma } + \frac{1}{2} I_{\mu \sigma } I_{\nu \rho } - \frac{\delta _{\mu \nu } \delta _{\rho \sigma } }{\dcft}  \right )+ \cdots 
\end{split}
\end{equation}
(the $I_{\mu \nu } $ are evaluated at $x_{23} $).
Let us take $\rho  \neq \sigma $ to avoid writing the last term. Then
\begin{equation}
\langle \cO (x_1) \cO (x_2) T_{\rho \sigma } (x_3) \rangle  = \alpha  \frac{c_T}{x_{12} ^{2\Delta -\dcft+2} x_{23} ^{2\dcft}} \left( x_{12} ^{\rho } -2 \frac{x_{12} \cdot x_{23} x_{23} ^{\rho } }{x_{23} ^2} \right) \left(x_{12} ^{\sigma } -2 \frac{x_{12} \cdot x_{23} x_{23} ^{\sigma } }{x_{23} ^2} \right) + \cdots .
\end{equation}
In order to compare to \eqref{eq:Lifshitz_T_in_OO_OOT}, we need $V_{\mu } $ for small $x_{12}$. To leading order it is given by
\begin{equation}
V^{\mu } = \frac{x_{12} ^{\mu } }{x_{23} ^2} - \frac{2x_{23 } \cdot x_{12} x_{23} ^{\mu } }{x_{23} ^4} +\cdots .
\end{equation}
Therefore, for $\rho  \neq \sigma $, substituting this in the full three point function \eqref{eq:Lifshitz_T_in_OO_OOT},
\begin{equation}
\begin{split}
\langle \cO (x_1) \cO (x_2) T_{\rho \sigma } (x_3) \rangle  &= \frac{c_{\cO \cO T} }{x_{12} ^{2\Delta -\dcft+2} x_{13} ^{\dcft-2} x_{23} ^{\dcft-2} } \frac{1}{x_{23} ^4} \left( x_{12} ^{\rho } -\frac{2x_{12} \cdot x_{23} x_{23} ^{\rho } }{x_{23} ^2} \right) \left(x_{12} ^{\sigma } - \frac{2 x_{12} \cdot x_{23} x_{23} ^{\sigma } }{x_{23} ^2} \right) + \cdots  = \\
&= \frac{c_{\cO \cO T} }{x_{12} ^{2\Delta -\dcft+2} x_{23} ^{2\dcft} } \left( x_{12} ^{\rho } -\frac{2x_{12} \cdot x_{23} x_{23} ^{\rho } }{x_{23} ^2} \right) \left(x_{12} ^{\sigma } - \frac{2 x_{12} \cdot x_{23} x_{23} ^{\sigma } }{x_{23} ^2} \right) + \cdots .
\end{split}
\end{equation}
Comparing the two expressions we get $\alpha =c_{\cO \cO T} /c_T$ and thus the OPE contains
\begin{equation} \label{eq:Lifshitz_OPE_OO_T}
\cO (x_1) \cO (x_2) \supset \frac{c_{\cO \cO T} }{c_T}   \frac{x_{12} ^{\mu } x_{12} ^{\nu } }{x_{12} ^{2\Delta -\dcft+2} } T_{\mu \nu } (x_2) .
\end{equation}

Note that the conformal Ward identity in these conventions is
\begin{equation} \label{eq:Lifshitz_WI_for_cOOT}
\frac{c_{\cO \cO T} }{c_{\cO \cO } } = - \frac{\dcft \Delta }{(\dcft -1)S_{\dcft -1} } =  - \frac{\dcft\Delta }{\dcft-1} \frac{\Gamma (\dcft/2)}{2\pi ^{\dcft/2} }\,,
\end{equation}
where $c_{\cO \cO } $ is defined through
\begin{equation} \label{eq:Lifshitz_appendix_cOO_def}
\langle \cO (x_1) \cO (x_2) \rangle = \frac{c_{\cO \cO } }{x_{12} ^{2\Delta } } .
\end{equation}

\section{Explicit evaluation of $\int d^{\dcft}x' \langle \cO (x) \cO (0) T_{00} (x')\rangle$} \label{section:Lifshitz_OOT_integral}

It will be useful to evaluate explicitly the integral over the position of the EM tensor of a correlation function. It will be used both in checking the GCS equations, and in comparing to expectations relating it to time dilation.

In this appendix $\cO $ will denote a general dimension $\Delta $ scalar primary operator in a Euclidean CFT in $\dcft$ dimensions. We would like to evaluate (see \eqref{eq:Lifshitz_T_in_OO_OOT})
\begin{equation}
\begin{split}
& \int d^d\vecx_1 dt_1 \langle \cO (\vecx,t) \cO (0) T_{00} (\vecx_1,t_1) \rangle = \\
& \qquad = \int d^d\vecx_1 dt_1 c_{\cO \cO T} \frac{\left( \frac{t-t_1}{(\vecx-\vecx_1)^2+(t-t_1)^2} + \frac{t_1}{\vecx_1^2+t_1^2} \right)^2 - \frac{1}{d+1} V_{\alpha } V_{\alpha } }{ (\vecx^2+t^2)^{(2\Delta -d+1)/2}  \left((\vecx-\vecx_1)^2+(t-t_1)^2 \right)^{\frac{d-1}{2} } (\vecx_1^2+t_1^2)^{\frac{d-1}{2} } }\,, 
\end{split}
\end{equation}
where $V_{\mu } \equiv \frac{(x-x_1)_{\mu } }{(x-x_1)^2} + \frac{x_{1,\mu } }{x_1^2}$, namely
\begin{equation}
\frac{x^{2\Delta -d+1} }{c_{\cO \cO T} } \int d^{\dcft}x_1 \langle \cO (x) \cO (0) T_{00} (x_1)\rangle = \int d^{\dcft}x_1 \left[ \frac{\left( \frac{t-t_1}{(x-x_1)^2} + \frac{t_1}{x_1^2} \right)^2}{x_1^{d-1} (x-x_1)^{d-1} } - \frac{1}{d+1}  \frac{x^2}{x_1^{d+1} (x-x_1)^{d+1} } \right].
\end{equation}

Note that if we were to evaluate the integral over the second term, it would manifestly have a logarithmic divergence. The same holds for the first term as well. We will relate this integral of the three point function to time dilation in the CFT, so we expect it to have no divergences and the logarithmic divergences just mentioned should cancel.

We will present briefly some steps in one way to evaluate this integral.
The most important ingredient is the calculation of the following integrals, which we computed in momentum space:
\begin{equation}
\begin{split}
& \int d^{\dcft } x' \frac{1}{(x')^a (x-x')^b} = \pi ^{\dcft /2} x^{\dcft - a - b} \frac{\Gamma \left(\frac{\dcft -a}{2} \right) \Gamma  \left( \frac{\dcft -b}{2} \right) \Gamma  \left( \frac{a+b - \dcft }{2} \right)}{\Gamma \left(\frac{a}{2} \right) \Gamma \left( \frac{b}{2} \right) \Gamma \left( \frac{2\dcft -a-b }{2} \right) } , \\
& \int d^{\dcft } x' \frac{x'_i (x-x')_i}{(x')^a (x-x')^b} =\frac{\pi ^{\dcft /2} }{4} (\dcft +4-a-b) \frac{\Gamma \left( \frac{\dcft -a}{2} +1\right)\Gamma \left(\frac{\dcft - b}{2} +1\right) \Gamma \left( \frac{a+b-\dcft -4}{2} \right)}{\Gamma \left( \frac{a}{2} \right) \Gamma \left( \frac{b}{2} \right) \Gamma \left( \frac{2\dcft +4-a-b}{2} \right)}  \cdot \\
& \qquad\qquad\qquad\qquad\qquad\qquad \cdot x^{\dcft -a -b } \left[ x^2+(\dcft +2-a-b)x_i^2 \right] , \\
& \int d^{\dcft } x' \frac{(x'_i)^2}{(x')^a (x-x')^b} =  \frac{\pi ^{\dcft /2} }{2} \frac{\Gamma \left(\frac{\dcft - a}{2} +1\right) \Gamma \left( \frac{\dcft -b}{2} \right) \Gamma \left(\frac{a+b-\dcft -2}{2} \right)}{\Gamma \left(\frac{a}{2} \right) \Gamma \left( \frac{b}{2} \right) \Gamma \left( \frac{2\dcft +2-a-b}{2} \right)} \cdot \\
& \qquad\qquad\qquad\qquad\qquad\qquad \cdot x^{\dcft + 2-a-b} \left[1 + \frac{a-\dcft -2}{2\dcft +2-a-b} \left(1+ (\dcft +2-a-b) \frac{x_i^2}{x^2} \right)\right] .
\end{split}
\end{equation}
Here $x_i$ is a fixed component out of the $\dcft $ components of $x$.

These are the ingredients in the integral that we want to evaluate. As was mentioned, we cannot directly use those since the separate integrals diverge. What we can do instead is to take $\dcft \to \dcft +\epsilon $ as an intermediate regularization, use the formulae just mentioned, and after summing the terms take the limit $\epsilon  \to 0$. This gives indeed a finite result
\begin{equation}
\int d^{\dcft}x_1 \left[ \frac{\left( \frac{t-t_1}{(x-x_1)^2} + \frac{t_1}{x_1^2} \right)^2}{x_1^{d-1} (x-x_1)^{d-1} } - \frac{1}{d+1}  \frac{x^2}{x_1^{d+1} (x-x_1)^{d+1} } \right] = 
- \frac{2d \pi ^{(d+1)/2} }{(d+1)\Gamma \left( \frac{d+3}{2} \right)} \frac{\vecx^2+t^2-(1+d)t^2}{(\vecx^2+t^2)^{(d+1)/2} }.
\end{equation}
Taking the canonical EM tensor, we can use the Ward identity \eqref{eq:Lifshitz_WI_for_cOOT} for $c_{\cO \cO T} $, and this gives for the integral over $\langle\cO \cO T_{00} \rangle$
\begin{equation}
c_{\cO \cO } \frac{2\Delta }{d+1} \frac{\vecx^2+t^2 - (1+d)t^2}{(\vecx^2+t^2)^{\Delta +1} } .
\end{equation}
Clearly if we were calculating $ \sum _{\mu } \langle \cO \cO T_{\mu \mu } \rangle$ we would get 0 here, consistent with the formula \eqref{eq:Lifshitz_T_in_OO_OOT} we started from.  However, \eqref{eq:Lifshitz_T_in_OO_OOT} does not include the contact terms
\begin{equation}
\langle T_{\mu \mu }(x) \cO (x_1) \cdots \cO (x_k) \rangle = - \sum _i \delta (x-x_i)\Delta  \langle \cO (x_1) \cdots \cO (x_k) \rangle .
\end{equation}
The contact term from each $\mu $ should be the same and therefore we expect in $\langle \cO (x) \cO (0) T_{00} (x_1)\rangle$ to have the contact term $ - \frac{\Delta }{d+1} \left( \delta (x_1-x)+\delta (x_1) \right) \langle \cO (x) \cO (0) \rangle$. The integral of this over $x_1$ gives the additional contribution $-\frac{2\Delta }{d+1} \cdot \frac{c_{\cO \cO } }{x^{2\Delta } } $. Together with the explicit integral we evaluated, we find
\begin{equation}\label{eq:eseven}
\int d^d\vecx_1 dt_1 \langle \cO (\vecx,t) \cO (0) T_{00} (\vecx_1,t_1) \rangle = - 2 \Delta c_{\cO \cO }   \frac{t^2}{(\vecx^2+t^2)^{\Delta +1} } .
\end{equation}

This is consistent with \eqref{eq:Lifshitz_correlators_relation_to_time_dilation}. The integral over the position of the EM tensor that we evaluated can be extracted from the coefficient of the term proportional to $\varepsilon $ in
\begin{equation}
\langle e^{\varepsilon \int d^d\vecx_1 dt_1\, T_{00} (\vecx_1,t_1)} \cO (\vecx,t) \cO (0) \rangle .
\end{equation}
The expectation value is evaluated in the CFT. As in \eqref{eq:Lifshitz_correlators_relation_to_time_dilation}, we expect the $\varepsilon $ term to be the same as
\begin{equation}
t \pder{}{t} \langle \cO (\vecx,t  ) \cO (0) \rangle = t \pder{}{t} \frac{c_{\cO \cO } }{(\vecx^2 + t^2 )^{\Delta } } = \frac{ - 2\Delta \cdot  t^2 c_{\cO \cO } }{(\vecx^2+t^2)^{\Delta +1 } } 
\end{equation}
which is indeed what we have found for the integrated three point function.

\bibliographystyle{utphys2.bst}
\bibliography{disorder}

\end{document}